\documentclass[journal=jacsat,manuscript=article, 
]{achemso}
\usepackage{graphicx} 
\usepackage{subcaption} 
\usepackage{float}
\usepackage{hyperref}
\usepackage{multirow}
\usepackage{chemformula} 
\usepackage[T1]{fontenc} 
\usepackage{braket}
\usepackage{booktabs}
\usepackage{makecell}
\usepackage{array}

\newcommand{\tallspace}{\rule{0pt}{18pt}}

\SectionNumbersOn

\author{Mariia Sapova} 
\affiliation{Department of Chemistry and Pharmaceutical Sciences, Faculty of
Sciences, Vrije Universiteit Amsterdam, De Boelelaan 1108, 1081 HZ Amsterdam, The Netherlands}

\author{Chandan Kumar} 
\affiliation{Van’t Hoff Institute for Molecular Sciences, University of Amsterdam, Science Park 904,
1098 XH Amsterdam, The Netherlands}

\author{Sahar Ashtari-Jafari} 
\affiliation{Department of Chemistry and Pharmaceutical Sciences, Faculty of
Sciences, Vrije Universiteit Amsterdam, De Boelelaan 1108, 1081 HZ Amsterdam, The Netherlands}

\author{Wybren J. Buma}

\email{W.J.Buma@uva.nl}

\affiliation{Van’t Hoff Institute for Molecular Sciences, University of Amsterdam, Science Park 904,
1098 XH Amsterdam, The Netherlands}

\affiliation{Van’t Hoff Institute for Molecular Sciences, University of Amsterdam, Science Park 904,
1098 XH Amsterdam, The Netherlands}
\alsoaffiliation[Radboud]{Radboud University, Institute for Molecules and Materials, FELIX Laboratory, Toernooiveld 7c, 6525 ED Nijmegen, The Netherlands}

\author{Lucas Visscher}

\email{L.Visscher@vu.nl}

\affiliation{Department of Chemistry and Pharmaceutical Sciences, Faculty of
Sciences, Vrije Universiteit Amsterdam, De Boelelaan 1108, 1081 HZ Amsterdam, The Netherlands}

\title{Quantitative agreement between experiment and theory for Vibrational Circular Dichroism enhanced by electronically excited states 
}

\begin{document}

\maketitle

\begin{abstract}

Intensity enhancement in vibrational circular dichroism (VCD) arises in open‐shell transition metal complexes from coupling between ground-state vibrational transitions and magnetic dipole-allowed transitions to low‐lying excited states (LLESs). In this work we apply Nafie’s vibronic coupling theory to M(II)-(-)-sparteine-Cl$_2$ (M=Zn, Co, Ni) complexes to investigate these enhancement effects. We show that the VCD intensity is extremely sensitive to the excitation energies that neither time-dependent density functional theory (TDDFT) nor state-averaged complete active space self consistent field (SA-CASSCF) calculations can predict with sufficient accuracy. We argue that instead of using more accurate quantum chemistry methods these excitation energies can be treated as parameters and optimized against experimental spectra. With this approach we obtain simulated VCD similarity scores above 0.4, a threshold considered reliable for absolute configuration assignment. The ability to quantitatively reproduce enhanced experimental spectra with computations opens up new research areas, offering amongst else unique possibilities for the study of chiral structure of systems such as transition metal complexes and metalloproteins that so far remained intractable.

\end{abstract} 

\section{Introduction}

Natural products and other molecules of interest for drug discovery often contain one or more chiral centers that are crucial for their biological activity. This characteristic makes it essential to be able to determine their absolute configuration (AC) or enantiomeric composition~\cite{Nafie:1976,Nafie:2011-1,BARRON2010,Magyarfalvi:2011,Stephens:2008}.
Electronic circular dichroism (ECD) spectroscopy is a spectroscopic method often employed to obtain insight, but its structural information content is relatively limited. Vibrational circular dichroism (VCD) offers in this respect significantly more information.\cite{Gorecki:2021}. 
VCD spectra are feature-rich, provide detailed information of the chemical bonds close to the chiral centers, and give insight into intermolecular interactions~\cite{Merten2017}. 
Many reviews cover both experimental and theoretical VCD methods \cite{BARRON2010,Nafie:2011,Nafie:2011-1,Magyarfalvi:2011,KUROUSKI201754,Ruud:2012}, but they typically focus on closed-shell systems for which the electronic ground state is well separated from excited states.
In those cases the VCD spectrum can be efficiently modeled using the magnetic field perturbation (MFP) theory of Stephens\cite{Stephens:1985} although the uncertainty in conformer energies can affect the reliability of the stereochemical assignment \cite{Koenis:2019}. 

For chiral molecules with low-lying excited states (LLES), as is typically the case in transition metal complexes, the situation is different. Such molecules are highly relevant in chiral material science and asymmetric catalysis~\cite{Bour:2015,CAO:2015} and there is therefore a growing interest in applying VCD spectroscopy to probe their chiral structure ~\cite{Merten:2012,Szilv:2013,Mazzeo:16,Pescitelli:2018,Sato:2018}. Due to the presence of these low-lying electronic states the vibrational and electronic degrees of freedom of the molecule are not well separated, which has been found to lead to significant enhancements of VCD signals. The phenomenon of enhanced VCD signals in transition metal complexes has been known since 1980, when the first experimental VCD spectra for ($-$)-sparteine complexes M(II)(sp)Cl$_2$ (M = Zn, Co, Ni) were reported in a narrow region of CH stretching modes~\cite{Barnett_1980}. Later He et al. extended the measurement to the fingerprint region, where the open-shell Co(II) and Ni(II) complexes have about ten times higher VCD intensity compared to the closed-shell Zn complex~\cite{He:2001}. Pescitelli and Di Bari recently published a comprehensive review compiling available data on VCD enhancement and classifying the systems that exhibit this phenomenon~\cite{Pescitelli_review}. They combined the systems with an open-shell metal center into a separate group, which contains mononuclear and multinuclear transition metal complexes, lanthanide complexes, and heme proteins. In most cases, the enhancement can be attributed to the coupling of vibrational and electronic excitations; however, other mechanisms can also come into play~\cite{Pescitelli_review}. 
Domingos et al. reported VCD spectra for ligands attached to paramagnetic Co(II) ~\cite{Domingos:12,Domingos:14} for which enhancements up to two orders of magnitude were observed. They showed that isostructural Co(III) complexes do not exhibit this intensity amplification, providing experimental evidence that the observed enhancement originates from vibronic coupling with LLES. In complexes with Schiff base ligands, a sign reversal effect was observed along with the enhanced VCD signals, and this phenomenon was also attributed to the presence of LLES in a series of studies by Pescitelli~\cite{Pescitelli_co_salen_2018, Pescitelli_co_salen_sym_2019, Gorecki_co_salen_sign_2021}. Enhancement due to LLES has also been observed in M(III) complexes: for example, Ru(III)(acac)$_3$ studied by More et al.~\cite{Mori_2011_Ru_lles}, and a tenfold enhancement was reported for [Co(III)stien(biur)$_2$]NBu$_4$~\cite{Johannessen_CoIII}.

The enhanced intensity provides two practical benefits. First, the overall sensitivity of VCD improves significantly. Arrico et al., for example.  applied this mechanism to enable a rapid and sensitive chirality recognition in $\alpha$-amino acids~\cite{Lorenzo:2017}. Second, the enhancement is distance-dependent as was shown in the studies of Domingos et al. who demonstrated that amplification occurs locally within specific regions of the molecule, providing a way to “zoom in” on particular parts of a biomolecular system~\cite{Domingos_2014_angewandte}. However, in order to be able to use these potential applications, a reliable computational approach is needed to analyze and predict the observed enhanced VCD spectra. As yet, such an approach has been notoriously lacking. 

Nafie developed the theoretical description of VCD for molecules with LLES, employing a sum‐over‐states (SOS) formalism that incorporates coupling between vibrational normal modes and LLES~\cite{Nafie:2004}.
Although Nafie's VCD vibronic theory ~\cite{Nafie:2004} provides a rigorous framework for vibronic enhancement, it has not been implemented in major quantum chemical software packages, and to date no \textit{ab initio} calculations have been published that successfully reproduce enhanced VCD spectra. Tomeček and Bour suggested an alternative approach and studied Co(II)-(-)(sparteine)Cl$_2$ and bis[(S)-N-(1-phenylethyl)salicylaldiminato]$\Delta$-Co(II) complexes. They employed a Herzberg-Teller expansion of the electronic wave function and incorporated nonadiabatic contributions in a model Hamiltonian, followed by full diagonalization of the Hamiltonian on the basis of adiabatic wave functions. Although their approach demonstrated clear enhancements, it was still unable to capture the correct signs for several VCD bands. In Co(sp)Cl$_2$, for example, the calculated spectrum reproduced the overall band shape reasonably well, but the predicted enhancement remained roughly three times smaller than what was observed experimentally. Similarly, for bis[(S)-N-(1phenylethyl)salicylaldiminato]$\Delta$-cobalt(II), the characteristic monosignate VCD profile observed in the experiment could not be reproduced by the calculations. 

In this work, we adopt Nafie’s vibronic coupling theory of VCD and argue that previous attempts to predict enhanced VCD failed because existing quantum chemistry methods cannot describe low‐lying excited states with the required accuracy. Moreover, we discuss that the commonly accepted assumption that the potential energy surfaces of the LLES do not differ significantly from that of the electronic ground state may not always hold and can influence significantly the predicted spectrum. We propose a computational protocol to address these limitations and demonstrate its effectiveness for Me(II)–($-$)‐sparteine complexes.
The article is organized as follows. In Section \ref{sec:VCD-MFP-method}, we review Stephens' MFP formalism for VCD. Nafie’s vibronic coupling theory is summarized in Section \ref{sec:intensity-enhanced}. Computational details are provided in Section \ref{sec:comp_details}. In Section \ref{sec:dicussion}, we present and discuss our results for Co(II) and Ni(II) sparteine complexes, and compare them with the spectrum predicted for the closed-shell Zn(II) sparteine complex. We therein show that -- despite the uncertainties in excitation energy prediction by state-of-the-art quantum chemical methods and neglect of Franck-Condon effects -- a reliable identification of absolute chirality is possible. This opens the way for application of VCD as a sensitive tool for structure determination of a much broader range of compounds than has been possible with MFP.

\section{Magnetic Field Perturbation theory (MFP)}\label{sec:VCD-MFP-method}

In Stephens MFP formalism for VCD~\cite{Stephens:1985}, the VCD rotational strength $R_i$ is written as~\cite{Stephens:1985,Stephens:1987,Nicu2008}
\begin{equation}
R_{i}= \textup{Im} [\mathbf{E_i}^{\textup{tot}} \cdot {\mathbf{M_i}}^{\textup{tot}}],
\end{equation}

\noindent where $i$ is the $i^{th}$ normal mode, and $\mathbf{E}^{\textup{tot}}$ and ${\mathbf{M}}^{\textup{tot}}$ are the total electric and magnetic dipole transition moments, respectively. The  $\mathbf{E}^{\textup{tot}}$ and ${\mathbf{M}}^{\textup{tot}}$ vectors are defined as

\begin{equation} \label{eq:EDTM}
E^{\textup{tot}}_{\beta,i}={\Bigg({\frac{\hbar}{{\omega}_{i}}}\Bigg)}^{\frac{1}{2}} \sum_{\lambda\alpha}P_{\alpha\beta}^{\lambda}S_{\lambda\alpha,i}
\end{equation}

\begin{equation} \label{eq:MDTM}
M^{\textup{tot}}_{\beta,i}  = -{({2{\hbar}^{3}{\omega}_{i}})}^{\frac{1}{2}} \sum_{\lambda\alpha}A_{\alpha\beta}^{\lambda}S_{\lambda\alpha,i},
\end{equation}

\noindent where $\mathbf{S}$ is the transformation matrix from Cartesian ($\lambda\alpha$) to normal ($i$) nuclear coordinates, and $\mathbf{P}$ and $\mathbf{A}$ are the atomic polar tensor (APT) and the atomic axial tensor (AAT), respectively. $\omega_{i}$ is the harmonic angular frequency of the $i^{th}$ mode, $\lambda$ is used to enumerate the nuclei and $\alpha$ and $\beta$ indicate Cartesian components of either the nuclear displacement vectors or the electromagnetic field, respectively. The APT and AAT tensors can be written as a sum of electronic and nuclear contributions. The APT tensor is then given by 

\begin{equation}
P_{\alpha\beta}^{\lambda} = E_{\alpha\beta}^{\lambda} + N_{\alpha\beta}^{\lambda}
\end{equation}

\noindent with the electronic contribution given by

\begin{equation}
E_{\alpha\beta}^{\lambda} = {\frac{\partial \langle \Psi_{g}(\mathbf{R}) | \mu_{\beta}| \Psi_{g}(\mathbf{R})\rangle }{\partial R_{\lambda\alpha}}\Bigg | }_{\mathbf{R}=\mathbf{R}_{0}}
\end{equation}
and the nuclear contribution by
\begin{equation}
N_{\alpha\beta}^{\lambda} = e Z_{\lambda} {\delta}_{\alpha\beta}.
\end{equation}

The AAT tensor is given by

\begin{equation}
A_{\alpha\beta}^{\lambda} = I_{\alpha\beta}^{\lambda} + J_{\alpha\beta}^{\lambda},
\end{equation}

\noindent where $\mathbf{I}^{\lambda}$ is the electronic contribution and $\mathbf{J}^{\lambda}$ the nuclear contribution. The electronic contribution is given as the overlap of the perturbed wave functions with respect to nuclear displacements and with respect to  magnetic field 

\begin{equation}
\label{AAT-stephan-eq}
I_{\alpha\beta}^{\lambda} = {\Bigg \langle \frac{\partial \Psi (\mathbf{R},\mathbf{B})}{ \partial R_{\lambda \alpha}} \Bigg| \frac{\partial \Psi(\mathbf{R},\mathbf{B})}{ \partial B_{\beta}}    \Bigg \rangle \Bigg |}_{\mathbf{R}=\mathbf{R}_{0}, \mathbf{B}=0}
\end{equation}

\noindent with a nuclear contribution given by

\begin{equation}
 J_{\alpha\beta}^{\lambda}= i \frac{e Z_{\lambda}}{4 \hbar c} \sum_{\gamma} {\epsilon}_{\alpha \beta \gamma} R_{\lambda \gamma}^{0},
\end{equation}

\noindent where ${\epsilon}_{\alpha \beta \gamma}$ is the Levi-Civita symbol.

\section{Nafie's vibronic theory of VCD with LLES}\label{sec:intensity-enhanced}

Stephens' efficient MFP formalism has been implemented in many quantum chemistry programs~\cite{Stephens:1985,Stephens:1987,Stephens:2008} but does not account for vibronic coupling with LLESs. Equations to express this coupling in terms of additional LLESs contributions to the APT and the AAT tensors have been derived by Nafie~\cite{Nafie:2004} and will be employed here. We refer to this paper for the full derivation and will only briefly discuss the SOS formulation of the APT and the AAT tensors. 

In Nafie's SOS formulation of vibronic coupling, the APT tensor ($E_{\alpha\beta}^{A}(\omega_{a})$) is written as
\begin{equation}
\label{APT_lles_sos}
E_{\alpha\beta}^{\lambda}(\omega_{a}) =  2\sum_{e\neq g} 
\Bigg [ \frac{\omega_{eg}^{2}}{\omega_{eg}^{2} - \omega_{a}^{2}} \Bigg ] 
\langle {\Psi_g} |  \mu_{\beta}|  \Psi_e \rangle \langle \Psi_{e} |  {\frac{\partial \Psi_{g}}{\partial R_{\lambda\alpha}} \rangle\Bigg | _{\mathbf{R}=\mathbf{R}_{0}}}.
\end{equation}

\noindent On the RHS of Eq.~\ref{APT_lles_sos}, ${\Psi_{g}}$ is the ground state electronic wave function and ${\Psi_{e}}$ refers to electronically excited-state wave functions. The frequency corresponding to the vertical electronic excitation energy of state $e$  is denoted as $\omega_{eg}$ while the normal mode frequency is given as $\omega_{a}$. In this formalism, the APT is dependent on the normal mode frequency, but it can be related to the one defined by Stephens by writing Eq.~\ref{APT_lles_sos} in terms of two contributions, 

\begin{equation}
\label{APT_cont}
\begin{aligned}
E_{\alpha\beta}^{\lambda}(\omega_{a}) & = 2\sum_{e\neq g} 
\Bigg [ 1 + \frac{\omega_{a}^{2}}{\omega_{eg}^{2} - \omega_{a}^{2}} \Bigg ]
\langle {\Psi_g} |  \mu_{\beta}|  \Psi_e \rangle \langle \Psi_{e} |  {\frac{\partial \Psi_{g}}{\partial R_{\lambda\alpha}} \rangle\Bigg | _{\mathbf{R}=\mathbf{R}_{0}}}  \\ 
& = 2\sum_{e\neq g} \langle {\Psi_g} |  \mu_{\beta}|  \Psi_e \rangle \langle \Psi_{e} |  {\frac{\partial \Psi_{g}}{\partial R_{\lambda\alpha}} \rangle\Bigg | _{\mathbf{R}=\mathbf{R}_{0}}} \\
& + {\underbrace{2\sum_{e\neq g} \frac{\omega_{a}^{2}}{\omega_{eg}^{2} - \omega_{a}^{2}} \langle {\Psi_g} |  \mu_{\beta}|  \Psi_e \rangle \langle \Psi_{e} |  {\frac{\partial \Psi_{g}}{\partial R_{\lambda\alpha}} \rangle\Bigg | _{\mathbf{R}=\mathbf{R}_{0}}} }_{\textup{enhancement term}}}.
\end{aligned}
\end{equation}

\noindent The first term on the RHS of Eq.~\ref{APT_cont} is equivalent to the APT defined in the MFP formalism,

\begin{equation}
\label{APT_equivalence}
\begin{aligned}
& 2\sum_{e\neq g} \langle {\Psi_g} |  \mu_{\beta}|  \Psi_e \rangle \langle \Psi_{e} |  {\frac{\partial \Psi_{g}}{\partial R_{\lambda\alpha}} \rangle\Bigg | _{\mathbf{R}=\mathbf{R}_{0}}} \\
& = 2 \langle {\Psi_g} |  \mu_{\beta}|  {\frac{\partial \Psi_{g}}{\partial R_{A\alpha}} \rangle\Bigg | _{\mathbf{R}=\mathbf{R}_{0}}} - 2 \langle {\Psi_g} |  \mu_{\beta}|  \Psi_g \rangle \langle \Psi_{g} |  {\frac{\partial \Psi_{g}}{\partial R_{\lambda\alpha}} \rangle\Bigg | _{\mathbf{R}=\mathbf{R}_{0}}} \\
& = {\frac{\partial \langle \Psi_{g} | \mu_{\beta}| \Psi_{g}\rangle }{\partial R_{\lambda\alpha}}\Bigg | }_{\mathbf{R}=\mathbf{R}_{0}},
\end{aligned}
\end{equation}

\noindent since $\langle \Psi_g | \frac{\partial \Psi_g}{\partial R_{\lambda\alpha}} \rangle = 0$ and the electron dipole moment operator carries no direct dependence on nuclear coordinates. The enhancement can thus be solely attributed to the second term of Eq.~\ref{APT_cont} which depends both on the normal mode frequency and the electronic excitation energy.

Like the APT, the AAT in the SOS formalism depends on both the normal mode frequency and the electronic excitation energy. It takes the form 

\begin{equation}
\label{AAT_sos_ref}
\begin{aligned} 
I_{\alpha\beta}^{\lambda}(\omega_{a})  & =  \sum_{e\neq g} 
\Bigg [ \frac{\omega_{eg}^{2}}{\omega_{eg}^{2} - \omega_{a}^{2}} \Bigg ]
\frac{ \langle  \frac{\partial \Psi_g}{\partial R_{\lambda\alpha}}  | \Psi_e \rangle
\langle \Psi_e |  m_{\beta} |  \Psi_g \rangle }{{\omega}_{eg}}
\Bigg | _{\mathbf{R}=\mathbf{R}_{0}}, \\
\end{aligned}
\end{equation}

\noindent where $\mathbf{m}$ is the magnetic dipole operator and we have multiplied the original definition of Nafie by a factor of $\frac{i}{2}$ to be consistent with the one of Stephens\cite{Nafie:2011}. Eq.~\ref{AAT_sos_ref} can be partitioned as

\begin{equation}
\label{AAT_cont}
\begin{aligned} 
I_{\alpha\beta}^{A}(\omega_{a})  & =  \sum_{e\neq g} 
\Bigg [1+ \frac{\omega_{a}^{2}}{\omega_{eg}^{2} - \omega_{a}^{2}} \Bigg ]
\frac{ \langle  \frac{\partial \Psi_g}{\partial R_{A\alpha}}  | \Psi_e \rangle
\langle \Psi_e |  m_{\beta} |  \Psi_g \rangle }{{\omega}_{eg}}
\Bigg | _{\mathbf{R}=\mathbf{R}_{0}}  \\
& =  \sum_{e\neq g} \frac{ \langle  \frac{\partial \Psi_g}{\partial R_{A\alpha}}  | \Psi_e \rangle
\langle \Psi_e |  m_{\beta} |  \Psi_g \rangle }{{\omega}_{eg}}
\Bigg | _{\mathbf{R}=\mathbf{R}_{0}} \\
& + {\underbrace{ \sum_{e\neq g} \Bigg [\frac{\omega_{a}^{2}}{\omega_{eg}^{2} - \omega_{a}^{2}} \Bigg ]
\frac{ \langle  \frac{\partial \Psi_g}{\partial R_{A\alpha}}  | \Psi_e \rangle
\langle \Psi_e |  m_{\beta} |  \Psi_g \rangle }{{\omega}_{eg}}
\Bigg | _{\mathbf{R}=\mathbf{R}_{0}} }_{\textup{enhancement term}}}.
\end{aligned}
\end{equation}

\noindent The first term of Eq.~\ref{AAT_cont} can be related to Eq.~\ref{AAT-stephan-eq}:

\begin{equation}
\label{AAT_equivalence}
\begin{aligned}
I_{\alpha\beta}^{A} & = {\Bigg \langle \frac{\partial \Psi_g }{ \partial R_{A \alpha}} \Bigg| \frac{\partial \Psi_g}{ \partial B_{\beta}}    \Bigg \rangle \Bigg |}_{\mathbf{R}=\mathbf{R}_{0}, \mathbf{B}=0}\\
& = \sum_{e\neq g} \frac{ \langle {\frac{\partial \Psi_{g}}{\partial R_{A\alpha}} | \Psi_e \rangle   \langle \Psi_e | m_{\beta}|  \Psi_g \rangle }}{{\omega}_{eg}}\Bigg | _{\mathbf{R}=\mathbf{R}_{0}},
\end{aligned}
\end{equation}

\noindent while the second term of Eq.~\ref{AAT_cont} represents the enhancement. 

These equivalences make it possible to compute the VCD rotational strength as a sum of standard and enhancement contributions and thereby allows for detailed analysis. We define   

\begin{equation}
\label{enhancement}
\begin{aligned}
& R_i^{\textup{enh}}= \textup{Im} [\mathbf{E}_i^{\textup{tot}} \cdot \mathbf{M}_i^{\textup{enh}} + \mathbf{E}_i^{\textup{enh}} \cdot \mathbf{M}_i^{\textup{tot}}] \\
& E_{i,\beta}^{\textup{enh}} = 2 {\Bigg({\frac{\hbar}{{\omega}_{i}}}\Bigg)}^{\frac{1}{2}} \sum_{e\neq g} \frac{\omega_{i}^{2}}{\omega_{eg}^{2} - \omega_{i}^{2}} \Bigg [  \langle {\Psi_g} |  \mu_{\beta}|  \Psi_e \rangle \langle \Psi_{e} |  \frac{\partial \Psi_{g}}{\partial R_{i}} \rangle\Bigg ] _{\mathbf{R}=\mathbf{R}_{0}}  \\
& M_{i,\beta}^{\textup{enh}} = - {({2{\hbar}^{3}{\omega}_{i}})}^{\frac{1}{2}} \sum_{e\neq g} \frac{\omega_{i}^{2}}{\omega_{eg}^{2} - \omega_{i}^{2}} \Bigg [  
\frac{\langle {\Psi_g} |  m_{\beta}|  \Psi_e \rangle }{\omega_{eg}}
\langle \Psi_{e} |  {\frac{\partial \Psi_{g}}{\partial R_{i}} \rangle\Bigg ] _{\mathbf{R}=\mathbf{R}_{0}}} 
\end{aligned}
\end{equation}

\noindent The enhancement corrections can be calculated separately and require  evaluation of excitation energies, electric and magnetic dipole transition moments for the LLES as are readily available in most quantum chemistry codes. The required non-adiabatic couplings (NACs) $ \langle \Psi_{e} |  \frac{\partial \Psi_{g}}{\partial R_{A\alpha}} \rangle$ can also be obtained from a number of programs.

Nafie’s formulation employs the vertical approximation in which the ground and excited state potential energy surfaces are assumed to have the same shape and to differ only by a constant energy shift. Within this approximation, one has $\phi^a_{ev}=\phi^a_{gv}$ for all vibrational states (labeled $v$) of normal mode ($a$) of the electronic ground ($g$) and excited states ($e$). Consequently, overlap integrals of the form $\braket{\phi^a_{g0}|\phi^a_{ev}}$ are zero for all transitions except the fundamental $0 \rightarrow 0$ transition. As we show in the results section, however, this vertical approximation does not hold for all low-lying excited states of sparteine complexes and is one of the reasons for previously observed differences between calculated and observed spectra.

\section{Computational details}\label{sec:comp_details}

\begin{figure}[H]
    \centering
    \begin{subfigure}{0.45\textwidth}
        \centering
        \includegraphics[width=\textwidth]{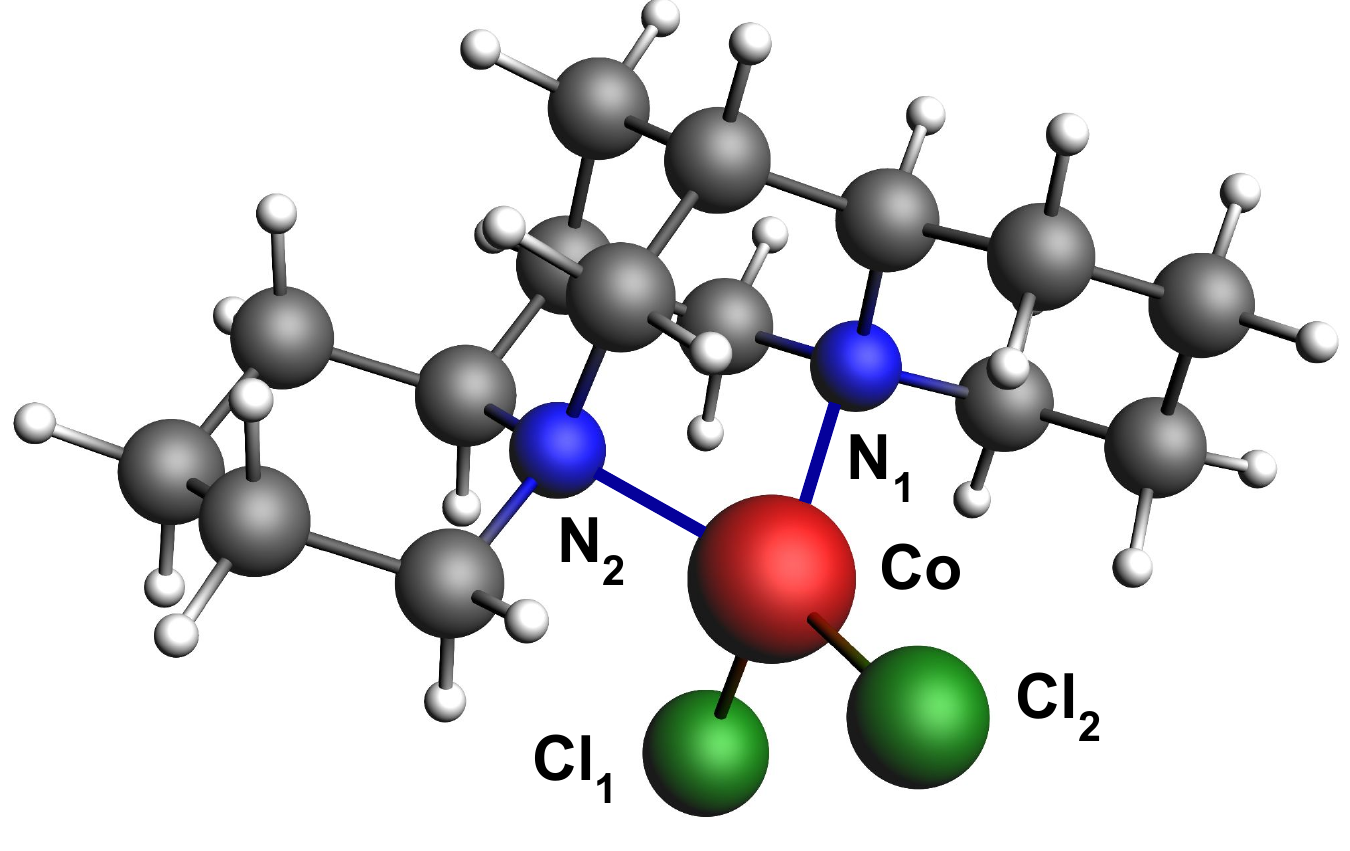}
    \end{subfigure}
    \begin{subfigure}{0.45\textwidth}
        \centering
        \includegraphics[width=\textwidth]{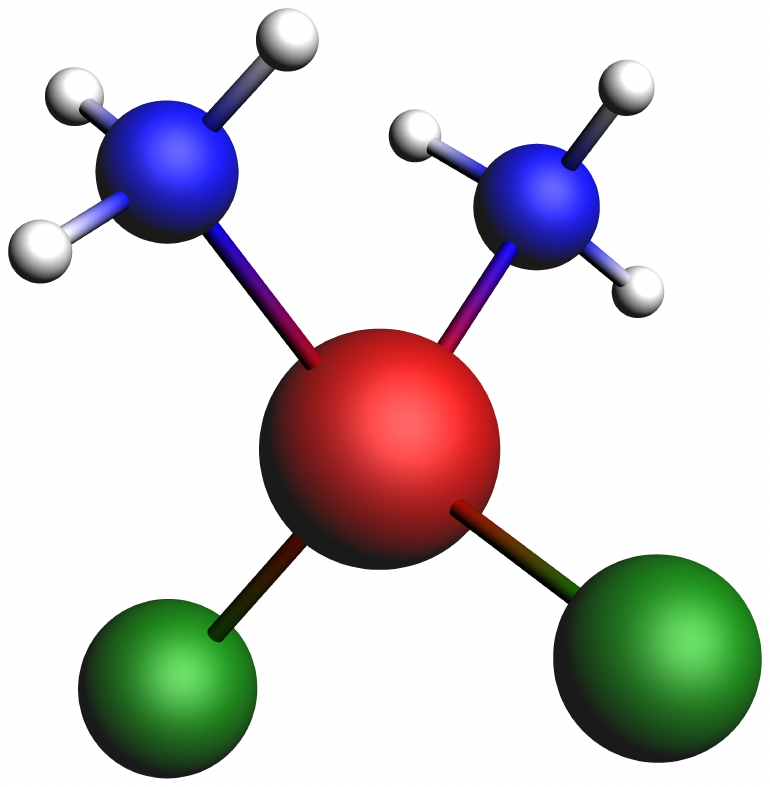}
    \end{subfigure}
    \caption{Co(sp)Cl$_2$ structure optimized on BP86-D3BJ/def2-TZVP level with CPCM treatment of CHCl$_3$ solvent effects (left) and the $C_{2v}$ Co(NH$_3$)$_2$Cl$_2$ model structure (right).}
    \label{fig:sparteine_complexes}
\end{figure}

We study the effects of low-lying excited states on the VCD spectra of the open-shell complexes Co(II)(sp)Cl$_2$ and Ni(II)(sp)Cl$_2$. The cobalt (d$^7$ electronic configuration) and nickel (d$^8$ electronic configuration) atoms form high-spin open-shell, four-coordinate complexes with the bidentate chiral (-)-sparteine ligand with the absolute configuration of 6R, 7S, 9S, 11S (~\autoref{fig:sparteine_complexes}). They can be compared with analogous zinc complexes where the electronic structure is that of a closed-shell configuration without unpaired spins with significantly higher excitation energies, and therefore negligible enhancement effects. 
To gain more insight about the electronic structure of the excited states, we also study model complexes with $C_{2v}$ symmetry.
These model complexes are constructed by truncating and symmetrizing the optimized sparteine complexes while preserving bond distances around the transition metal.

According to previous studies~\cite{Wiberg_sparteine_conformers}, the sparteine complexes are inherently rigid as a result of the ligand structure itself. To provide an independent verification of this finding, we use CREST~\cite{crest} within the Amsterdam Modeling Suite (AMS)~\cite{AMSlink} to generate initial conformers of Zn(sp)Cl$_2$ employing the GFN1-xTB method~\cite{GFN1-xTB}. Conformers within 5 kcal/mol of the lowest energy structure are subsequently reoptimized at the DFT level using BP86~\cite{BP86_becke, BP86_perdew}-D3BJ~\cite{grimme2010_d3, grimme2011_damping}/def2-TZVP with the  conductor-like polarizable continuum model (CPCM)~\cite{cpcm} to model CHCl$_3$ solvent effects. All the IR and MFP VCD calculations are performed on the same level of theory for closed-shell Zn(sp)Cl$_2$ and open-shell Co(sp)Cl$_2$ and Ni(sp)Cl$_2$ complexes.

We calculate excited states of open-shell complexes using CASSCF and TDDFT, including dipole transition moments (EDTM/MDTM) and non-adiabatic couplings.
TDDFT calculations are performed with ORCA 6.0~\cite{ORCA}. We use a def2-TZVP~\cite{def2_basis_set} basis set with def2/JK~\cite{def2_jk} auxiliary sets and extract electric and magnetic transition dipole moments (EDTM/MDTM) and non-adiabatic couplings for further enhancement calculations. CASSCF calculations are performed with both ORCA 6.0 and Openmolcas v23.06~\cite{openmolcas}. In ORCA, we perform state-averaged (SA-) and state-specific (SS-) CASSCF on sparteine and model $C_{2v}$ complexes with the def2-TZVP basis set and the def2/JK auxiliary fitting. OpenMolcas is employed specifically for its implementation of non-adiabatic coupling vectors at the SA-CASSCF level. There we use the ANO-RCC-VTZP basis on the metal~\cite{ano_rcc_tm}, N, and Cl atoms ~\cite{ano_rcc_main}, and the minimal ANO-RCC-MB basis on C and H.

We calculate enhanced VCD spectra using our developed Python implementation of Nafie's vibronic coupling theory described in Section \ref{sec:intensity-enhanced}. To quantify the agreement of the calculated IR and VCD spectra with the observed ones we use Shen~\cite{Shen_2010} simIR and simVCD metrics. To this end, we digitized the experimental IR and VCD spectra for all sparteine complexes from the thesis of He~\cite{He_thesis} using a Python script.
A frequency scaling factor is obtained by maximizing the simIR between the computed and experimental IR spectra over the 950–1500 cm$^{-1}$ window. The same factor is then applied to the calculated VCD spectra.

\section{Results and discussion}\label{sec:dicussion}

\subsection{Magnetic field perturbation calculations}

Zn(II) has a  closed‑shell $d^{10}$ configuration with high-lying excited states. Enhancement effects are in this case thus negligible and the MFP approximation is fully justified. As a result, this complex serves as a good benchmark for MFP‐based VCD calculations. The IR spectra of all three studied M(sp)Cl$_2$ (M = Zn, Co, Ni) are almost identical, so the Zn(sp)Cl$_2$ complex can be used to select the most appropriate DFT functional and to evaluate conformational effects. Conformational analysis reveals an energy difference between the two lowest energy conformers of Zn(sp)Cl$_2$ of 4.2 kcal/mol which exceeds the value of 3 kcal/mol typically used to select contributing conformers. This is in agreement with previous studies of the sparteine ligand ~\cite{Wiberg_sparteine_conformers} where it was also found that Me(II)(sp)Cl$_2$ complexes exist predominantly in a single conformation, which  makes the systems ideal for analysing their VCD spectra in detail.

Several benchmark studies have shown that the generalized gradient approximation (GGA) functional BP86~\cite{BP86_becke, Perdew:1986}
outperforms B3LYP~\cite{b3lyp_stephens}
in reproducing experimental equilibrium geometries of small first-row transition metal complexes \cite{Buhl_tm_benchmark_2006, Jensen_tm_bond_length_benchmark}. Minenkov et al.\cite{Minenkov_ru_complexes_benchmark_2012} furthermore reported that B3LYP systematically overestimates the metal-ligand bonds in transition metal complexes of the first and second row. A more recent benchmark for metal–dinitrogen complexes \cite{Zhao_dft_benchmark_MetN2} further showed that the performance of DFT is highly sensitive to the nature of the N$_2$ coordination mode at the metal center. In view of these observations and to establish the optimal computational approach for the present system we compare in the following the performance of three DFT functionals (BP86, B3LYP, PBE0)~\cite{pbe0}
with and without Grimme’s D3(BJ) dispersion correction ~\cite{grimme2010_d3, grimme2011_damping} in the gas phase and with CPCM of CHCl$_3$ ~\cite{cpcm} (see detailed plots in the SI).

\begin{figure}[H]
    \centering
    \begin{subfigure}{0.49\textwidth}
        \centering
        \includegraphics[width=\textwidth]{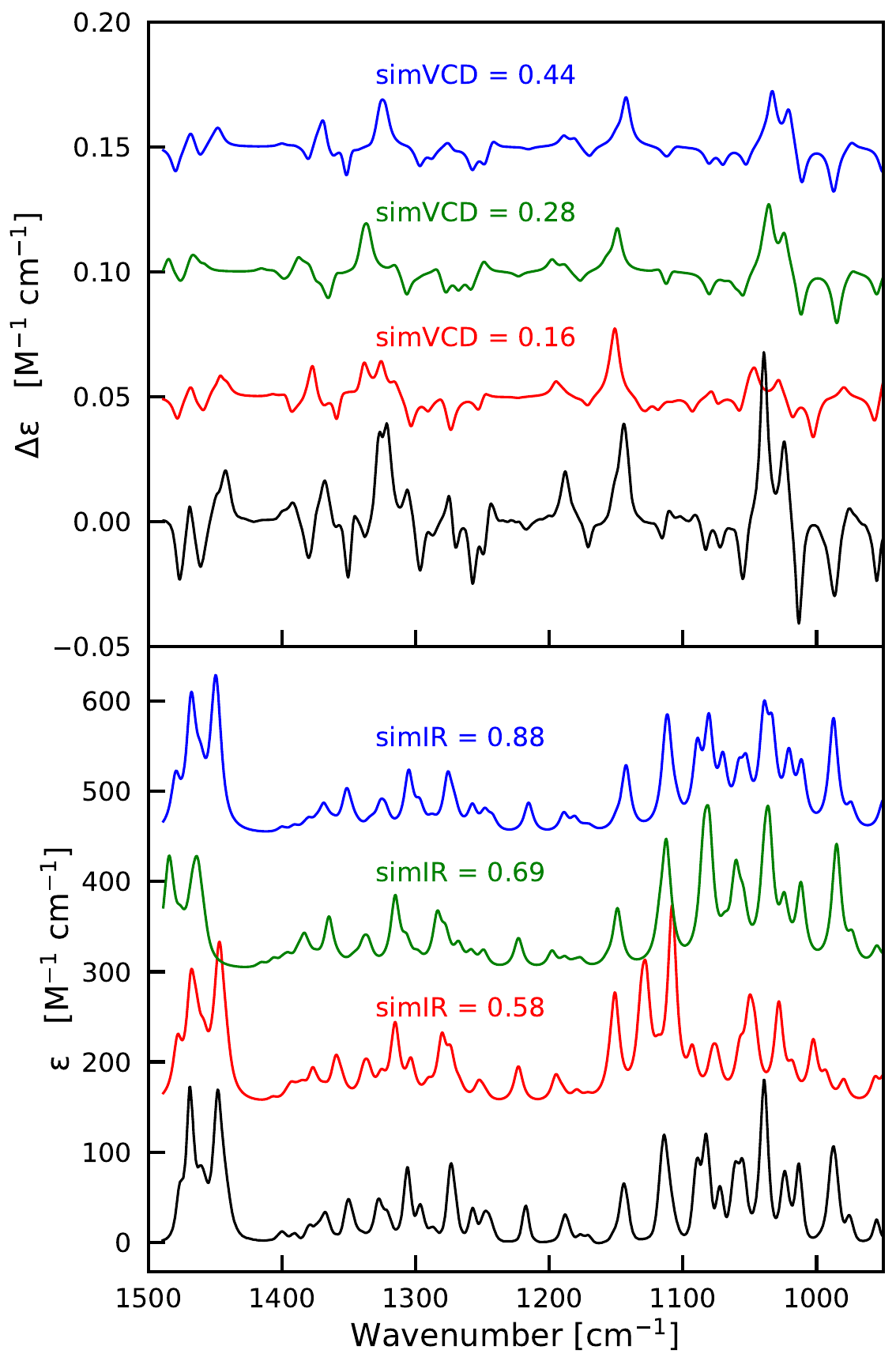}
    \end{subfigure}
    \caption{Comparison of VCD (top) and IR (bottom) spectra of Zn(sp)Cl$_2$. Experimental spectra are given in black, the spectra are calculated with CPCM of CHCl$_3$ with PBE0/def2-TZVP (red, simVCD = 0.16), B3LYP/def2-TZVP (green, simVCD = 0.28), and BP86/def2-TZVP (blue, simVCD = 0.44).}
    \label{fig:zn_cpcm}
\end{figure}

\begin{figure}[H]
    \centering
    \begin{subfigure}{0.45\textwidth}
        \centering
        \includegraphics[width=\textwidth]{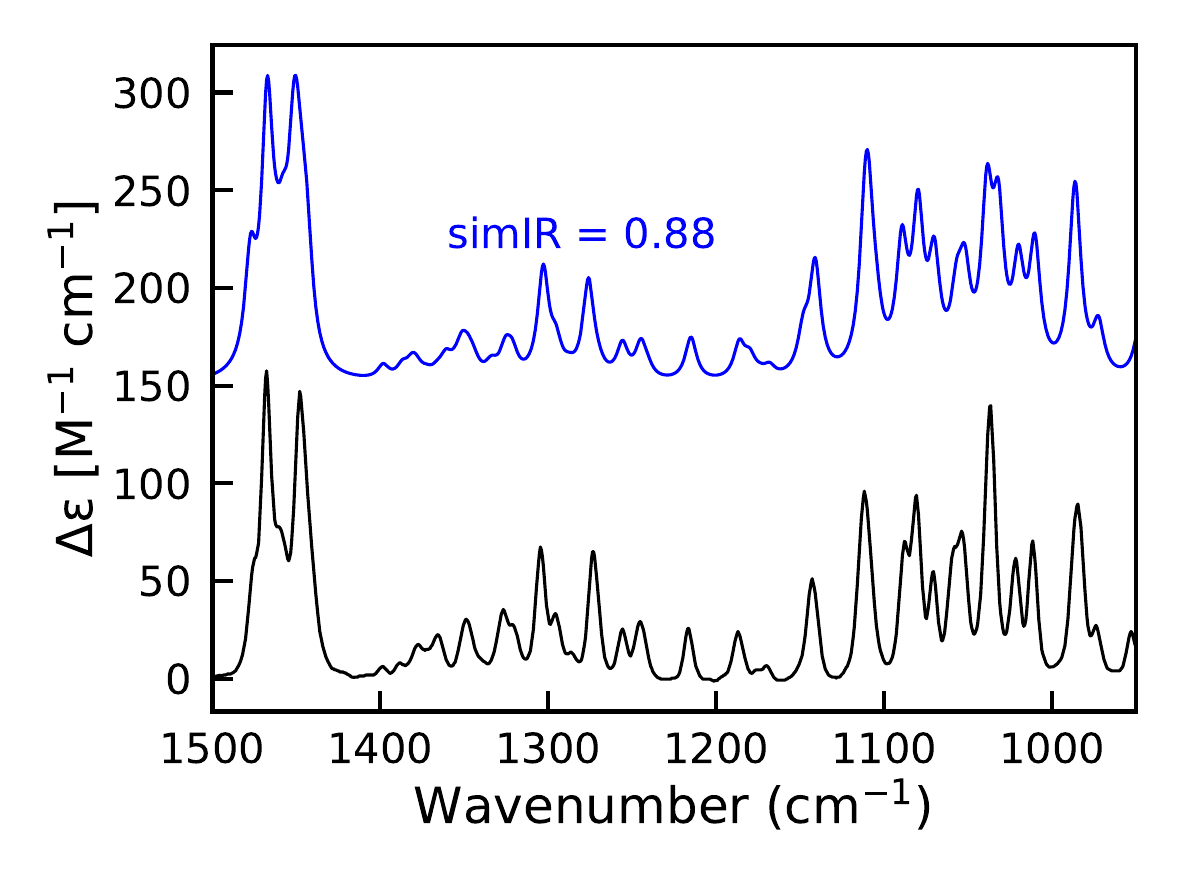}
    \end{subfigure}
    \begin{subfigure}{0.45\textwidth}
        \centering
        \includegraphics[width=\textwidth]{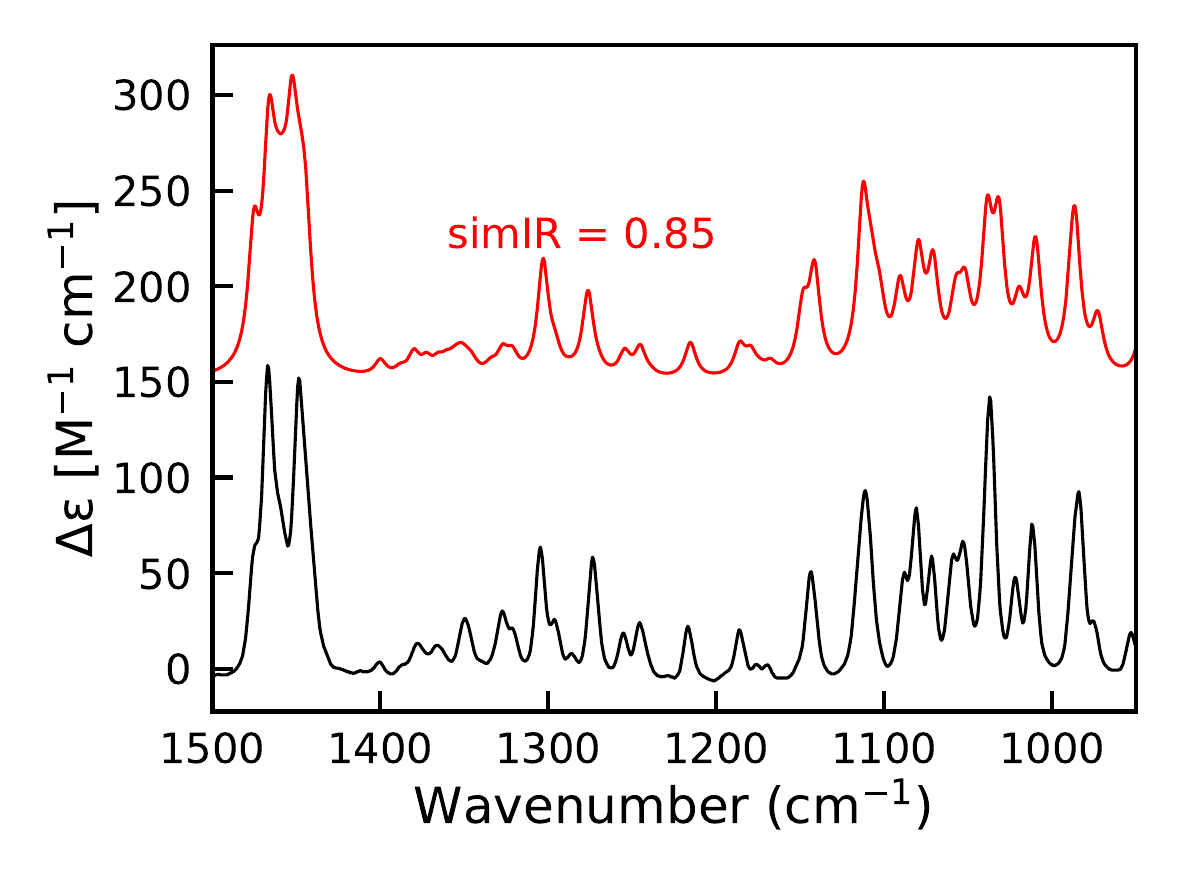}
    \end{subfigure}
    \caption{Left panel: calculated (blue, simIR = 0.88) and observed (black) IR spectra of Co(sp)Cl$_2$. Right panel:   calculated (red, simIR = 0.85) and observed (black) IR spectra of Ni(sp)Cl$_2$. All calculations have been performed with BP86-D3BJ/def2-TZVP with CPCM of CHCl$_3$.}
    \label{fig:ir_spectra}
\end{figure}

\autoref{fig:zn_cpcm} shows calculated IR and MFP-VCD spectra of Zn(sp)Cl$_2$ with the considered functionals. The addition of the dispersion correction improves both simIR and simVCD for BP86 and B3LYP (see Figure S1 in SI). BP86‑D3BJ/def2‑TZVP with CPCM demonstrates the best agreement with the experiment with good similarities for both IR and VCD (simIR = 0.88, simVCD = 0.44), noticing that simVCD values larger than 0.4 are typically considered to be secure to assign the absolute configuration~\cite{Koenis_conformer_energies}. ~\autoref{fig:ir_spectra} shows that this level of theory predicts IR spectra for the open-shell Co(sp)Cl$_2$ and Ni(sp)Cl$_2$ systems  that also have a high similarity to the experiment. We thus conclude that calculations at the BP86‑D3BJ/def2‑TZVP with CPCM level of theory are suitable to optimize the ground state structure of these compoundsand compute their IR spectra .

\subsection{Excited state calculations}

We next investigate with TDDFT and SA-CASSCF the excitation energies and nature of LLES for the Co(II)(sp)Cl$_2$ and Ni(II)(sp)Cl$_2$ complexes. 
The contributions of each excited state to the enhanced VCD spectra depend on the EDTM, MDTM, excitation energy, nonadiabatic coupling matrix elements, and vibrational frequencies, so the quality of all these quantities is important to obtain a reliable estimate of the enhancement that can be expected in VCD. 

\begin{figure}[H]
    \centering
    \begin{subfigure}{0.9\textwidth}
        \centering
        \includegraphics[width=\textwidth]{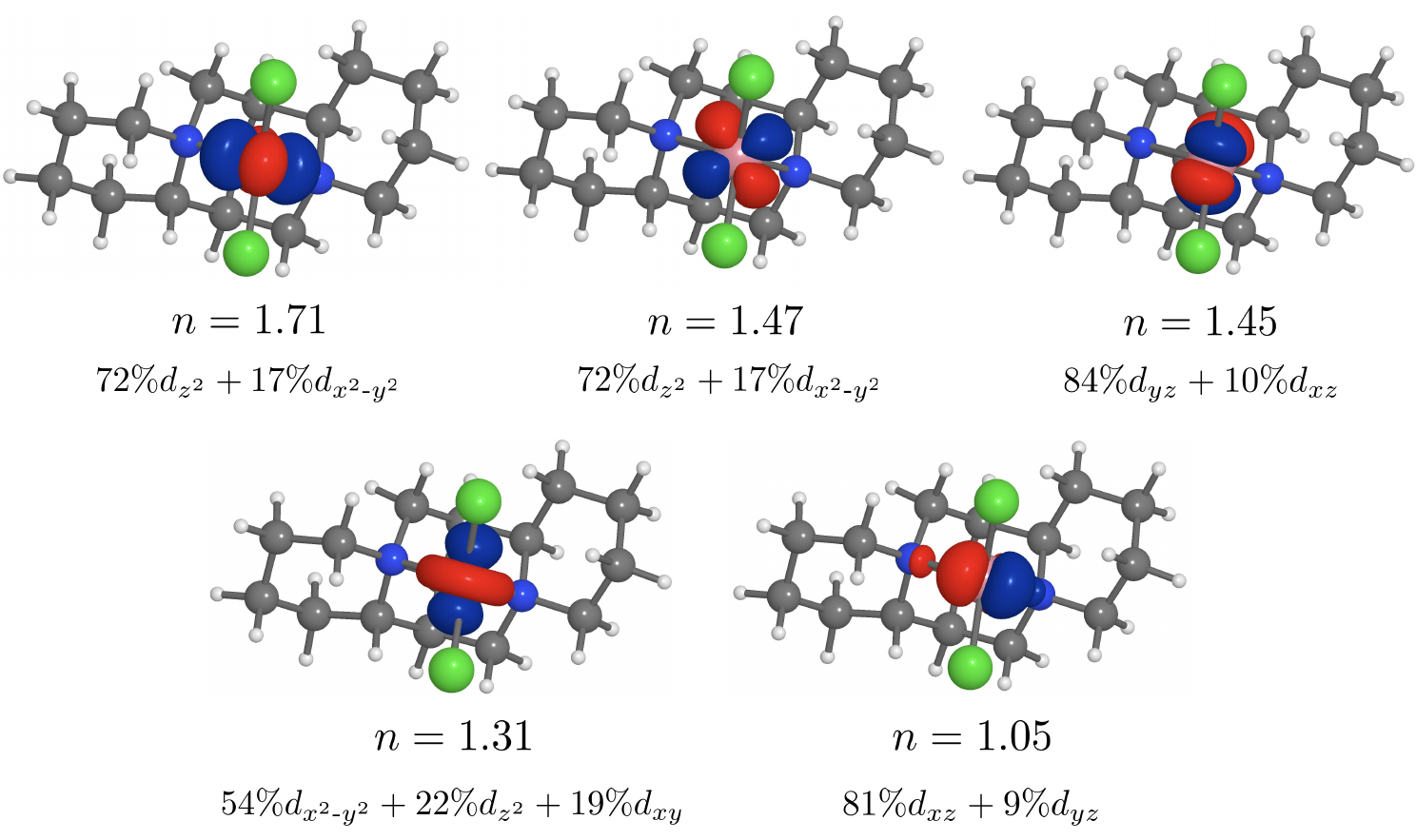}
    \end{subfigure}
    \caption{SA-CASSCF(7, 5) natural active orbitals of Co(sp)Cl$_2$ with occupation numbers ($n$). Löwdin orbital compositions are shown for atomic orbitals contributing more than 5\%. }
    \label{fig:co_orbitals}
\end{figure}

According to experimental studies, pseudotetrahedral M(sp)Cl$_2$ complexes exhibit low-lying, dipole-forbidden $d$-$d$ transitions that underlie their enhancement effect~\cite{He:2001}. Our SA-CASSCF calculations on Co(II)(sp)Cl$_2$ and Ni(II)(sp)Cl$_2$ confirm the dominant $d$ character of active orbitals with Löwdin orbital compositions 94\% and higher (see \autoref{fig:co_orbitals}).
In the main text, we present results for minimal active spaces of (7,5) for Co(II) and (8,5) for Ni(II), while data for the larger active spaces for Co(sp)Cl$_2$ is provided in the SI (Table S2). Since the transitions of interest are highly localized, we also construct a smaller model complex (M(II)(NH$_3$)$_2$Cl$_2$) with $C_{2v}$ symmetry. This higher symmetry allows state-specific CASSCF calculations, thereby avoiding possible artifacts introduced by orbital averaging in SA-CASSCF.

\begin{figure}[H]
    \centering
    \begin{subfigure}{0.49\textwidth}
        \centering
        \includegraphics[width=\textwidth]{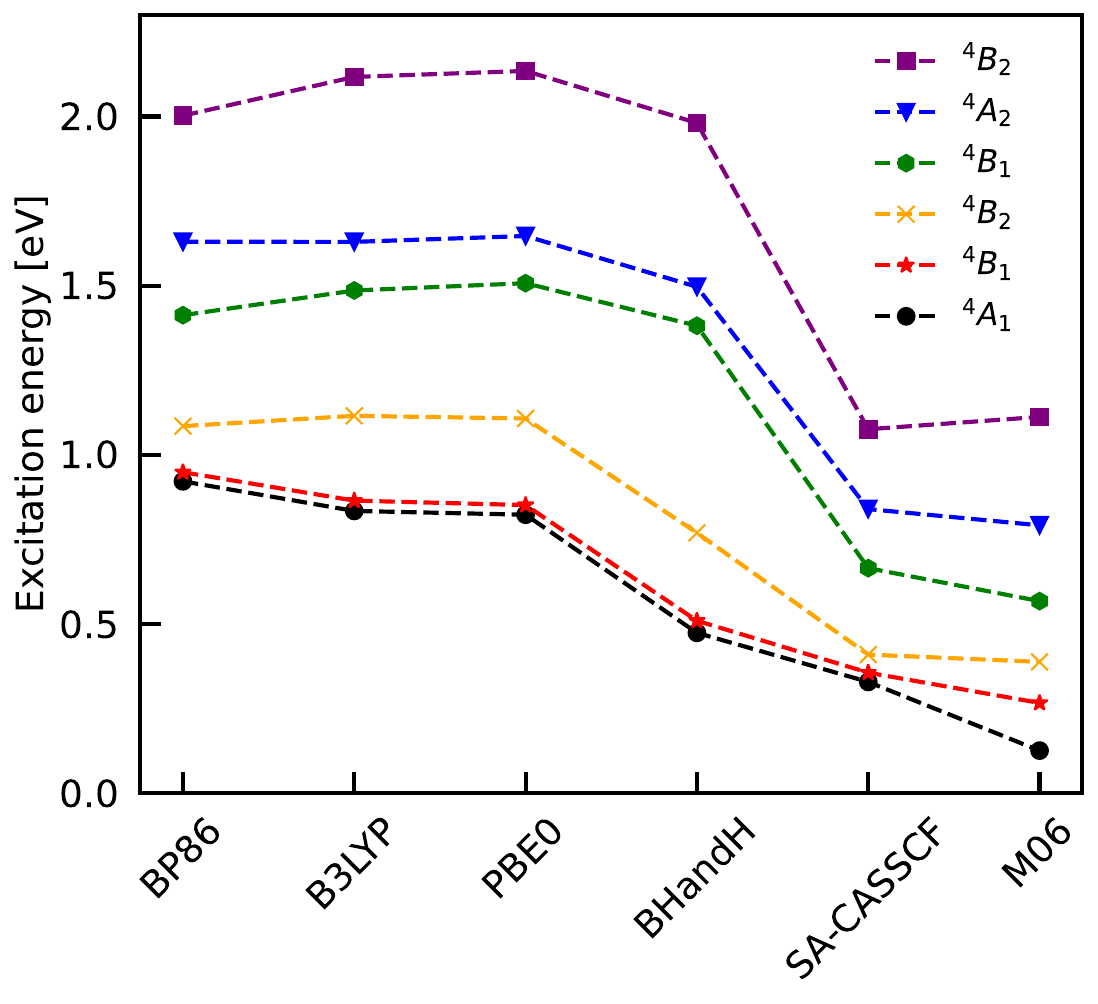}
        \label{fig:tddft_co}
    \end{subfigure}
    \begin{subfigure}{0.49\textwidth}
        \centering
        \includegraphics[width=\textwidth]{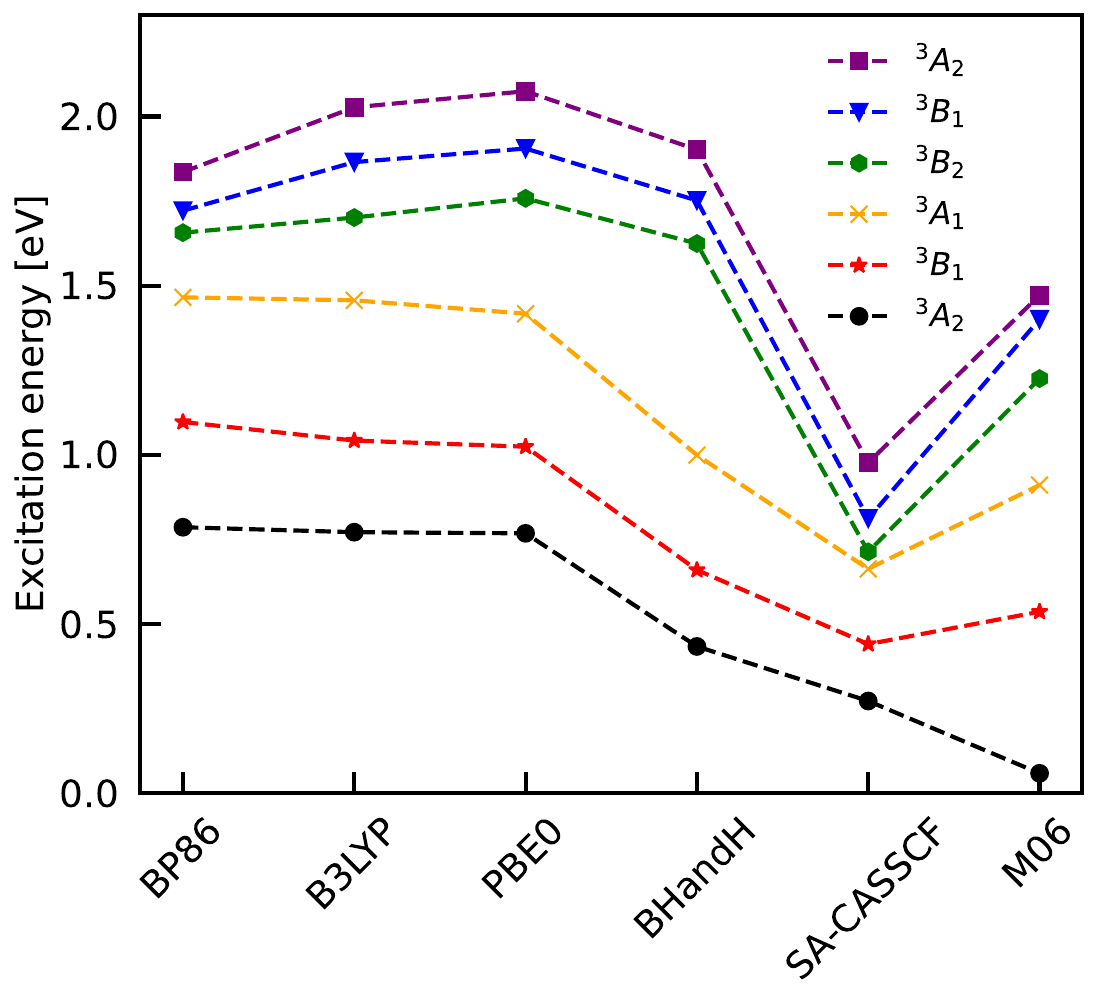}
        \label{fig:tddft_ni}
    \end{subfigure}
    \caption{Excitation energies of the first 6 excited states of Co(sp)Cl$_2$ (left) and Ni(sp)Cl$_2$ (right) calculated with TDDFT and SA-CASSCF in Orca~\cite{ORCA}. To distinguish the excited states we use for non-symmetric sparteine the symmetry labels of the $C_{2v}$ model complex (see \autoref{tab:excit_energies}. The functionals are ordered from GGA (BP86) to hybrid functionals with increasing Hartree-Fock exchange: B3LYP (20\%), PBE0 (20\%), BHandH (50\%). The Hybrid Meta-GGA functional M06 yields a first excitation energy that is too low, and is put to the right of the reference results to facilitate the comparison.}
    \label{fig:tddft}
\end{figure}

Co(sp)Cl$_2$ SA-CASSCF(7, 5) calculations show that the first three excited states lie approximately 0.2 eV below the next higher state (see ~\autoref{fig:tddft}). Since the contribution of each state scales inversely with $w_{eg} - w_a$, only excited states with energies that nearly coincide with the frequency of a vibrational mode yield significant enhancements. Contributions from the fourth and higher excited states are thus for all practical purposes negligible. For Ni(sp)Cl$_2$ we find on the other hand that there is already a significant energy gap between the first and second excited state (see ~\autoref{fig:tddft}), and for this complex we therefore only need to consider the first excited state in the enhancement calculations. 

In order to assess to what extent excitation energies calculated at the TDDFT level are able to reproduce these SA-CASSCF energies, we compare in ~\autoref{fig:tddft} the latter energies with excitation energies computed with representative GGA, hybrid, and hybrid meta-GGA functionals. From this Figure it becomes clear that pure GGAs substantially overestimate the energies of the low-lying states for a high-spin complex. Introducing Hartree–Fock exchange systematically improves the agreement: B3LYP (20\% HF exchange) reduces the error, and BHandH (50\% HF exchange) performs even better. In contrast, the hybrid meta-GGA M06~\cite{M06} yields a first excitation energy that is too low relative to the SA-CASSCF benchmark values.

In order to determine whether the use of state-averaged orbitals for the sparteine complexes does not introduce mixing of the character of the electronic states, we compare in \autoref{tab:excit_energies} CASSCF/NEVPT2 excitation energies for the sparteine and symmetric model complexes. Notably, the SA-CASSCF(7,5) results for Co(sp)Cl$_2$ and Co(NH$_3$)$_2$Cl$_2$ are nearly identical, confirming the highly localized character of these transitions. For the $C_{2v}$ complex the ground ($^4A_2$) and excited states ($^4A_1$, $^4B_1$ and $^4B_2$) belong each to different irreducible representations, which permits state-specific CASSCF optimizations. Here, the first two excited states are almost degenerate, while the third state lies slightly higher in energy. For the Ni(II)(NH$_3$)$_3$Cl$_2$ complex, the ground state is $^3B_2$, while the three lowest excited states are the $^3A_2$, $^3B_1$, and $^3B_2$ states. Overall, the close agreement between the SS-CASSCF energies of the high-symmetry model and the SA-CASSCF results confirms that the state-averaged orbitals provide a reliable description of these local excitations.

\begin{table*}[h]

\setlength{\tabcolsep}{6pt}       
\renewcommand{\arraystretch}{1.5} 
\caption{Vertical (\textbf{v}) and adiabatic (\textbf{a}) excitation energies (eV) of sparteine and model complexes calculated at the CASSCF level. NEVPT2 corrected energies are given in brackets.}

\label{tab:excit_energies}
\begin{tabular}{c|cc|cc}

                    & \multicolumn{2}{c|}{\textbf{Co(II)(NH$_3$)$_2$Cl$_2$}} & \multicolumn{2}{c}{\textbf{Co(II)(sp)Cl$_2$}}      \\ \hline
Symmetry           & \makecell{\tallspace SS-CASSCF (\textbf{v})\\(NEVPT2)}    & \makecell{\tallspace SA-CASSCF (\textbf{v})\\(NEVPT2)}       & \makecell{\tallspace SA-CASSCF (\textbf{v})\\(NEVPT2)}     & \makecell{\tallspace SA-CASSCF (\textbf{a})\\(NEVPT2)}  \\  \hline
$^4A_2$ $\rightarrow$ $^4A_1$ & 0.35 (0.39)            & 0.34 (0.44)            & 0.34 (0.42)           & 0.22 (0.34)    \\
$^4A_2$ $\rightarrow$ $^4B_1$ & 0.34 (0.42)            & 0.37 (0.45)           & 0.37 (0.48)            & 0.32 (0.51)    \\
$^4A_2$ $\rightarrow$ $^4B_2$ & 0.41 (0.50)            & 0.41 (0.52)           & 0.42 (0.53)            & 0.26 (0.45)  \\
\multicolumn{5}{c}{} \\ 
                    & \multicolumn{2}{c|}{\textbf{Ni(II)(NH$_3$)$_2$Cl$_2$}} & \multicolumn{2}{c}{\textbf{Ni(II)(sp)Cl$_2$}}       \\ \hline
Symmetry            & \makecell{\tallspace SS-CASSCF (\textbf{v})\\(NEVPT2)}       & \makecell{\tallspace SA-CASSCF (\textbf{v})\\(NEVPT2)}    & \makecell{\tallspace SA-CASSCF (\textbf{v})\\(NEVPT2)}     & \makecell{\tallspace SA-CASSCF (\textbf{a})\\(NEVPT2)}     \\ \hline
$^3B_2$ $\rightarrow$ $^3A_2$ & 0.30 (0.35)            & 0.29 (0.37)             & 0.30 (0.39)  & 0.25 (0.34) \\
$^3B_2$ $\rightarrow$ $^3B_1$ & 0.48 (0.52)          & 0.43 (0.57)             & 0.47 (0.61)  & 0.45 (0.72) \\
$^3B_2$ $\rightarrow$ $^3A_1$ & 0.63 (0.72)            & 0.62  (0.83)           & 0.70 (0.92)  & 0.53 (0.91)\\
\end{tabular}
\end{table*}

As discussed above, one of the key approximations in the SOS formalism is that the potentials energy surfaces of ground and excited states are not displaced with respect to each other. To check whether this assumption holds for the present systems, we show in the last column of \autoref{tab:excit_energies} the adiabatic excitation energies of the Co(sp)Cl$_2$ and Ni(sp)Cl$_2$ complexes. For the ground state we retain the BP86 geometry as it yields geometries in best agreement with experimental data (see Figure S2 in SI). 
For the excited states, however, we decided to optimize geometries using the BHandH functional, because it provides a state description that resembles more closely the one obtained from SA-CASSCF than the description obtained with TDDFT employing BP86 (see additional results in Table S3 in SI). We thereby assume that -- since the calculation of the TDDFT excited-state gradients requires evaluation of many terms shared with a non-adiabatic couplings (NACs) calculation within the linear response framework -- a functional that yields more accurate NACs is likely to provide a more reliable description of excited-state gradients. For Co(sp)Cl$_2$ we find that  geometry relaxation of the first excited state relaxation lowers the excitation energy by more than 0.10 eV, which is more than a third of the vertical excitation energy and thus quite significant. For the second excited state, on the other hand, the relaxation energy is much smaller though still more than 10\% of its vertical excitation energy. Finally, for the third excited state a significant relaxation energy is observed as well. For Ni(sp)Cl$_2$ similar observations are made although in this case the relaxation energy of the first excited state is somewhat lower. 

The important conclusion that we thus need to draw is that applying the vertical approximation within the SOS formalism will have a significant impact on the outcome of the calculations.\footnote{In passing by, we note that a similar assumption is made as well in the MFP formalism but in that case its repercussions can be expected to be much less.} To achieve quantitatively accurate enhanced VCD predictions by fully \textit{ab initio} calculations would require explicit calculations of vibrational overlap integrals and their inclusion in the calculation. This will lead, however, to impractically large computational demands, and alternative solutions are needed. In the next sections we will therefore introduce a readily usable semi-empirical approach and demonstrate that such an approach allows for calculations that quantitatively reproduce experimental VCD spectra.

\subsection{SA-CASSCF calculations of enhancements}

\begin{figure}[H]
    \centering
    \begin{subfigure}[t]{0.45\textwidth}
        \centering
        \includegraphics[width=\textwidth]{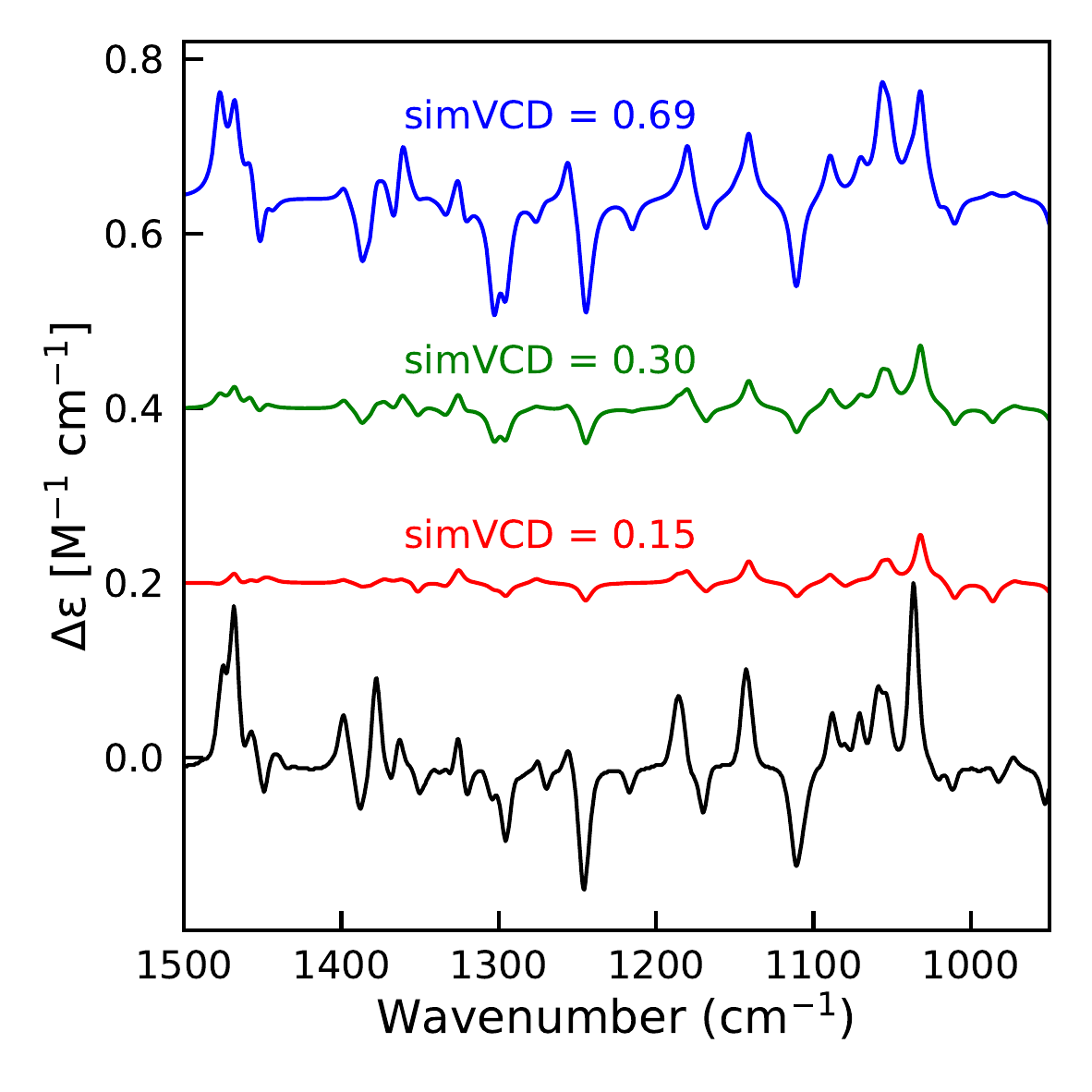}
        \label{fig:enh_casscf_co_vcd}
    \end{subfigure}
    \begin{subfigure}[t]{0.45\textwidth}
        \centering
        \includegraphics[width=\textwidth]{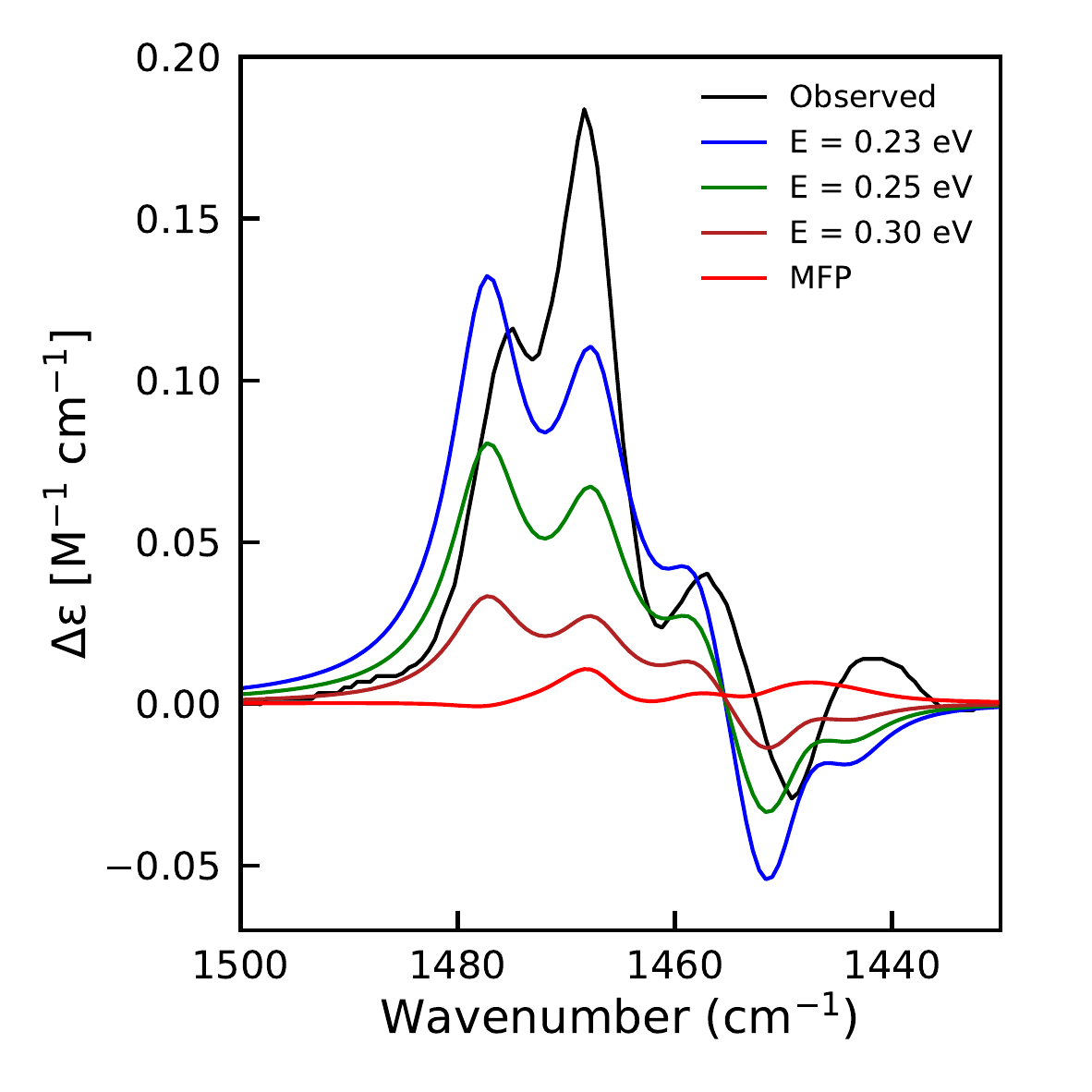}
        \label{fig:enh_casscclef_co_sens}
    \end{subfigure}
    \caption{Left panel: enhanced VCD spectra of Co(sp)Cl$_2$ calculated using MFP (red, simVCD = 0.15), SA-CASSCF (green, simVCD = 0.30), and optimized excited state energies (blue, simVCD = 0.69). Right panel: dependence of the VCD intensity in the 1430-1500 cm$^{-1}$ region on the excitation energies of the first two excited states. 
    }
    \label{fig:enh_casscf_co}
\end{figure}

We compute enhanced VCD spectra using our Python implementation of \autoref{enhancement}. To this purpose the APT and AAT tensors are evaluated at the BP86-D3BJ/CPCM level, and corrections to these tensors are obtained from OpenMolcas SA-CASSCF calculations averaged over 4 states. The results of VCD calculations for Co(sp)Cl$_2$ are shown in the left panel of  \autoref{fig:enh_casscf_co}. Inclusion of three excited states with unmodified vertical excitation energies produces a slight increase in simVCD compared to the MFP calculation, but fails to reproduce the pronounced amplification observed experimentally. Given the uncertainty due to the accuracy of the computed excitation energies and the previously discussed caution that needs to be taken in using the vertical approximation, we decided to treat the excited state energies as tunable parameters and optimize them as to maximize the simVCD value. Such an approach closely follows the approach that we used in earlier work to deal with uncertainties in determining Boltzmann weights in conformational averaging\cite{Koenis_conformer_energies}. We thus take the transition dipole moments and non-adiabatic couplings directly from the SA-CASSCF calculations and leave them unchanged, but use the excited-state energies of the three lower excited states as tunable parameters. Using this approach we can achieve excellent agreement with the experimental spectrum of Co(sp)Cl$_2$ with a simVCD value of 0.69. The energies of the first two excited states converge to a value of 0.23 eV, which is approximately 0.1 eV lower than the value obtained with SA-CASSCF calculations. Such a difference is quite acceptable and is well within the expected error of this method. We find that the energy of the third state does not change by optimization and remains at 0.42 eV, thereby reducing the number of tunable parameters required to just two. 

The right panel in ~\autoref{fig:enh_casscf_co} illustrates the sensitivity of the simulated VCD enhancement to the energies of the first two excited states. The calculated bands are slightly shifted relative to the experimental spectrum because we applied only a single frequency scaling factor from the IR calculations. More important is that the Figure convincingly demonstrates that even a minor change of 0.02 eV in excitation energy affects dramatically the amplification of the 1430–1500 cm$^{-1}$ bands. Such a small energy difference far exceeds the typical uncertainty of even the highest-level quantum chemistry methods available. Consequently, fitting low‐lying excitation energies to experimental VCD data would be more than justified and is a quite practical route to reproduce the observed enhancements.

\begin{figure}[H]
    \centering
    \begin{subfigure}{0.49\textwidth}
        \centering
        \includegraphics[width=\textwidth]{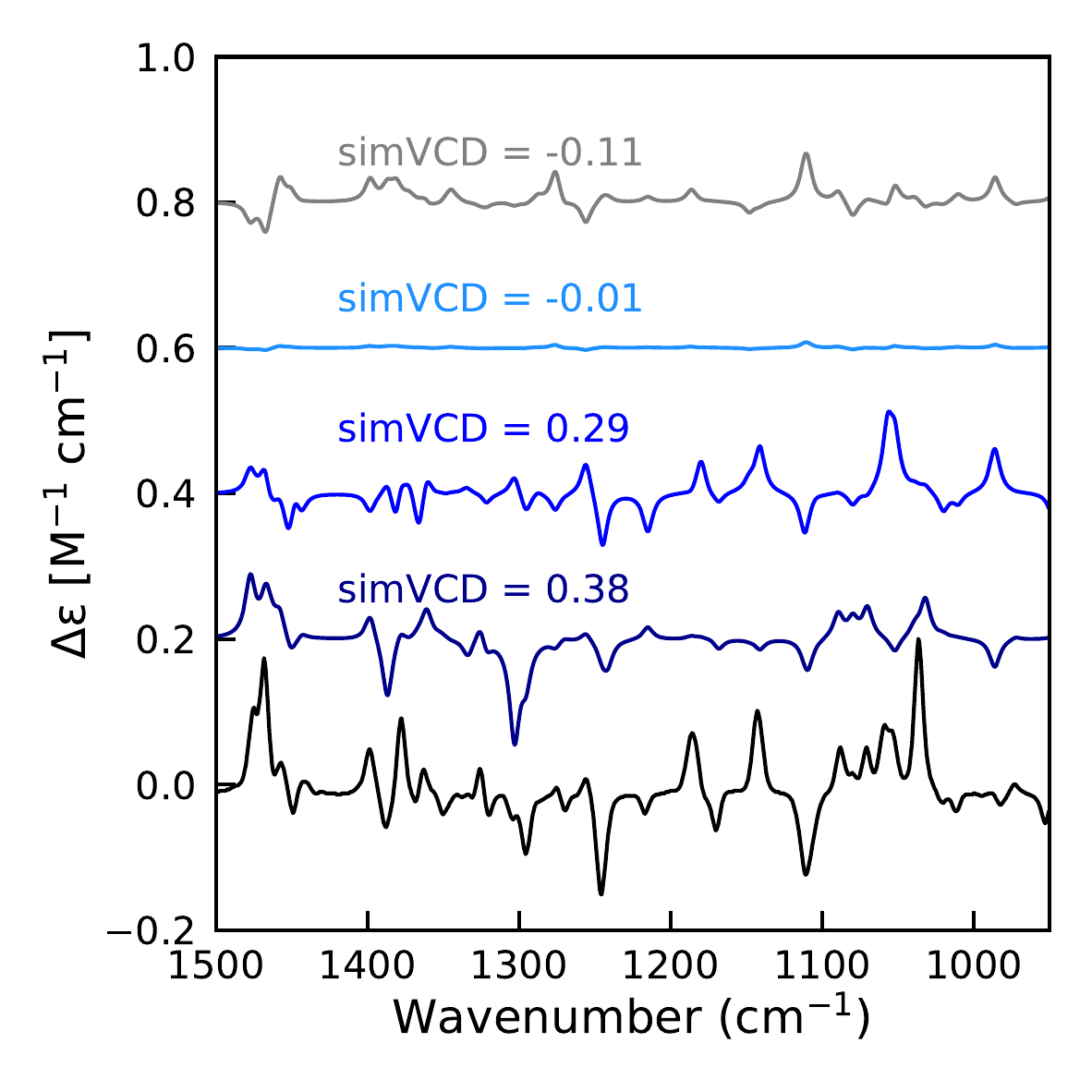   }
    \end{subfigure}
    \caption{Individual contributions of excited states with optimized energies to the enhancement of VCD bands in the VCD spectrum of Co(sp)Cl$_2$. The first two states both have excitation energies 0.23 eV and give rise to a simVCD value of 0.38 (dark blue) and 0.29 (blue), respectively. The excitation energy of the third state is 0.42 eV and its contribution is almost zero (light blue). The grey line corresponds to the excitation energy of the third state shifted to 0.23 eV in which case a simVCD value of -0.11 is obtained.}
    \label{fig:enh_casscf_co_contrib}
\end{figure}

The individual contributions of each excited state to the enhanced VCD spectrum are shown in ~\autoref{fig:enh_casscf_co_contrib}. Considering only the first or second excited state increases the simVCD value to 0.38 and 0.29, respectively. In contrast, using the third excited state, which is at 0.42 eV, gives virtually no similarity to the observed spectrum (simVCD = –0.01, light-blue trace), indicating that its energy lies above the resonance window required for enhancement. When the energy of the third state is artificially shifted to 0.23 eV (gray trace) its simVCD even has an inverted sign, giving further confirmation that it does not contribute to the experimentally observed spectrum. From these results, we conclude that only the two lowest excited states are responsible for the vibronic enhancement in Co(sp)Cl$_2$. One final aspect that is important to recognize is that the individual contributions  show that each state contributes differently both in sign and amplitude to the enhancement of the various bands in the VCD spectrum. The enhancements observed in the experimental spectrum thus actually provide a fingerprint of the extent to which each excited state is of importance in determining the enhancement of that band.

\begin{figure}[H]
    \centering
    \begin{subfigure}{0.49\textwidth}
        \centering
        \includegraphics[width=\textwidth]{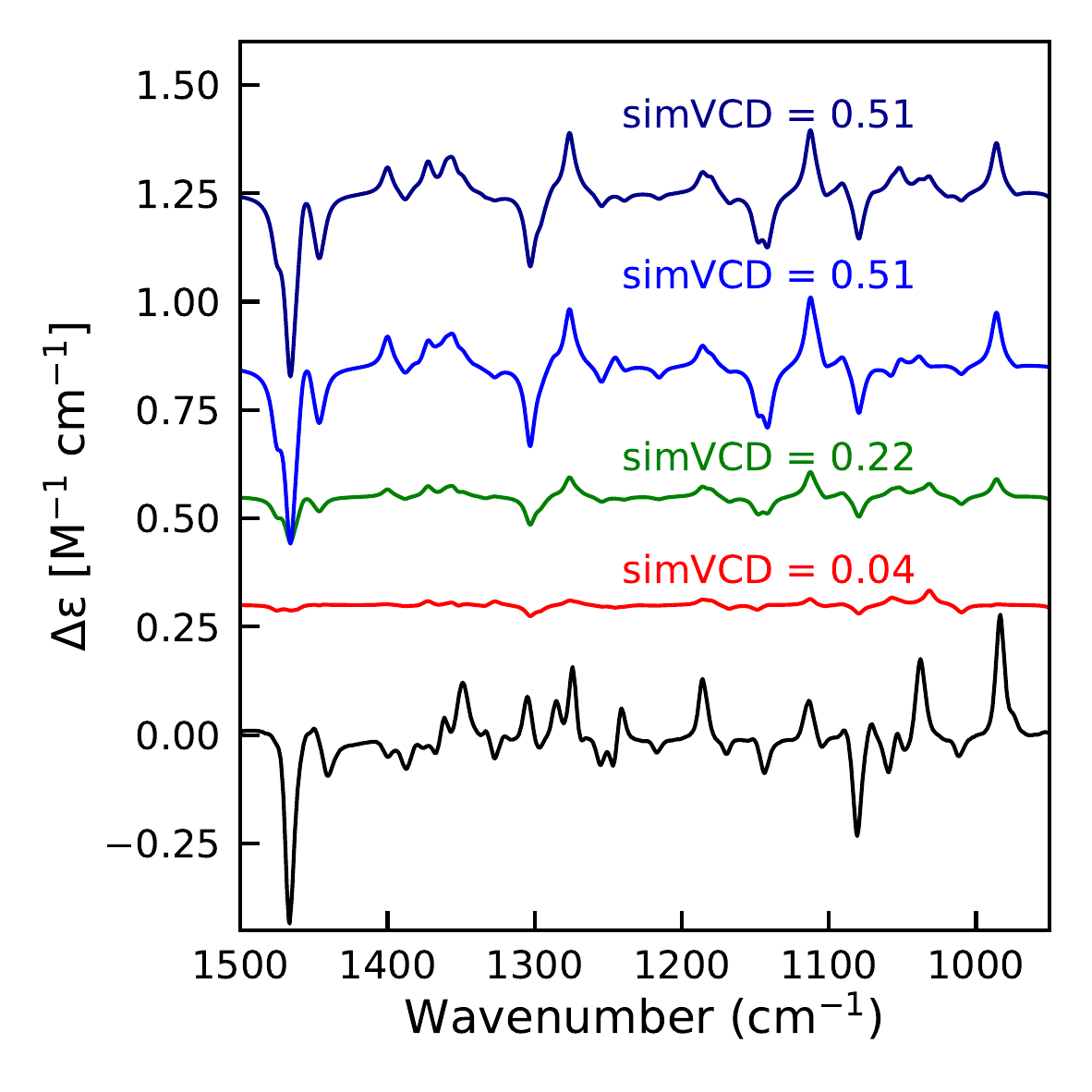}
    \end{subfigure}
    \caption{Enhanced VCD spectra of Ni(sp)Cl$_2$ calculated with MFP (red, simVCD = 0.04), SA-CASSCF (green, simVCD = 0.22) and SA-CASSCF with optimized energies including one (blue, simVCD = 0.51) and three states (dark blue, simVCD = 0.51). 
    }
    \label{fig:enh_casscf_ni}
\end{figure}

For Ni(sp)Cl$_2$ the enhanced spectra with unmodified SA-CASSCF energies also show only a slight enhancement compared to MFP spectra ~\autoref{fig:enh_casscf_ni}). Optimization of the excitation energies of the 3 lowest excited states gives a simVCD value of 0.51 with optimized energies of 0.24, 0.31, and 0.37 eV. Including only the lowest state in the optimization yields virtually identical spectra, indicating that in this case only one state contributes to the enhancement. This observation is consistent with the calculated SA-CASSCF excitation energies which predict that the second and third excited states lie significantly higher than the first one. After the optimization the calculated Ni(sp)Cl$_2$ spectrum shows one noticable difference with the experimentally observed spectrum in that in the calculated enhanced spectra an intense negative peak appears around 1300 cm$^{-1}$. This discrepancy already appears in the MFP spectra and might be due to artifacts arising from the implicit modeling of solvent effects. If we exclude this 1300-1320 cm$^{-1}$ region, a simVCD value of 0.57 is obtained. Irrespective of whether this region is included or not, the overall conclusion remains, however, that our approach enables a reliable assignment of the absolute configuration of the complex.

We conclude that by tuning the lowest excitation energies, SA-CASSCF is able to predict VCD spectra of sparteine complexes that are in excellent agreement with the experimental spectra. However, SA-CASSCF calculations of non-adiabatic couplings will present a serious bottleneck for applying this approach to larger systems. To overcome this computational bottleneck, we will in the next section demonstrate that TDDFT can fruitfully be employed as a more scalable alternative to model VCD intensity enhancements.

\subsection{TDDFT calculations of enhancements}

\begin{figure}[H]
    \centering
    \begin{subfigure}{0.43\textwidth}
        \centering
        \includegraphics[width=\textwidth]{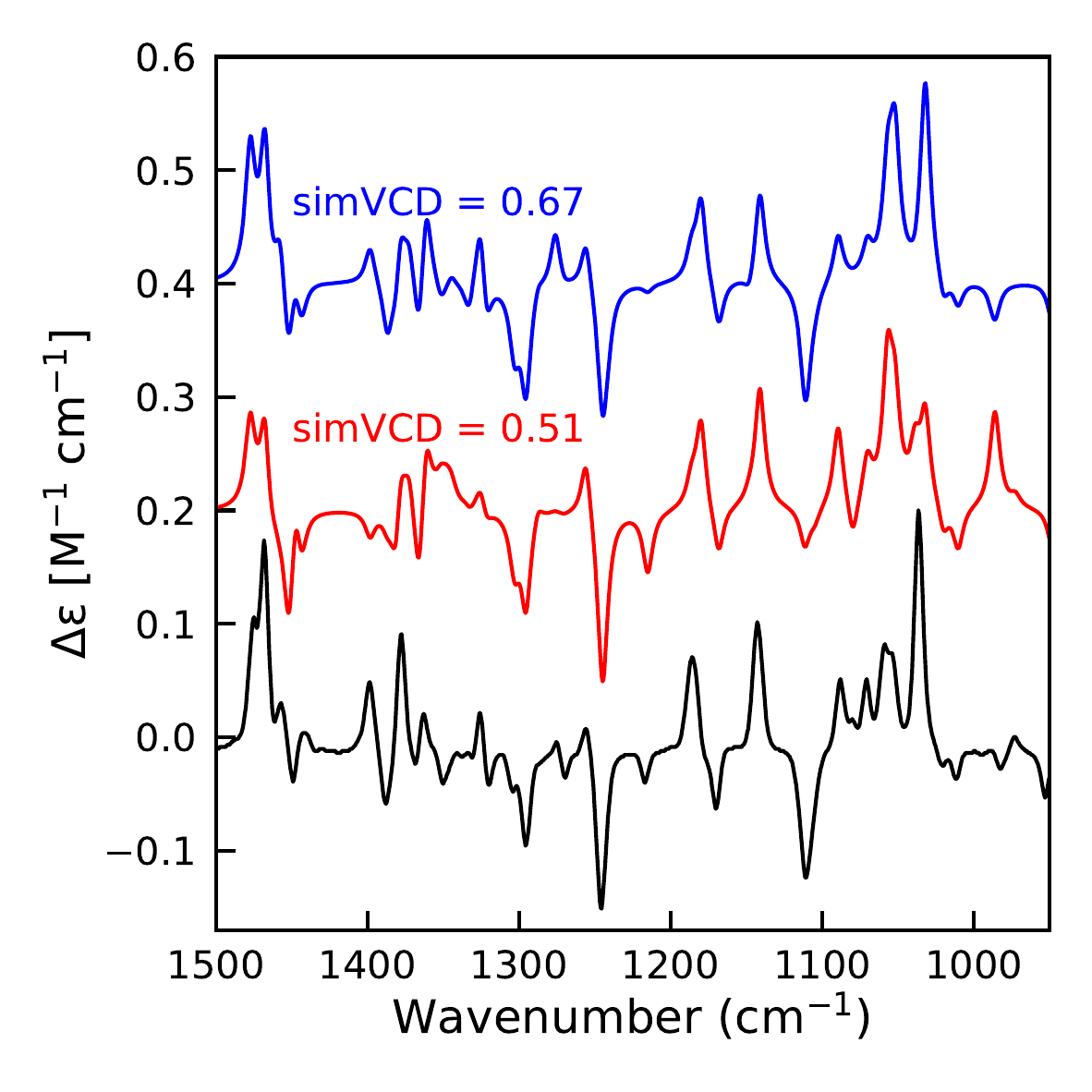}
    \end{subfigure}
        \begin{subfigure}{0.43\textwidth}
        \centering
        \includegraphics[width=\textwidth]{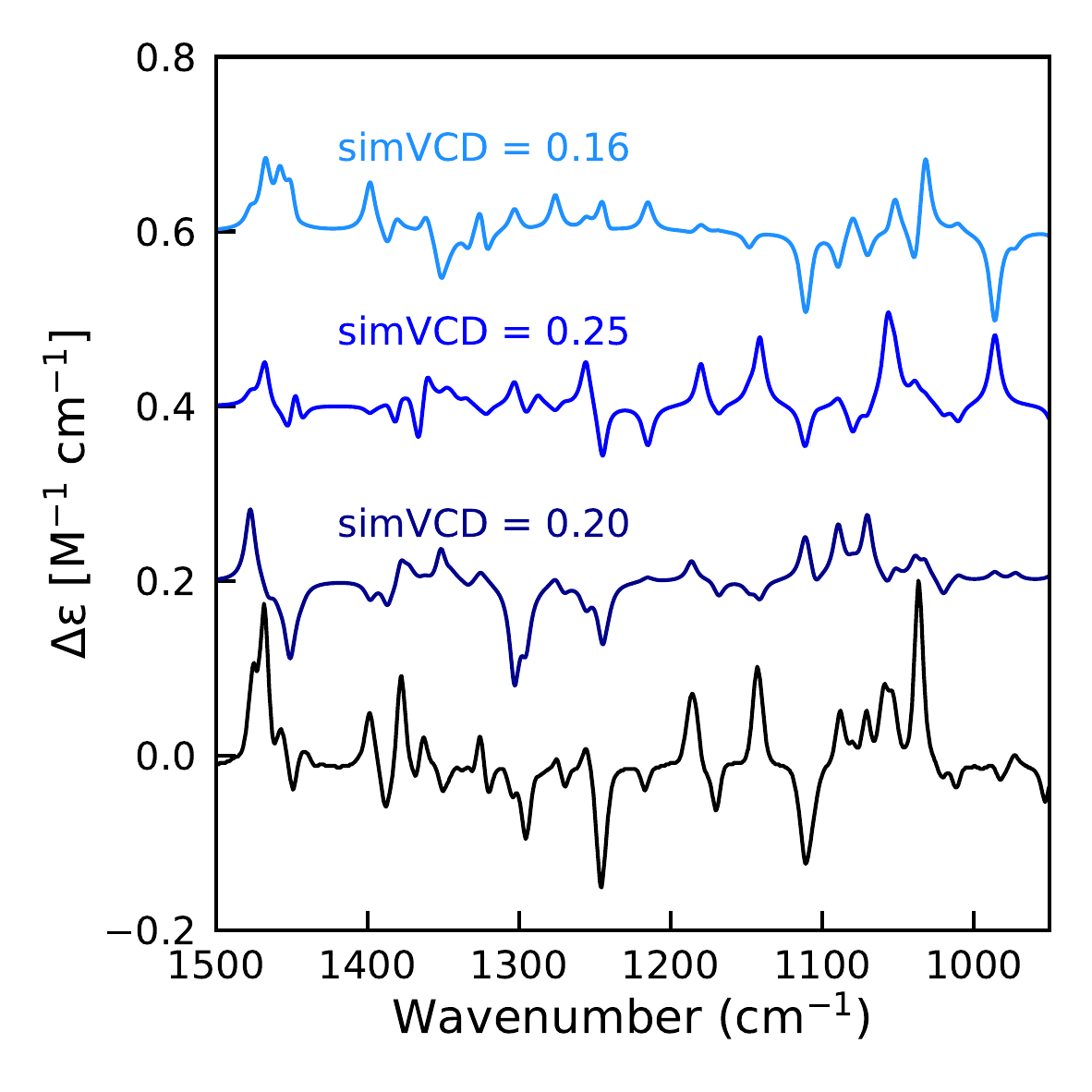}
    \end{subfigure}
    \caption{Left panel: Enhanced VCD spectra of Co(sp)Cl$_2$ calculated with the BHandH functional including two (red, simVCD = 0.51) and three excited states (blue, simVC = 0.67). Right panel: Individual contributions of the first three excited states to the enhancements in the VCD spectrum of Co(sp)Cl$_2$ using BHandH leading to simVCD values of 0.20, 0.25 and 0.16, respectively.
    }
    \label{fig:enh_tddft_co}
\end{figure}

Most density functional approximations that we studied predict excitation energies that are too high to contribute to the enhancements (see additional results in Table S4 and S5 in SI). To remedy this, we optimize the excitation energies in the same way as in the SA-CASSCF approach, while retaining the transition dipole moments and non-adiabatic couplings from the TDDFT calculations. We find that calculations using BHandH and B3LYP yield nearly identical simVCD values, and we therefore report here the BHandH results (see ~\autoref{fig:enh_tddft_co}). Data for B3LYP and other DFAs is available in the SI.

\begin{figure}[H]
    \centering
    \begin{subfigure}{0.45\textwidth}
        \centering
        \includegraphics[width=\textwidth]{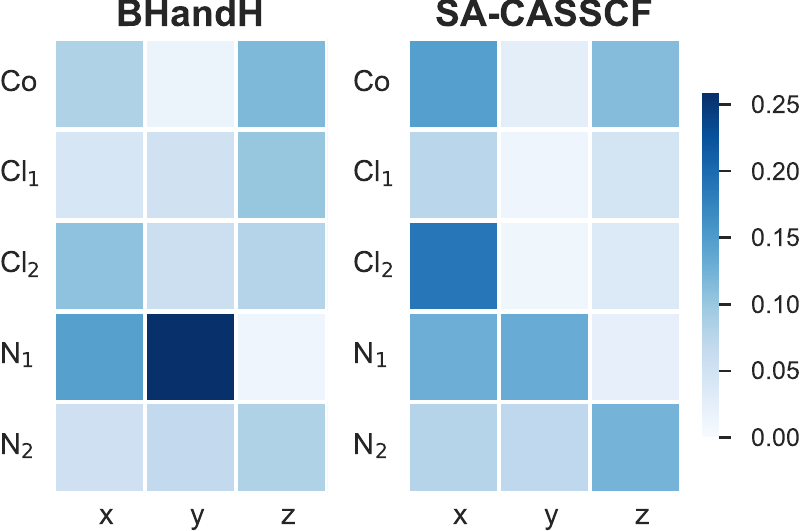}
        \subcaption{NACs between ground and first excited state}
    \end{subfigure}
    \begin{subfigure}{0.45\textwidth}
        \centering
        \includegraphics[width=\textwidth]{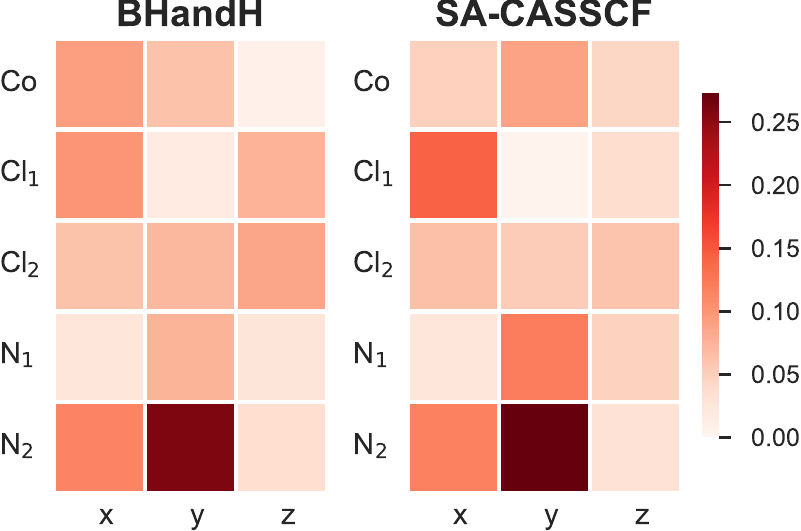}
        \subcaption{NACs between ground and second excited state}
    \end{subfigure}
        \begin{subfigure}{0.45\textwidth}
        \centering
        \includegraphics[width=\textwidth]{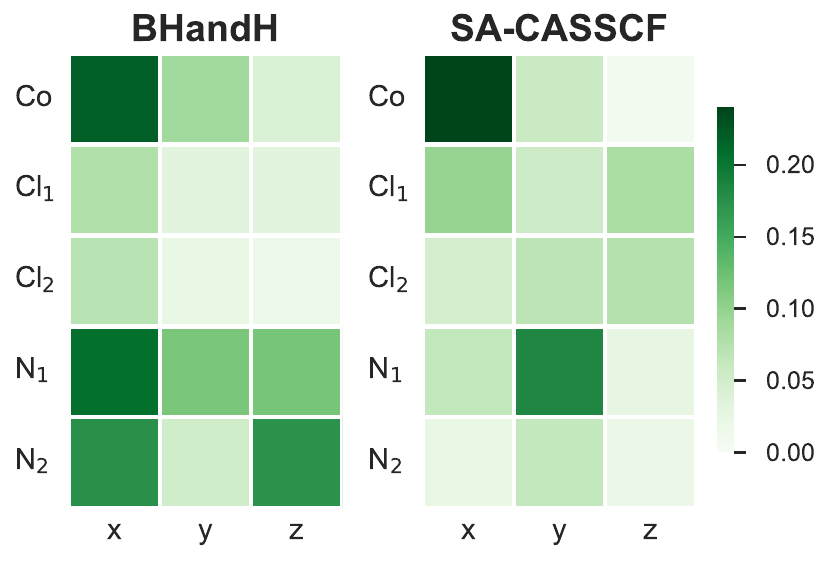}
        \subcaption{NACs between ground and third excited state}
    \end{subfigure}
    \caption{Non-adiabatic couplings between ground and first three excited states of Co(sp)Cl$_2$ calculated with BHandH and SA-CASSCF for atoms around Co(II).}
    \label{fig:nac_bhandh_co}
\end{figure}

For the Co(sp)Cl$_2$ complex, optimization of the excitation energies of the three lowest excited states leads for all three states to an excitation energy around 0.23 eV. 
This differs from the SA-CASSCF results, where the third state remains higher and does not contribute to the enhancement. We attribute these differences to differences in non-adiabatic couplings (NACs) (see \autoref{fig:nac_bhandh_co}). The only significant coupling values correspond to atoms directly bonded to the transition metal, again confirming the distance‐dependence of the enhancement effect. Comparison of the BHandH and SA-CASSCF NACs between the ground and second excited states shows that they are quite similar. This is not the case for the first and third excited state. In particular we notice that NACs between the ground and third excited state involving the N atoms are noticeably stronger with BHandH, which likely enables its contribution to the enhancement. We find that the non‐adiabatic couplings computed with B3LYP closely match those obtained with BHandH, which accounts for the nearly identical enhanced VCD profiles predicted by both methods. NACs, difference plots and mean errors for other DFAs can be found in the SI.

When we restrict the optimization to just the two lowest excited states, the agreement with the experiment remains very good. The simVCD value is reduced to 0.54 compared to 0.67 when all three states are adjusted -- which is still quite acceptable in terms of reliability of the assignment of the absolute configuration --  but equally important is that the overall spectral profile is still well reproduced. 
Most of the discrepancy arises in the 1250–1400 cm$^{-1}$ region, and from the more pronounced negative peak around 1100 cm$^{-1}$. 
The righthand panel of ~\autoref{fig:enh_tddft_co} shows the contributions of each excited state. Comparison of these contributions emphasizes once again the observation made previously when discussing the SA-CASSCF results that each state has a unique enhancement fingerprint both in sign and magnitude. For instance, the first excited state produces a negative feature near 1300 cm$^{-1}$, while both the second and third excited states give positive peaks at that frequency. Moreover, different states dominate distinct spectral regions: in the 1500–1200 cm$^{-1}$ region a superposition of the first and second excited state contributions reproduces the overall band shape, the 1200-1100 cm$^{-1}$ region is well described by the second state, and finally 1100-1000 cm$^{-1}$ is closer to the shape of the third state contribution. 

\begin{figure}[H]
    \centering
    \begin{subfigure}{0.49\textwidth}
        \centering
        \includegraphics[width=\textwidth]{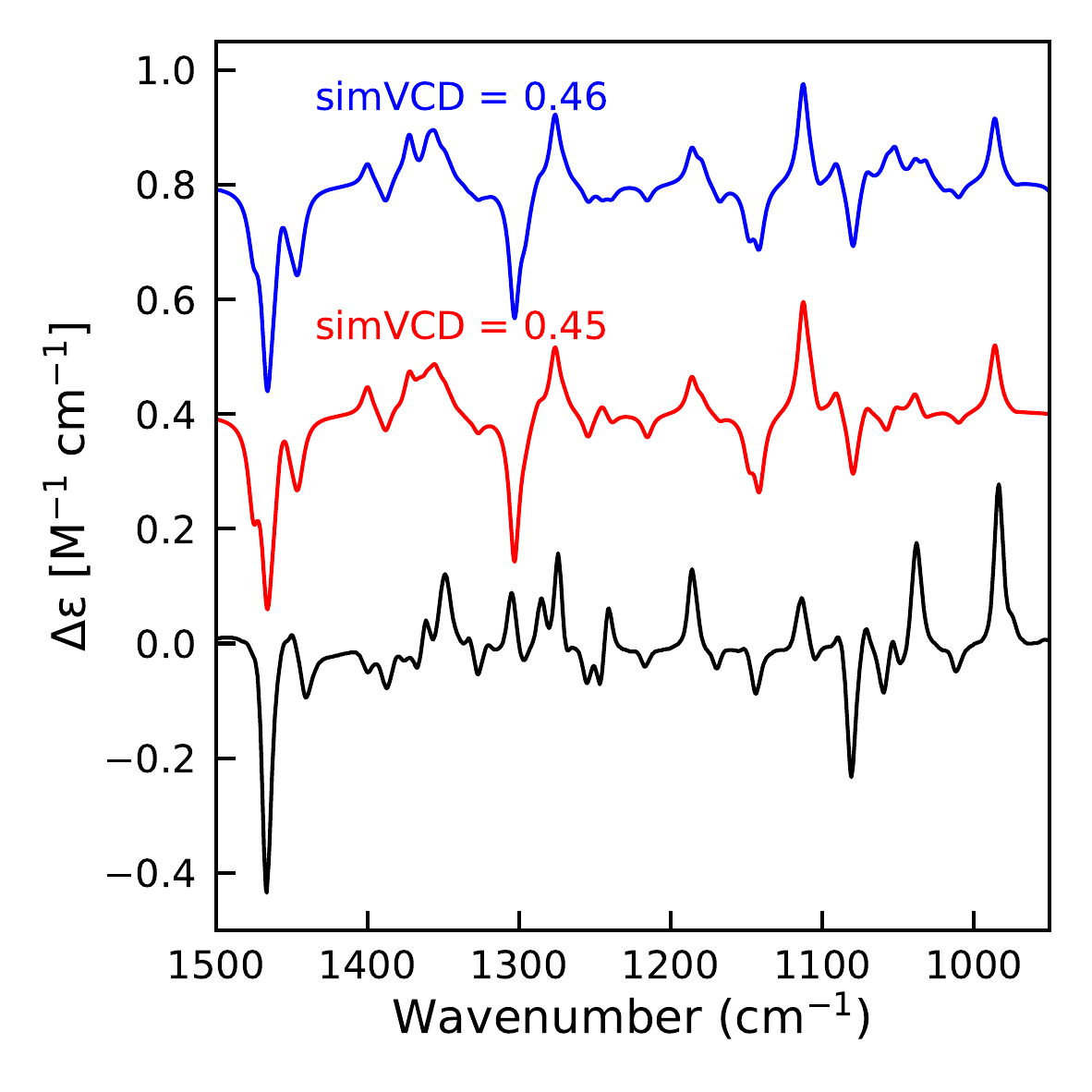}
    \end{subfigure}
    \caption{Enhanced VCD spectra of Ni(sp)Cl$_2$ calculated with BHandH functional with one (red, simVCD = 0.45) and three  (blue, simVCD = 0.46) excited states.}
    \label{fig:enh_tddft_ni}
\end{figure}

\begin{figure}[H]
    \centering
    \begin{subfigure}{0.45\textwidth}
    
        \centering
        \includegraphics[width=\textwidth]{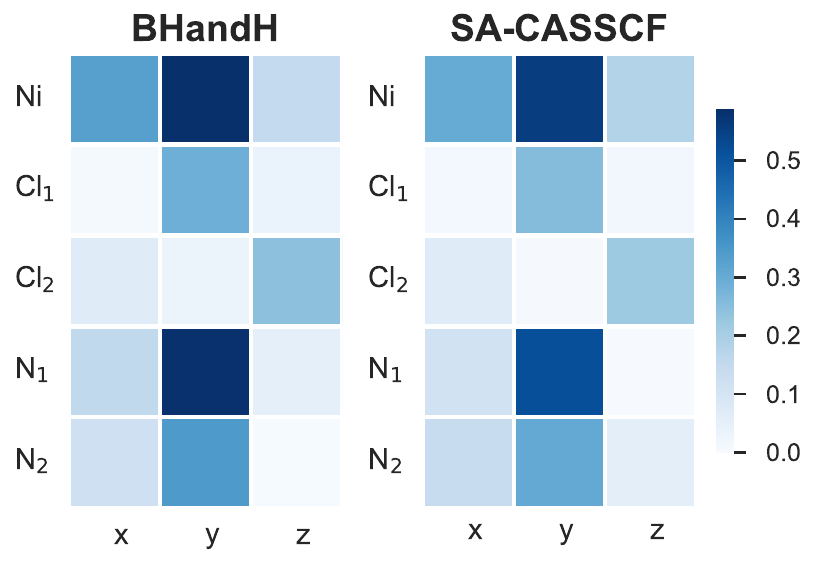}
        \subcaption{NACs between ground and first excited state}
    \end{subfigure}
    \begin{subfigure}{0.45\textwidth}
        \centering
        \includegraphics[width=\textwidth]{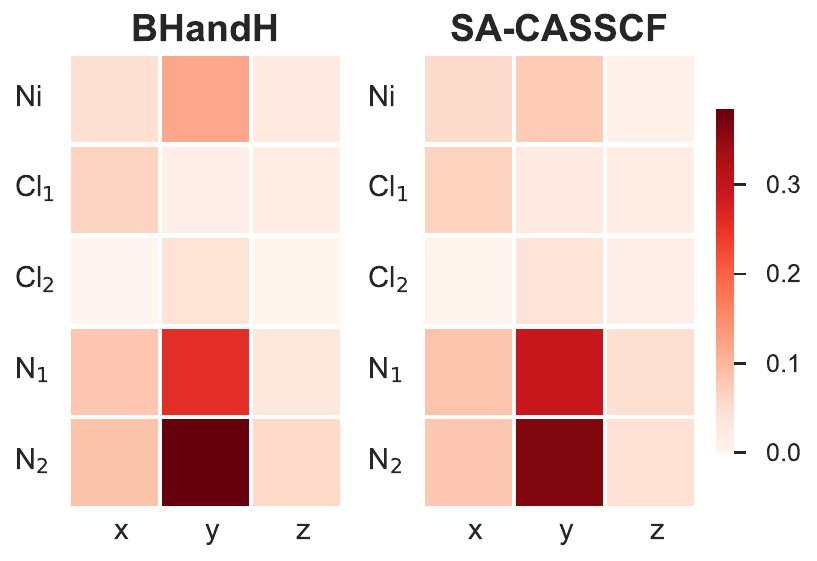}
        \subcaption{NACs between ground and second excited state}
    \end{subfigure}
        \begin{subfigure}{0.45\textwidth}
        \centering
        \includegraphics[width=\textwidth]{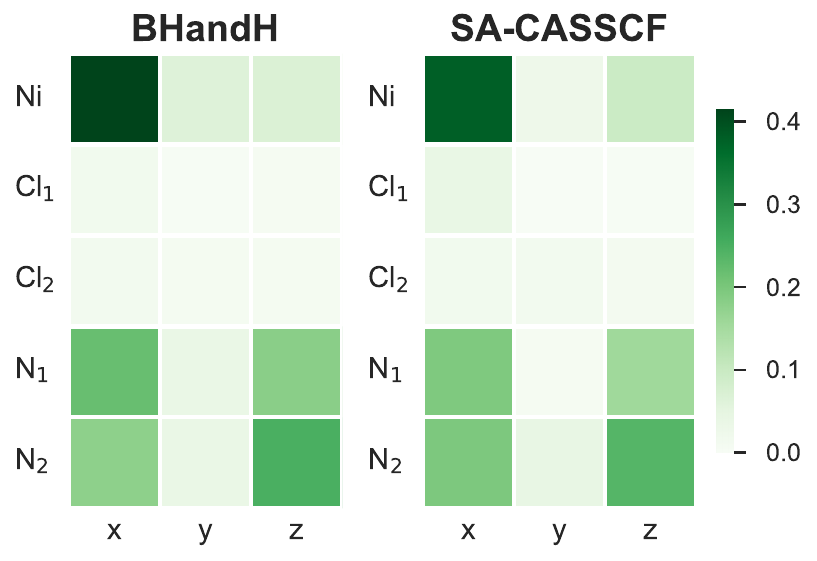}
        \subcaption{NACs between ground and third excited state}
    \end{subfigure}
    \caption{Non-adiabatic couplings between ground and first three excited states of Ni(sp)Cl$_2$ calculated with BHandH and SA-CASSCF.}
    \label{fig:nac_bhandh_ni}
\end{figure}

For Ni(sp)Cl$_2$, the TDDFT results using the BHandH functional are essentially identical to those obtained with SA-CASSCF: only the first excited state contributes to the enhancement, giving a final simVCD value of 0.45 (see ~\autoref{fig:enh_tddft_ni}) and an optimized energy of 0.23 eV. The NACs computed with BHandH closely match those from SA-CASSCF for Ni(sp)Cl$_2$ (~\autoref{fig:nac_bhandh_ni}), which explains that both approaches lead to the same result and confirms our explanation for the differences found for the Co complex. Similar to what was found with SA-CASSCF, we also observe with TDDFT a large negative peak at 1300 cm$^{-1}$.

\subsection{Assignment of absolute configuration}

\begin{figure}[H]
    \centering
    \begin{subfigure}{0.45\textwidth}
        \centering
        \includegraphics[width=\textwidth]{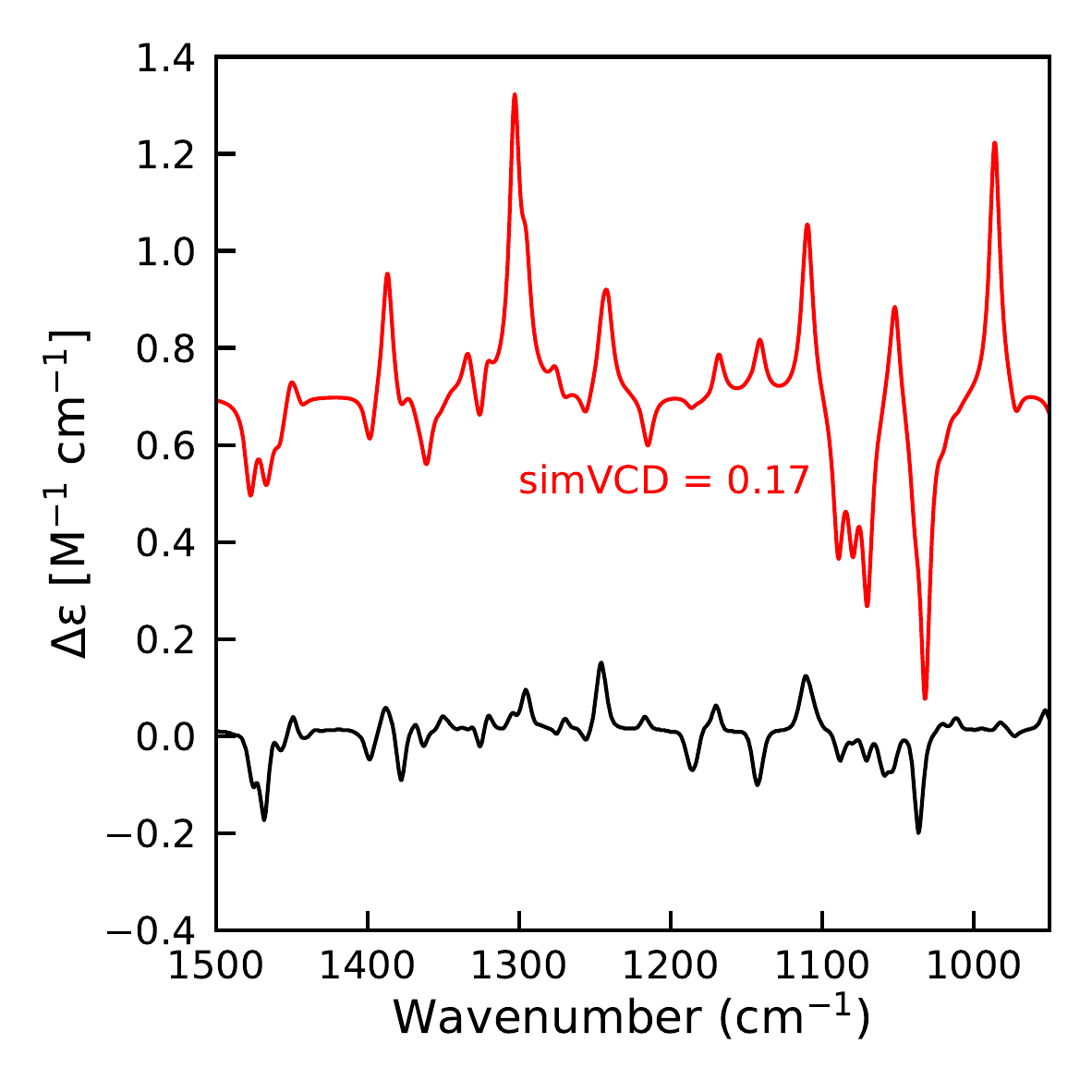}
    \label{fig:enh_casscf_ni_enan_vcd}  
    \end{subfigure}
    \begin{subfigure}{0.45\textwidth}
        \centering
        \includegraphics[width=\textwidth]{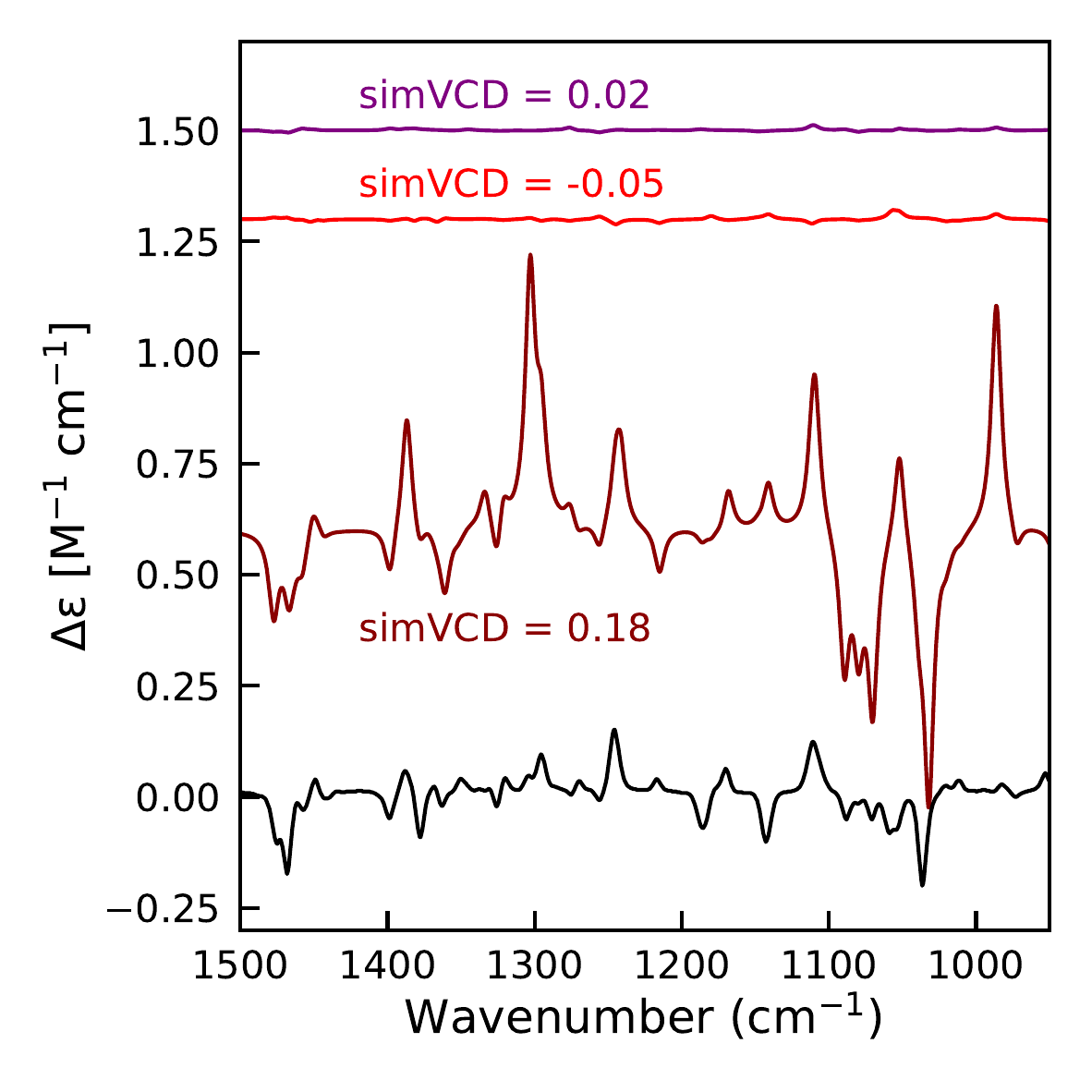}
    \label{fig:enh_casscf_ni_enan_contr}    
    \end{subfigure}
    \caption{Left panel: Enhanced VCD spectra for the enantiomer of Co(sp)Cl$_2$ (left). The calculated spectrum is obtained using SA-CASSCF with three excited states and optimized excitation energies (red); the enantiomeric VCD spectrum is obtained by inversion of the original spectrum (black). Right panel: Individual contributions of each state. The second and third state have negligible contributions and give rise to spectra with a simVCD value of -0.05 (red) and 0.02 (purple), respectively.}
    \label{fig:enh_casscf_co_enan}
\end{figure}

A critical test of the applicability of our approach is its ability to come to a reliable assignment of the absolute configuration. In order to explore this important aspect, we inverted the experimental VCD spectrum and repeated the energy optimization procedure using our SA-CASSCF results. For Co(sp)Cl$_2$ the results of such an optimization are shown in \autoref{fig:enh_casscf_co_enan}. The first and foremost conclusion that can be drawn is that such an optimization leads to a simVCD value of 0.17 which is (i) clearly much lower than obtained for the correct configuration (0.69) and (ii) too low to come to a reliable assignment of the absolute configuration. Looking into more detail into the results of the optimization, we find that it leads to an nonphysically low value of the excitation energy of the first excited state of 0.08 eV, a value for which perturbation theory would in any case no longer be valid. Furthermore, it is only this state that contributes to the enhancements; the excitation energies of the other two states converge to 0.37 eV and these states contribute negligibly. The VCD spectrum that then comes out is a spectrum in which band  intensities are dramatically overestimated, especially in the 1100–1000 cm$^{-1}$ region, which is a further indication that it is not possible to fit the observed experimental spectrum with the calculated input of the opposite enantiomer. 

\begin{figure}[H]
    \centering
    \begin{subfigure}{0.45\textwidth}
        \centering
        \includegraphics[width=\textwidth]{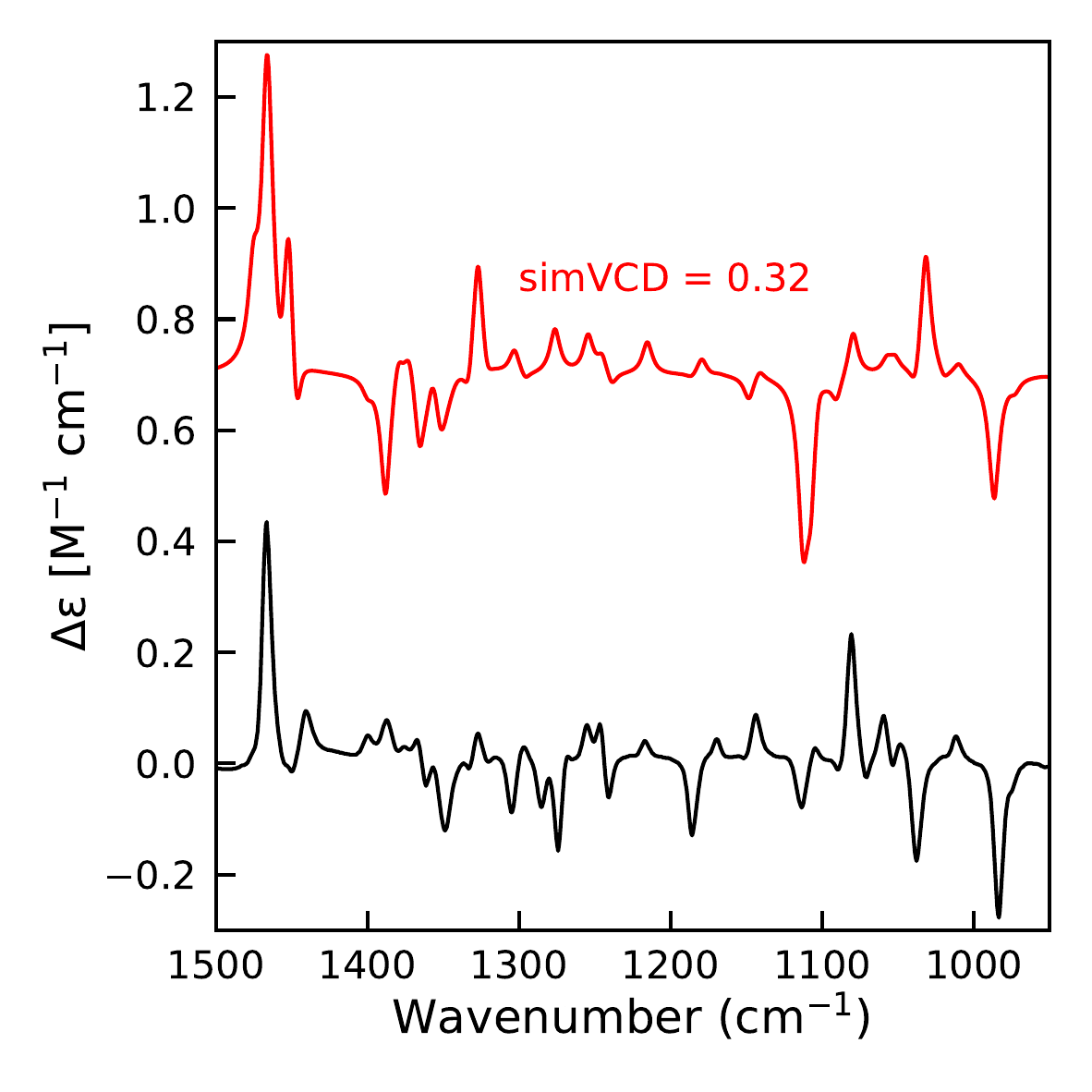}
    \label{fig:enh_casscf_ni_enan_vcd}  
    \end{subfigure}
    \begin{subfigure}{0.45\textwidth}
        \centering
        \includegraphics[width=\textwidth]{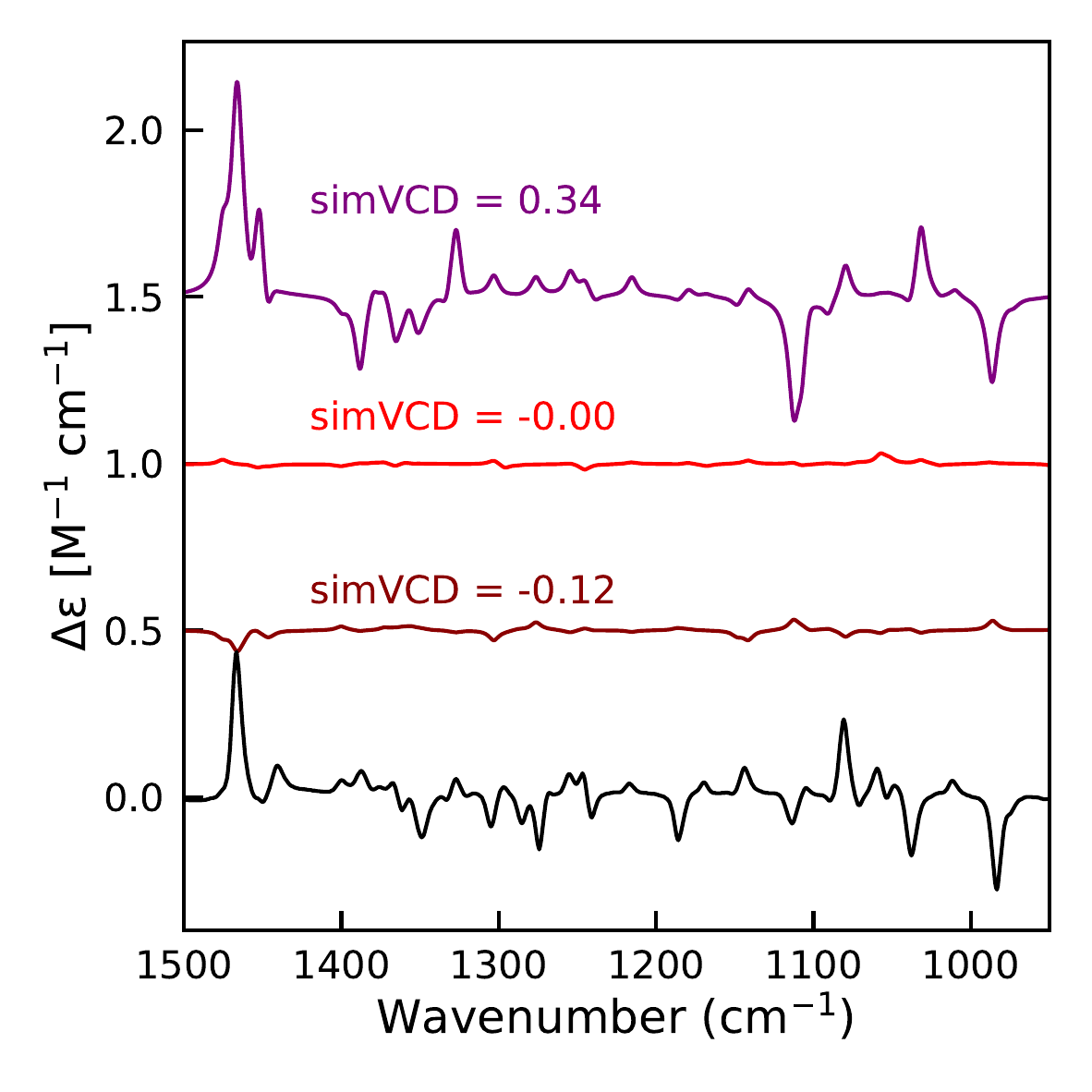}
    \label{fig:enh_casscf_ni_enan_contr}    
    \end{subfigure}
    \caption{Left panel: Enhanced VCD spectra for the enantiomer of Ni(sp)Cl$_2$ (left). The calculated spectrum is obtained using SA-CASSCF states and optimized excitation energies (red); the enantiomeric VCD spectrum is obtain by inversion of the original spectrum (black). Right panel: Individual contributions of each state. The first state gives rise to a spectrum with a simVCD value of -0.12 (dark red); the overall positive simVCD value is primarily originating from the third state (purple, simVCD = 0.34)}
    \label{fig:enh_casscf_ni_enan}
\end{figure}

For Ni(sp)Cl$_2$ (\autoref{fig:enh_casscf_ni_enan}) energy optimization similarly leads to a smaller simVCD value though less pronounced than observed for the Co complex (0.32 versus 0.51). We also find that the overall shape of the fitted spectrum does not deviate as much as found for the Co complex, although there are regions where there are significant  discrepancies with the experimental spectrum, especially in 1300-1100 cm$^{-1}$ range. Further support for the conclusion that the observed spectrum cannot be attributed to the enantiomer -- as one would conclude if one would not know the absolute configuration of the complex -- is the observation that optimization reduces the excitation energy of the third excited state to a low value of 0.18 eV while the first two states end up with a similar energy of 0.37 eV. Such an inversion would be at odds with a priori expectations on the ordering of these states. A consequence of these state energies is that only the third excited state contributes constructively to the fit.
 
We thus conclude that fitting the spectrum of the opposite enantiomer with the proposed optimization procedure leads to results that can readily be identified as being inconsistent with the experiment. This makes it possible to rely on the results of the tuned enhanced calculation to come to a robust and unambiguous assignment of the absolute configuration, provided that the set of low-lying states used in the tuning is carefully selected and validated.

\subsection{Understanding the enhancement mechanism}

We conclude our analysis by pointing out that another advantage of the formalism presented in this work is that it offers the possibility to come to a more detailed understanding of the enhancement mechanism. To this end we notice that we can distinguish enhancement due to the magnetic and electric dipole components of the electronic transition moments:
$$
R_i^{\textup{enh}}= \textup{Im} [\mathbf{E}_i^{\textup{tot}} \cdot \mathbf{M}_i^{\textup{enh}} + \mathbf{E}_i^{\textup{enh}} \cdot \mathbf{M}_i^{\textup{tot}}]
$$

\begin{figure}[H]
    \centering
    \begin{subfigure}{0.49\textwidth}
        \centering
        \includegraphics[width=\textwidth]{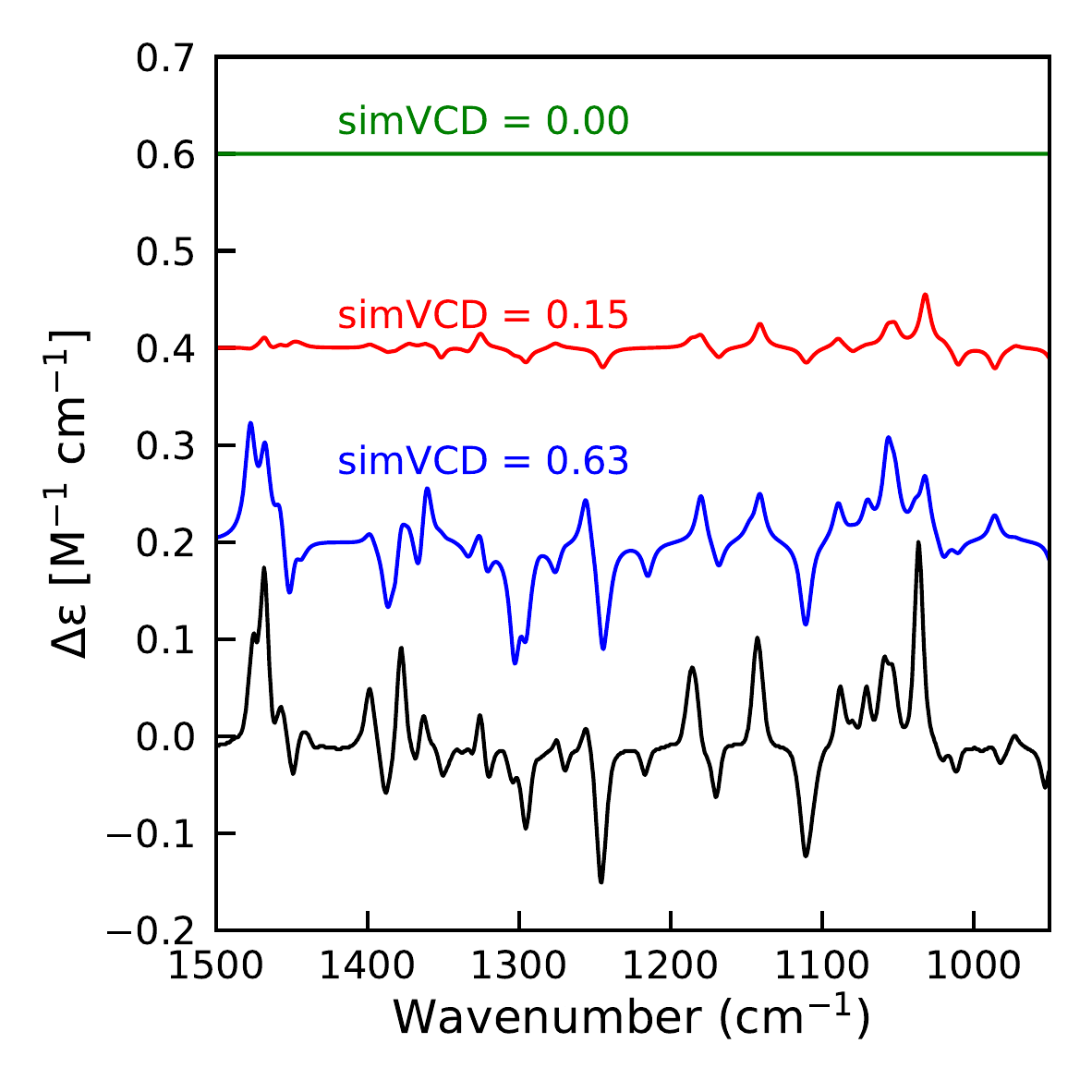}
    \end{subfigure}
    \begin{subfigure}{0.49\textwidth}
        \centering
        \includegraphics[width=\textwidth]{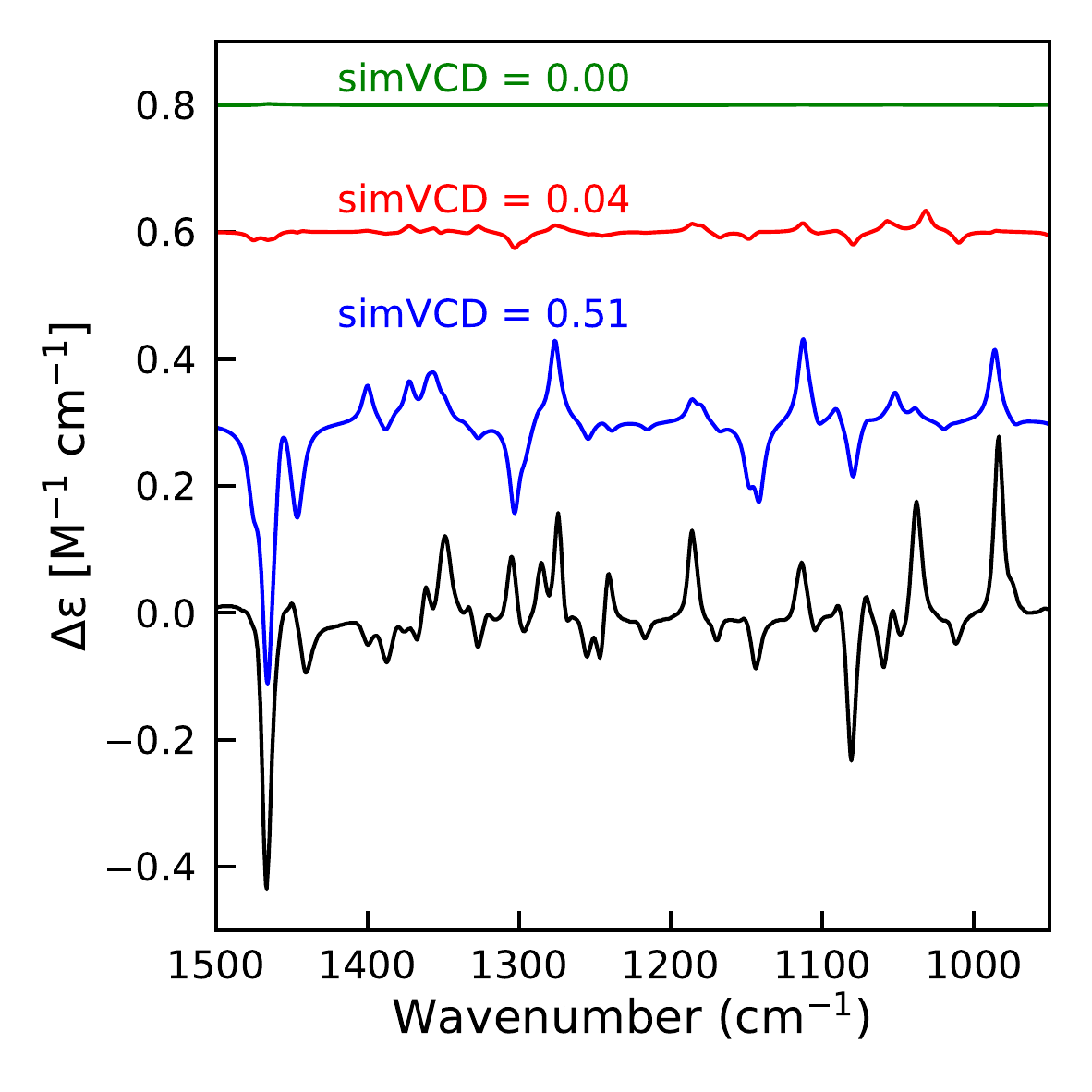}
    \end{subfigure}
    \caption{Decomposition of enhanced VCD spectra into MFP (red), electric dipole enhancement (green) and magnetic dipole enhancement (blue) for Co(sp)Cl$_2$ (left) and Ni(sp)Cl$_2$ (right) calculated with SA-CASSCF excited states properties and optimized geometries.}
    \label{fig:enh_casscf_ni_decomp}
\end{figure}

\noindent For the case at hand we find for all the calculated enhanced spectra that the electric dipole contribution is negligible. This is as expected since electric dipole transition moments associated with pure $d$-$d$ transitions are formally forbidden (Laporte selection rule).
The decomposition of the enhanced spectra into contributions for Co(sp)Cl$_2$ and Ni(sp)Cl$_2$ is demonstrated in \autoref{fig:enh_casscf_ni_decomp}. We clearly observe the magnetic dipole enhancement dominating the signal, making the accuracy of the APT tensor (from the MFP calculation) and the magnetic dipole transition moments between electronic states more critical than the AAT tensor and the electric dipole transition moments. Although for the present case such a result is as expected, for a more general case such a decomposition will clearly contribute to a further elucidation of the factors that contribute to an observed enhancement.

\section{Conclusions}\label{sec:conclusions}

In this work, we have applied the sum‐over‐states approach introduced by Nafie to calculate enhancements of band in VCD spectra of transition metal complexes. We find that these enhancements are extremely sensitive to the excitation energies, making it challenging for quantum chemistry methods to predict these values with sufficient accuracy. Rather than pushing for even more accurate excited state calculations, we propose treating these key excitation energies as adjustable parameters. By optimizing just a few values, we can reproduce experimental VCD spectra with a high similarity, achieving simVCD values over 0.4 for both SA-CASSCF and TDDFT.

At the CASSCF level, only two excited states for the cobalt complex and one for the nickel complex are needed to reproduce the main features of the enhanced VCD spectra, demonstrating that the sum‐over‐states expansion converges rapidly in these systems. Including different excited states can change the sign of a given vibrational transition, which highlights the importance of selecting the appropriate states in the enhancement calculation.

Our results indicate that TDDFT can perform comparably to CASSCF, provided that an appropriate functional is chosen. In our study, BHandH delivers the best VCD enhancements, likely due to its high fraction of Hartree–Fock exchange, which is crucial for accurately describing $d$-$d$ transitions in transition metal complexes. In fact, for Ni(sp)Cl$_2$, the non-adiabatic couplings calculated with BHandH closely match those of SA-CASSCF, leading to nearly identical spectra with SA-CASSCF states. For Co(sp)Cl$_2$, we observe some differences in the couplings to the first and third excited states between the two methods, accounting for the slight differences in their respective contributions. 

An important observation noted in earlier studies as well is the local character of the VCD enhancement. Here, we show that significant non-adiabatic couplings occur only between the transition metal and its close neighbors: two nitrogens and two chlorines in the sparteine case. Consequently, we assume that for larger systems, one could compute these couplings on a truncated model without substantial loss of accuracy. We plan to explore this idea in future work. A further direction we are working on concerns the fact that in the present study we have studied rigid sparteine complexes as a proof-of-concept application of vibronic coupling theory. The next step will be to extend this approach to flexible complexes, where conformational variability introduces additional complications.

\section{Acknowledgements}\label{sec:acknowledgements}

We acknowledge financial support from NWO in the framework of the Fund New Chemical Innovations (NWO Project Nr.731.014.209) and the National Grow Fund Quantum Delta NL (NWO Project Nr. NGF.1582.22.036). We furthermore acknowledge NWO for providing access to Snellius, hosted by SURF through the Computing Time on National Computer Facilities call for proposals.

\newpage
\providecommand{\latin}[1]{#1}
\makeatletter
\providecommand{\doi}
  {\begingroup\let\do\@makeother\dospecials
  \catcode`\{=1 \catcode`\}=2 \doi@aux}
\providecommand{\doi@aux}[1]{\endgroup\texttt{#1}}
\makeatother
\providecommand*\mcitethebibliography{\thebibliography}
\csname @ifundefined\endcsname{endmcitethebibliography}  {\let\endmcitethebibliography\endthebibliography}{}

\end{document}


\newpage

\section{Magnetic field perturbation calculations}

\begin{figure}[H]
    \centering
    \begin{subfigure}{0.45\textwidth}
        \centering
        \includegraphics[width=\textwidth]{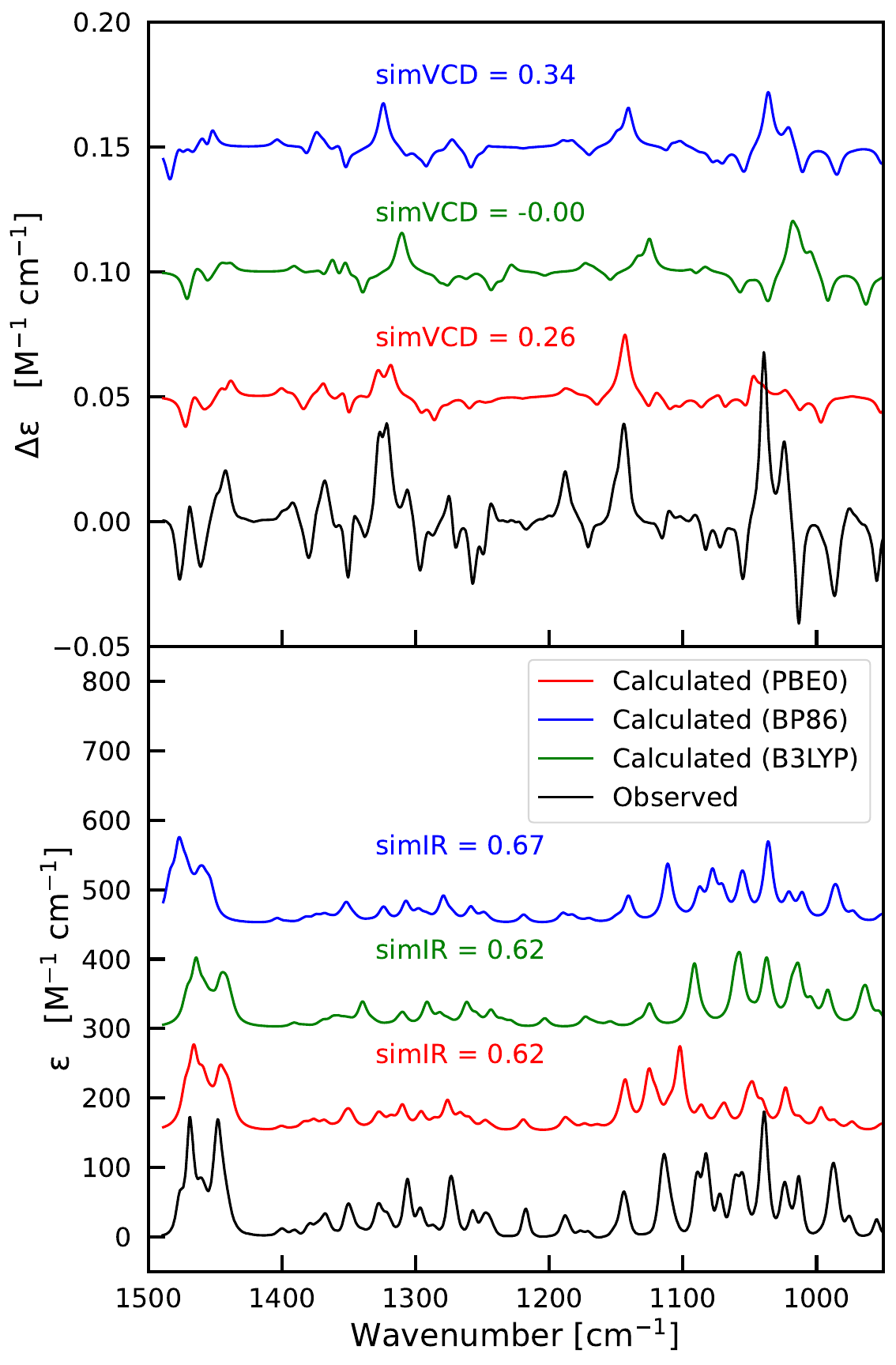}
    \end{subfigure}
    \begin{subfigure}{0.45\textwidth}
        \centering
        \includegraphics[width=\textwidth]{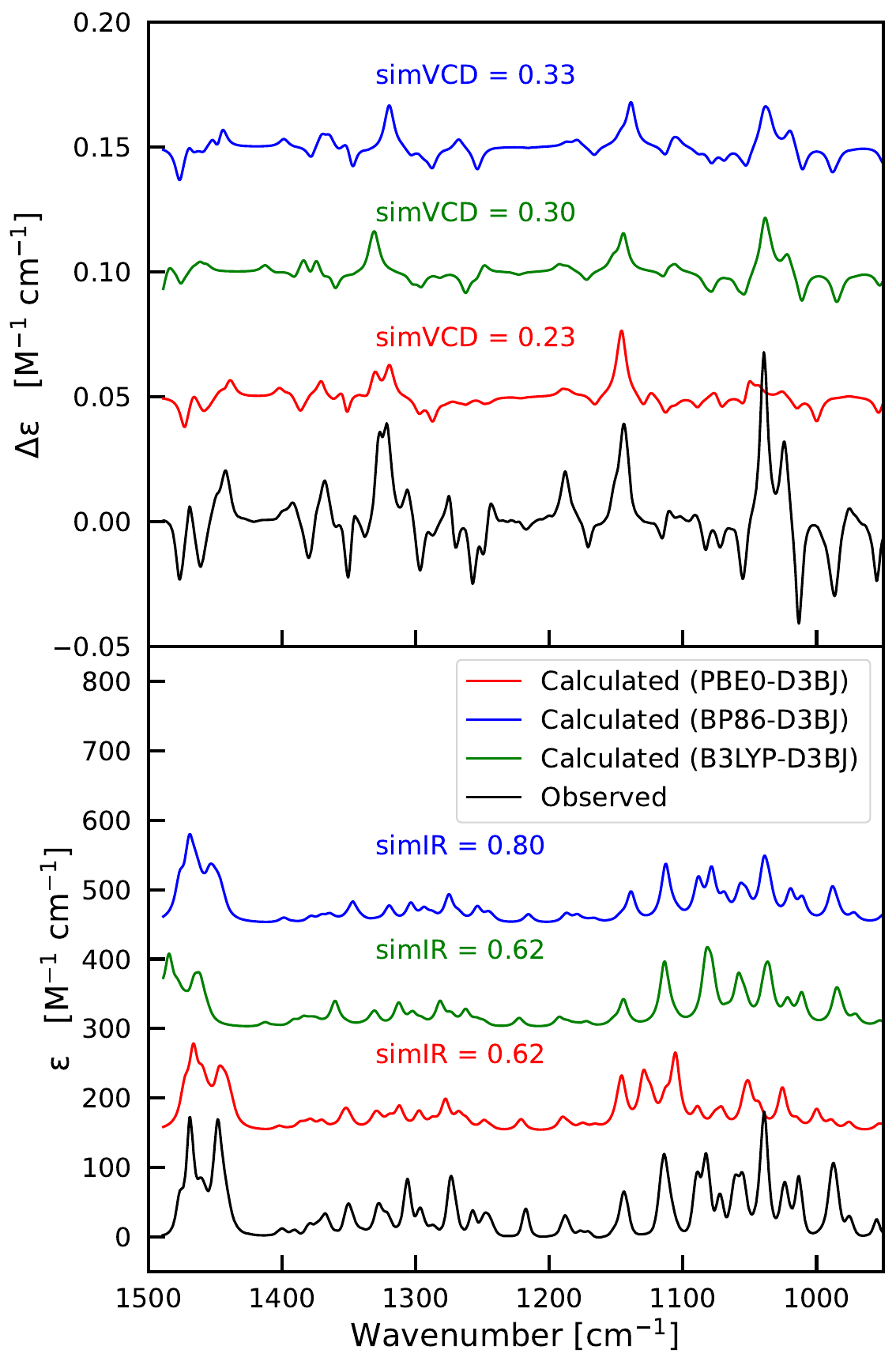}
    \end{subfigure}
    \caption{Comparison of VCD (top) and IR (bottom) spectra of Zn(sp)Cl$_2$. Experimental spectra are given in black. All VCD spectra are calculated using frequency scaling factors from maximizing simIR values. For the B3LYP~\cite{b3lyp_stephens} functional simVCD is approximately zero, however, using a separate frequency scaling factor for VCD would give simVCD = 0.22. Calculation are performed with Orca~\cite{ORCA} using the def2-TZVP~\cite{def2_basis_set} basis sets with def2/JK~\cite{def2_jk}.}
    \label{fig:ir_spectra}
\end{figure}

\begin{figure}[H]
    \centering
    \begin{subfigure}{0.7\textwidth}
        \centering
        \includegraphics[width=\textwidth]{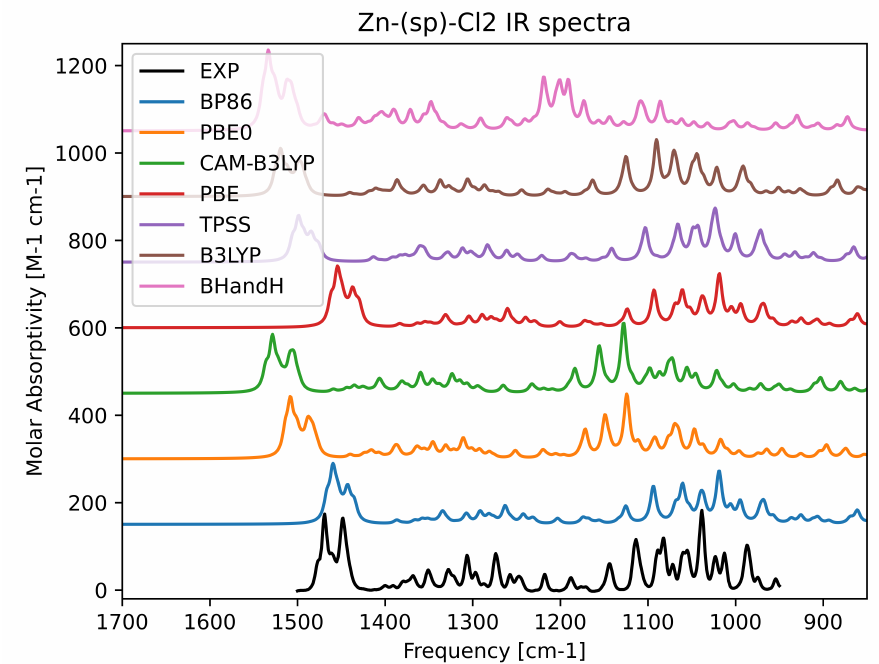}
    \end{subfigure}
    \caption{IR spectra calculated for Zn(sp)Cl$_2$ with various DFAs. All the calculations are performed with ADF~\cite{AMSpaper} (version 2023) on DFA/TZP preoptimized geometries. No frequency scaling factor is applied.}
    \label{fig:ir_spectra_adf}
\end{figure}

~\newpage

\section{Excited state calculations}

\begin{table}[H]
\centering
\begin{tabular}{c|ccccc|ccccc}
\toprule
 & \multicolumn{5}{c|}{\textbf{Co(II)(sp)Cl$_2$}} & \multicolumn{5}{c}{\textbf{Ni(sp)Cl$_2$}} \\
\#. & BP86 & B3LYP & PBE0 & BHandH & M06 & BP86 & B3LYP & PBE0 & BHandH & M06 \\ \hline
1 & 0.92 & 0.83 & 0.82 & 0.47 & 0.13 & 0.79 & 0.77 & 0.77 & 0.43 & 0.06 \\
2 & 0.95 & 0.86 & 0.85 & 0.51 & 0.27 & 1.10 & 1.04 & 1.02 & 0.66 & 0.54 \\
3 & 1.08 & 1.12 & 1.11 & 0.77 & 0.39 & 1.47 & 1.46 & 1.42 & 1.00 & 0.91 \\
4 & 1.41 & 1.49 & 1.51 & 1.38 & 0.57 & 1.66 & 1.70 & 1.76 & 1.62 & 1.23 \\
5 & 1.63 & 1.63 & 1.65 & 1.50 & 0.79 & 1.72 & 1.87 & 1.91 & 1.75 & 1.40 \\
6 & 2.00 & 2.12 & 2.13 & 1.98 & 1.11 & 1.84 & 2.03 & 2.07 & 1.90 & 1.47 \\
\bottomrule
\end{tabular}
\caption{Comparison of excitation energies for Co(II) and Ni(II) sparteine complexes with various DFT functionals calculated with ORCA~\cite{ORCA} on the BP86-D3(BJ)-CPCM(CHCl$_3$)~\cite{grimme2010_d3, grimme2011_damping} optimized geometry. All energy values are given in eV.}
\label{tab:CoNi-DE}
\end{table}

\begin{table}[H]
\begin{tabular}{c|c|c|c}
 \toprule
\#. & CAS(7, 5)/NEVPT2 & CAS(7, 10)/NEVPT2 & CAS(11, 12)/NEVPT2 \\ \hline
1 & 0.33 (0.42)             & 0.34 (0.39)              & 0.35 (0.39)               \\
2 & 0.36 (0.47)             & 0.37 (0.43)              & 0.42 (0.46)               \\
3 & 0.41 (0.52)             & 0.42 (0.47)              & 0.49 (0.50)               \\
4 & 0.67 (0.86)             & 0.70 (0.80)              & 0.77 (0.83)               \\
6 & 0.84 (1.05)             & 0.87 (0.97)              & 0.96 (1.02)               \\
7 & 1.08 (1.36)            & 1.13 (1.27)              & 1.24 (1.34) \\
\bottomrule
\end{tabular}
\caption{Comparison of excitation energies for Co(II)(sp)Cl$_2$ calculated with Orca on SA-CASSCF/NEVPT2 level with different active space size. All energy values are given in eV.}
\label{tab:casscf_diff_cas}
\end{table}

\begin{table}[]
\begin{tabular}{c|c|cc|c|cc}
 \toprule
          & BP86 & adiab  & vert  & BHandH & adiab  & vert  \\ \hline
 & 1             & 0.16 (0.25) & 0.34 (0.42) & 1          & 0.16 (0.26) & 0.32 (0.40) \\
Co(sp)Cl$_2$          & 2             & 0.43 (0.56) & 0.37 (0.48) & 2          & 0.26 (0.44) & 0.37 (0.48) \\
          & 3             & 0.29 (0.39) & 0.42 (0.53) & 3          & 0.19 (0.38) & 0.41 (0.51) \\ \hline
 & BP86 & adiab  & vert  & BHandH & adiab  & vert  \\ \hline
        & 1             & 0.30 (0.38) & 0.30 (0.39) & 1          & 0.12 (0.24) & 0.31 (0.39) \\
 Ni(sp)Cl$_2$           & 2             & 0.77 (1.04) & 0.47 (0.62) & 2          & 0.32 (0.62) & 0.49 (0.65) \\
          & 3             & 0.55 (0.86) & 0.70 (0.92) & 3          & 0.40 (0.81) & 0.75 (0.99) \\
\bottomrule
\end{tabular}
\caption{Comparison of adiabatic (adiab) and vertical (vert) excitation energies for Co(II)(sp)Cl$_2$ and Ni(II)(sp)Cl$_2$ calculated with SA-CASSCF and NEVPT2 corrections (given in parenthesis).
Adiabatic energies are calculated with SA-CASSCF/NEVPT2 on optimized ground and excited state geometries from BP86-D3BJ (BP86 in table) and BHandH functionals. All energy values are given in eV.}
\end{table}

~\newpage

\subsection{GGA functionals} 

\begin{table}[H]
\caption{Excitation energies of Co((II)sp)Cl$_2$ complex in eV calculated with GGA DFAs (see \href{https://www.scm.com/doc/ADF/Input/Density_Functional.html}{Density Functional on ADF website}). The calculations were performed with ADF~\cite{AMSpaper} (2019 version) for the lowest 10 electronic transitions on BP86/TZP preoptimized geometry using TZP basis sets.} 
\label{table:1_GGA_exci_energy}
\centering
\begin{tabular}{L{0.500cm} C{2.400cm} C{2.000cm} C{2.000cm} C{2.000cm} C{2.000cm} C{2.000cm} }
 \toprule

\#. & BP86  & PBE &  mPBE  & revPBE & OPBE  &  PW91     \\
 \midrule
 1 & 1.086  & 1.048 &  1.079  &  1.133& 1.033  &  1.069\\ 
 2 & 1.108  & 1.066 &  1.097  &  1.148& 1.043  &  1.090 \\
 3 & 1.187  & 1.146 &  1.176  &  1.230& 1.127  &  1.169 \\
 4 & 1.626  & 1.584 &  1.614  &  1.667& 1.548  &  1.608 \\
 5 & 1.778  & 1.729 &  1.758  &  1.811& 1.681  &  1.755 \\
 6 & 1.998  & 1.952 &  1.982  &  2.036& 1.910  &  1.975 \\
 7 & 2.864  & 2.879 &  2.896  &  2.926& 2.995  &  2.856 \\
 8 & 3.002  & 3.017 &  3.035  &  3.067& 3.128  &  2.994 \\
 9 & 3.148  & 3.161 &  3.177  &  3.208& 3.260  &  3.140 \\
10 & 3.179  & 3.185 &  3.205  &  3.240& 3.268  &  3.168 \\

\bottomrule \\

\toprule
\#. & mPw  & BLYP &  OLYP  & LB94 &  KT1 & KT2     \\
 \midrule
 1 & 1.118  & 1.074 &  1.022  & 0.384 &  1.496 &  1.644    \\
 2 & 1.140  & 1.100 &  1.028  & 0.463 &  1.541 &  1.685   \\
 3 & 1.219  & 1.171 &  1.106  & 0.502 &  1.631 &  1.772 \\
 4 & 1.658  & 1.611 &  1.529  & 0.885 &  2.083 &  2.225 \\
 5 & 1.807  & 1.753 &  1.654  & 0.972 &  2.257 &  2.390 \\
 6 & 2.027  & 1.975 &  1.885  & 1.142 &  2.460 &  2.594 \\
 7 & 2.869  & 2.805 &  2.915  & 1.885 &  3.079 &  3.139 \\
 8 & 3.009  & 2.948 &  2.927  & 1.964 &  3.228 &  3.296 \\
 9 & 3.154  & 3.086 &  3.054  & 2.151 &  3.386 &  3.441 \\
10 & 3.188  & 3.123 &  3.073  & 2.174 &  3.443 &  3.522 \\
 \bottomrule
\end{tabular}
\end{table}

\newpage
\subsection{Global Hybrid and range-separated functionals}

\begin{table}[H]
\caption{Excitation energies of Co(sp)Cl$_2$ complex in eV calculated with hybrid and range-separated DFAs (see \href{https://www.scm.com/doc/ADF/Input/Density_Functional.html}{Density Functional on ADF website}). The calculations were performed with ADF~\cite{AMSpaper} (2019 version) for the lowest 10 electronic transitions on the BP86/TZP optimized geometry using TZP basis sets.} 
\label{table:1_GGA_exci_energy}
\centering
\begin{tabular}{L{0.500cm} C{2.400cm} C{2.000cm} C{2.000cm} C{2.000cm} C{2.000cm} C{2.000cm} }
 \toprule

\#. & PBE0  & OPBE0 &  B3LYP  & O3LYP & X3LYP  &  BHandH     \\
 \midrule
 1 &  0.937 &  0.895& 0.959   &  0.485& 0.939  & 0.258 \\ 
 2 &  1.015 &  0.978& 1.030   &  0.553& 1.011  & 0.334  \\
 3 &  1.141 &  1.107& 1.134   &  0.605& 1.125  & 0.514  \\
 4 &  1.593 &  1.540& 1.576   &  1.050& 1.567  & 1.144  \\
 5 &  1.719 &  1.690& 1.688   &  1.156& 1.682  & 1.301  \\
 6 &  2.103 &  2.057& 2.060   &  1.467& 2.061  & 1.707  \\
 7 &  4.486 &  4.647& 3.967   &  3.230& 4.090  & 6.155  \\
 8 &  4.707 &  4.864& 4.188   &  3.415& 4.312  & 6.236  \\
 9 &  4.717 &  4.881& 4.197   &  3.479& 4.319  & 6.358  \\
10 &  4.765 &  4.915& 4.259   &  3.494& 4.380  & 6.381  \\
\bottomrule \\

\toprule

\#. & BHandHLYP  & B1PW91 &  mPW1PW  & CAMY-B3LYP & WB97X  &  CAM-B3LYP    \\
 \midrule
 1 & 0.710  & 0.982 &  1.005  & 0.939 & 0.785  & 0.902     \\
 2 & 0.773  & 1.061 &  1.084  & 1.019 & 0.868  & 0.986    \\
 3 & 0.948  & 1.187 &  1.209  & 1.130 & 0.977  & 1.099  \\
 4 & 1.604  & 1.640 &  1.664  & 1.572 & 1.403  & 1.537  \\
 5 & 1.744  & 1.768 &  1.790  & 1.685 & 1.506  & 1.648  \\
 6 & 2.158  & 2.152 &  2.175  & 2.061 & 1.876  & 2.025  \\
 7 & 6.379  & 4.484 &  4.476  & 4.427 & 4.882  & 4.639  \\
 8 & 6.457  & 4.708 &  4.700  & 4.636 & 5.048  & 4.837  \\
 9 & 6.583  & 4.714 &  4.707  & 4.678 & 5.167  & 4.901  \\
10 & 6.607  & 4.765 &  4.759  & 4.721 & 5.275  & 4.936  \\
 \bottomrule
\end{tabular}
\end{table}

~\newpage
\section{Enhanced VCD calculations}

\subsection{B3LYP functional}

\begin{figure}[H]
    \centering
    \begin{subfigure}[t]{0.45\textwidth}
        \centering
        \includegraphics[width=\textwidth]{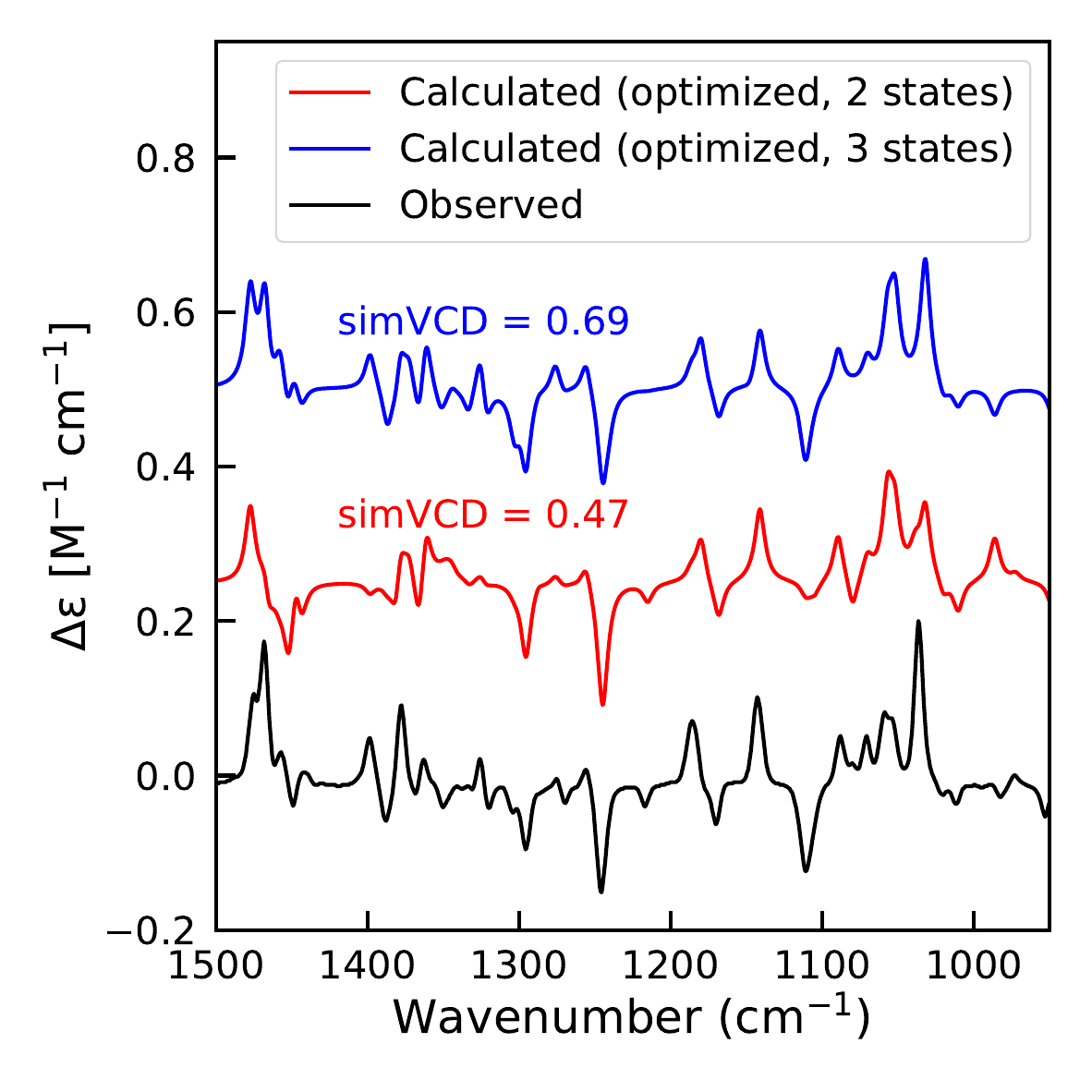}
    \end{subfigure}
        \begin{subfigure}[t]{0.45\textwidth}
        \centering
        \includegraphics[width=\textwidth]{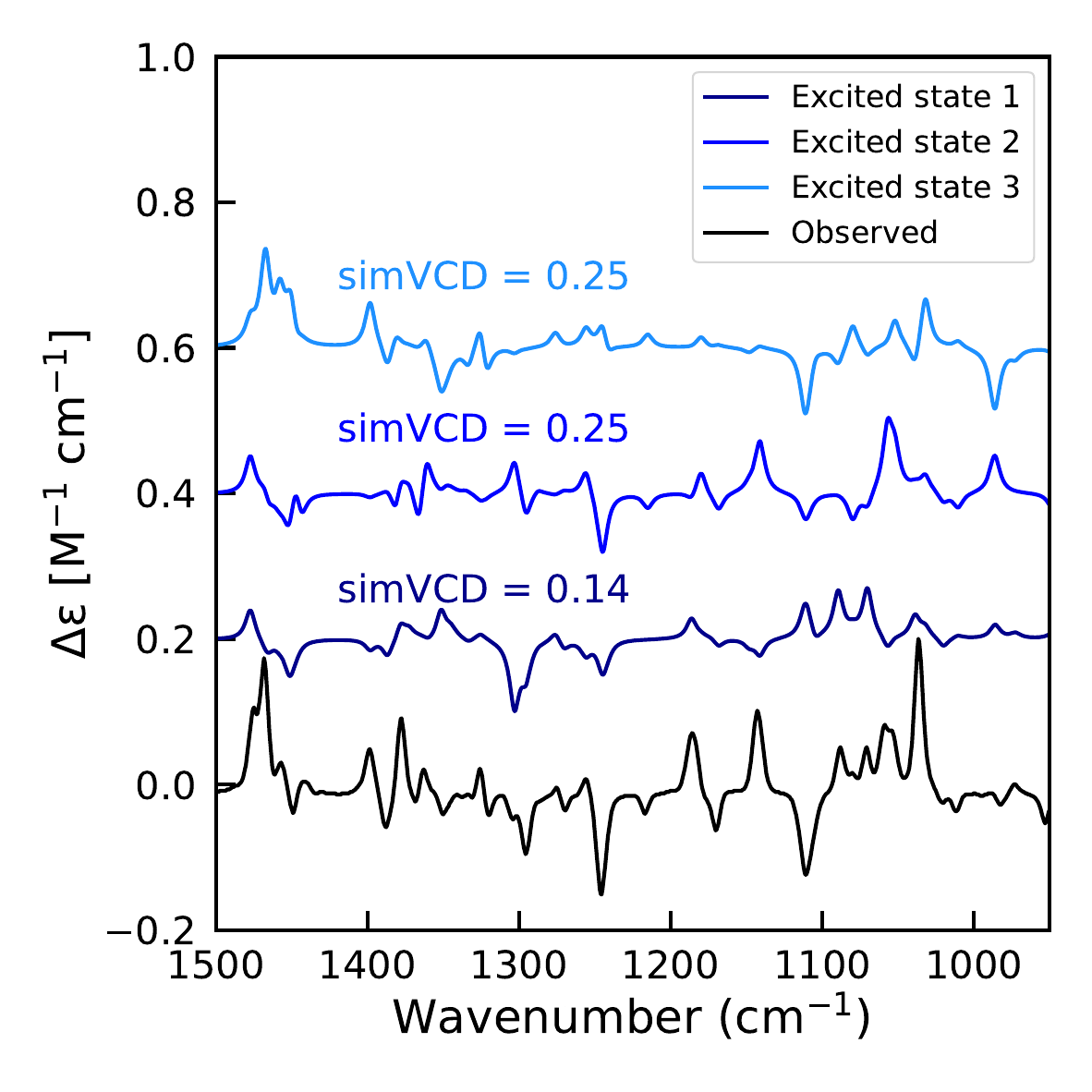}
    \end{subfigure}
    \caption{Enhanced VCD spectra of Co(sp)Cl$_2$ calculated with the B3LYP functional including two and three excited states (left). Individual contributions of the three excited states to the enhancement for Co(sp)Cl$_2$ from B3LYP calculation (right).}
    \label{fig:enh_b3lyp_co}
\end{figure}

\begin{figure}[H]
    \centering
    \begin{subfigure}{0.45\textwidth}
        \centering
        \includegraphics[width=\textwidth]{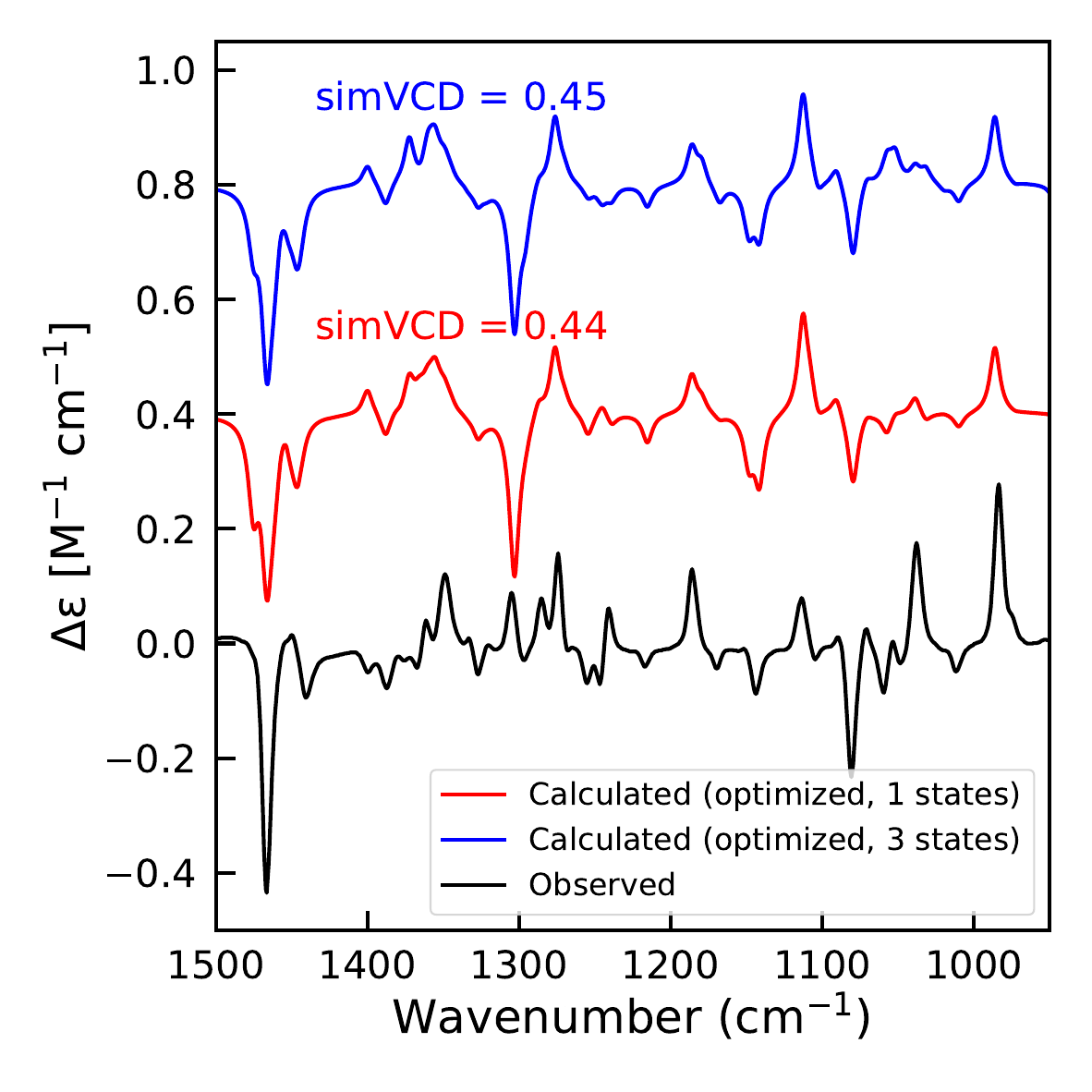}
    \end{subfigure}
    \label{fig:enh_b3lyp_ni}
    \caption{Enhanced VCD spectra of Ni(sp)Cl$_2$ calculated with the B3LYP functional including one (red) and three (blue) excited states.}
\end{figure}

\subsection{BP86 functional}

\begin{figure}[H]
    \centering
    \begin{subfigure}[t]{0.4\textwidth}
        \centering
            \includegraphics[width=\textwidth]{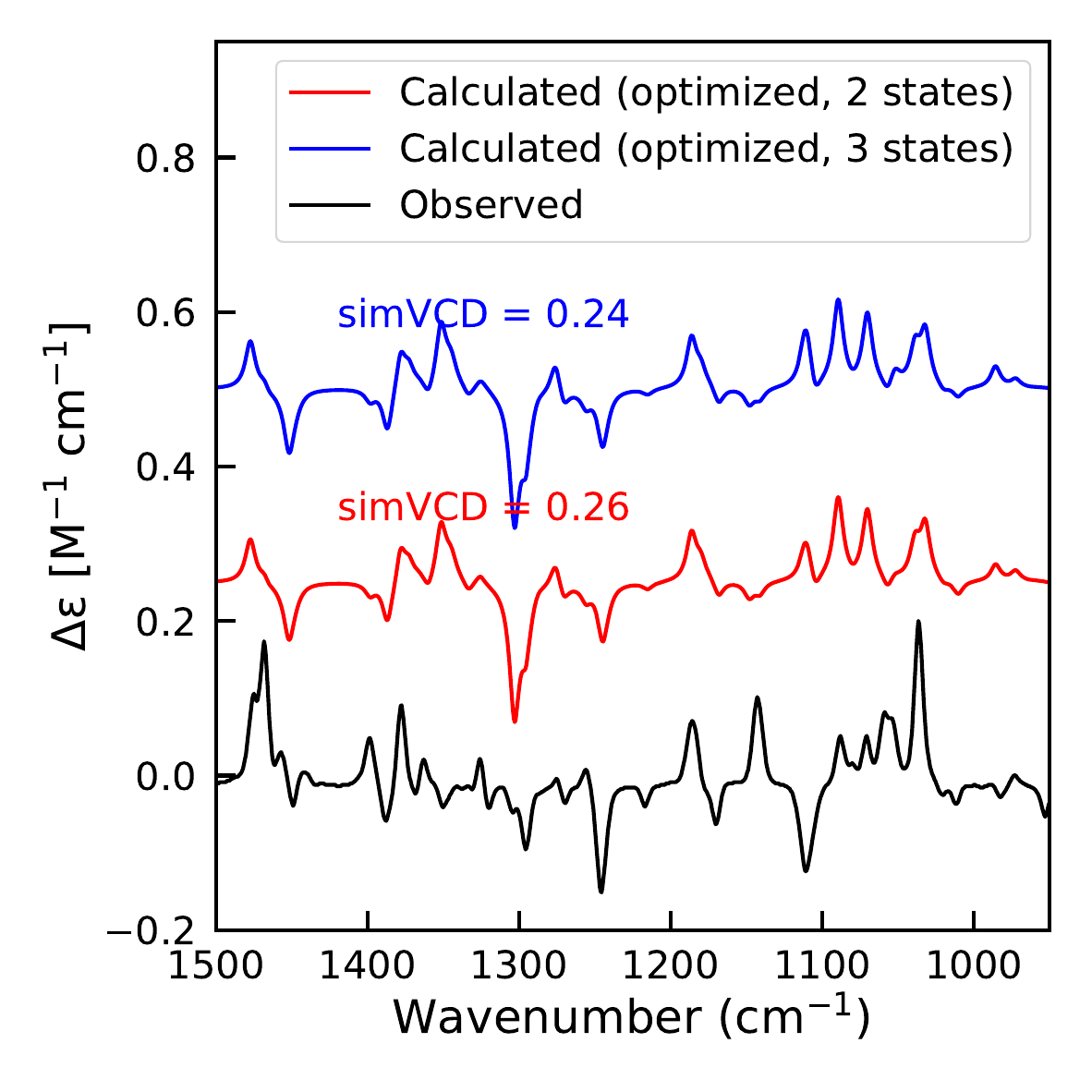}
    \end{subfigure}
        \begin{subfigure}[t]{0.4\textwidth}
        \centering
        \includegraphics[width=\textwidth]{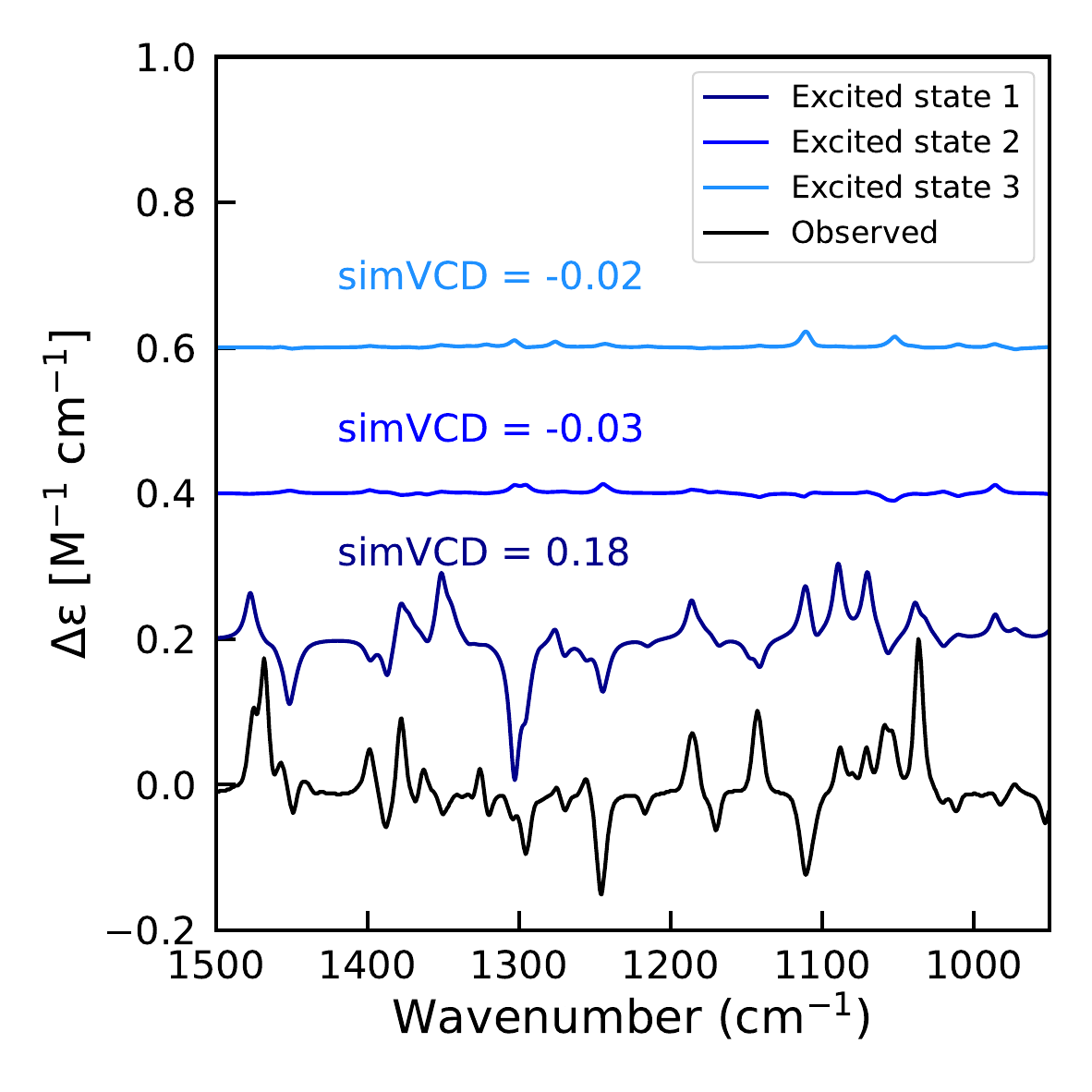}
    \end{subfigure}
    \caption{Enhanced VCD spectra of Co(sp)Cl$_2$ calculated with the BP86 functional. The excitation energies of the three states after optimization are 0.20, 0.37 and 0.37 eV. The first state energy is similar to the SA-CASSCF result, however, since there is no degeneracy with the second state, in fact only the first state is responsible for the enhancement effect. Here the 1200-1100 cm$^{-1}$ is significantly different from the observed spectra, and the fine structure is missing in 1500-1300 cm$^{-1}$, which overall sums into simVCD smaller that with BHandH and B3LYP.}
    \label{fig:enh_bp86_co}
\end{figure}

\begin{figure}[h]
    \centering
    \begin{subfigure}[t]{0.4\textwidth}
        \centering
            \includegraphics[width=\textwidth]{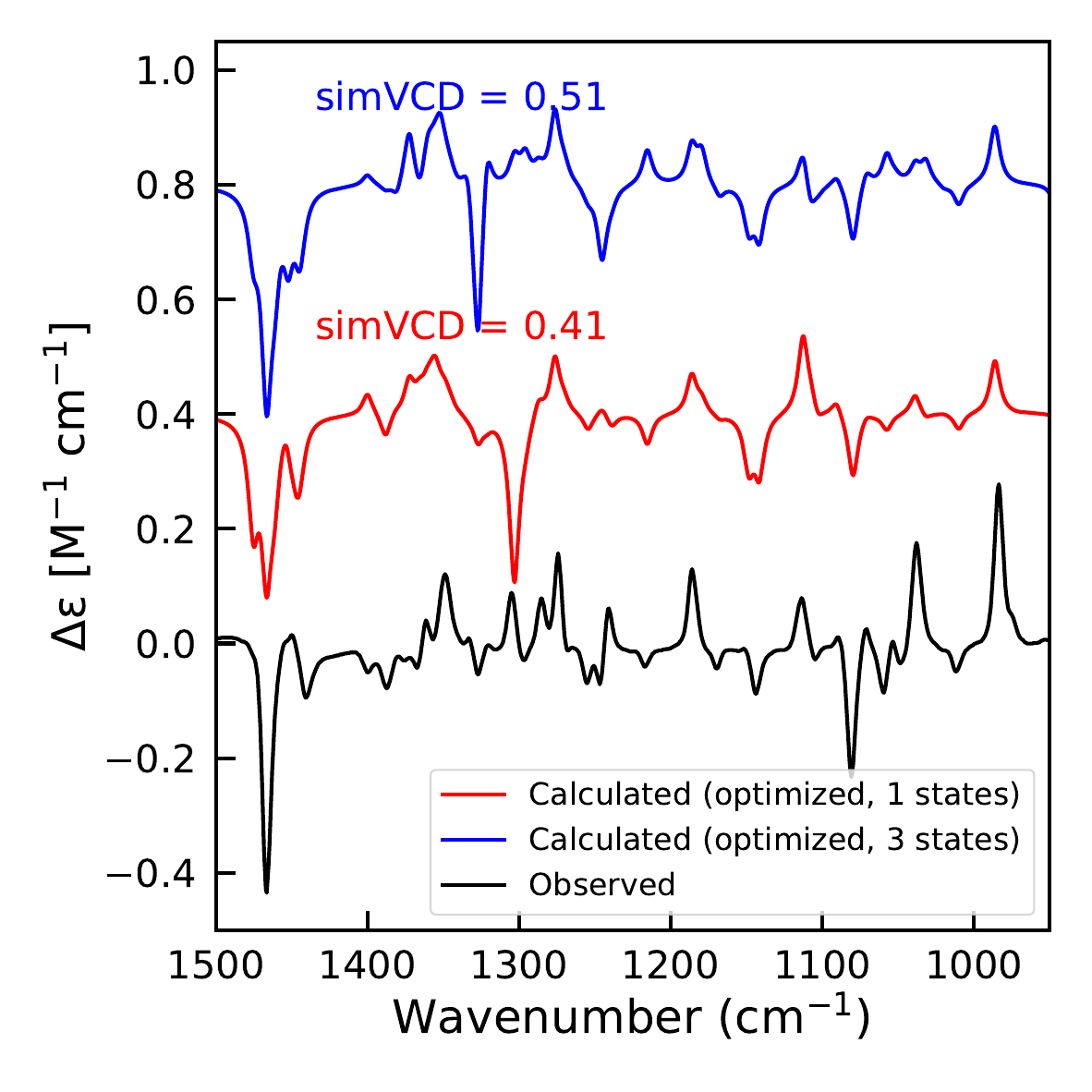}
    \end{subfigure}
        \begin{subfigure}[t]{0.4\textwidth}
        \centering
        \includegraphics[width=\textwidth]{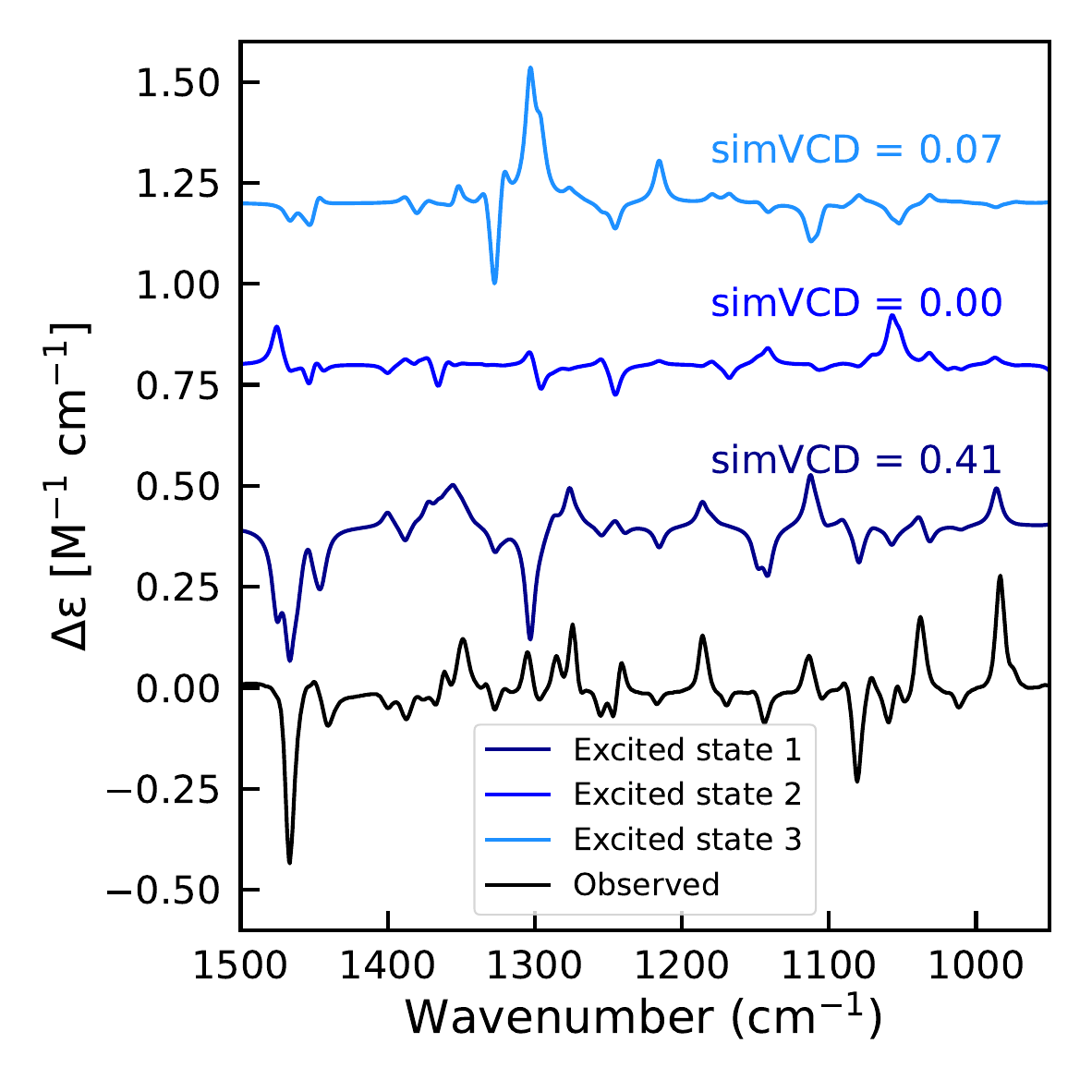}
    \end{subfigure}
    \caption{Enhanced VCD spectra of Ni(sp)Cl$_2$ calculated with BP86 functional. The excitation energies of the three states after optimization are 0.20, 0.22 and 0.16 eV. The optimization procedure leads to significant lowering of the third excited state, which contradicts the SA-CASSCF result. The improved simVCD value comes from a significant positive peak in the third state at 1300 cm$^{-1}$, which compensates the negative peak from the first excited state.}
    \label{fig:enh_bp86_ni}
\end{figure}

\newpage

\section{Non-adiabatic couplings}
\subsection{Co(sp)Cl$_2$}

\begin{figure}[H]
    \centering
    \begin{subfigure}{0.45\textwidth}
        \centering
        \includegraphics[width=\textwidth]{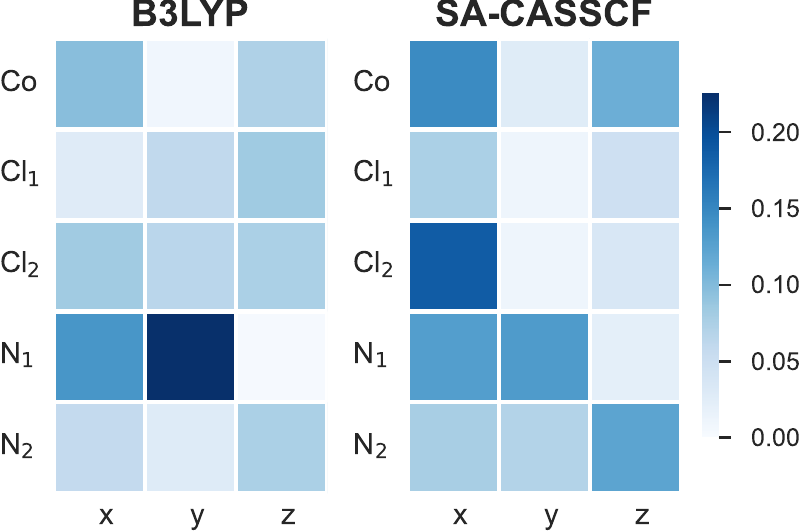}
        \subcaption{NACs between the ground and first excited state}
    \end{subfigure}
    \begin{subfigure}{0.45\textwidth}
        \centering
        \includegraphics[width=\textwidth]{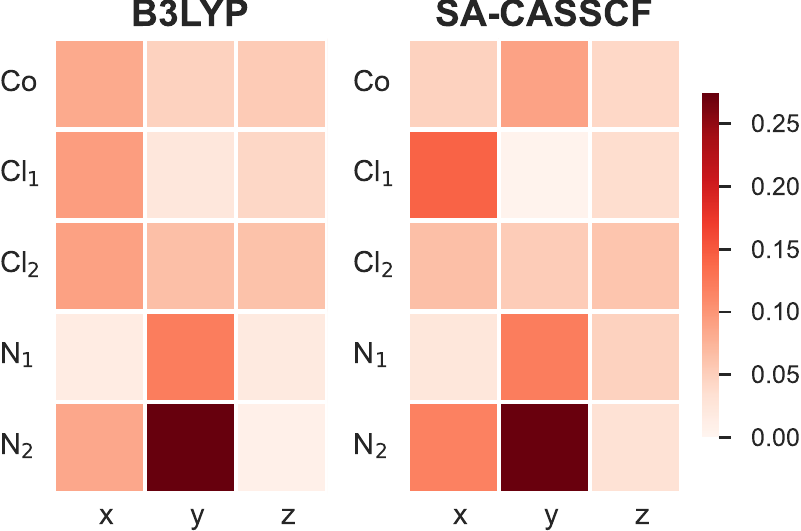}
        \subcaption{NACs between the ground and second excited state}
    \end{subfigure}
        \begin{subfigure}{0.45\textwidth}
        \centering
        \includegraphics[width=\textwidth]{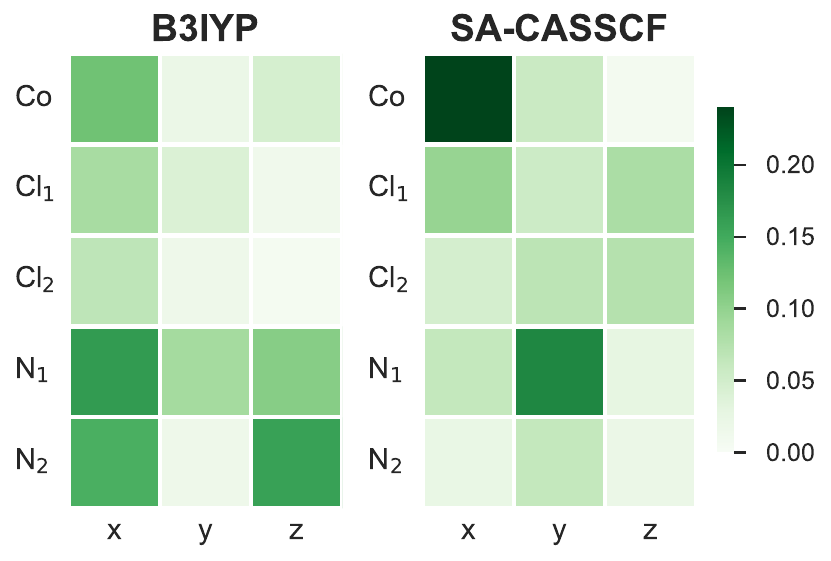}
        \subcaption{NACs between the ground and third excited state}
    \end{subfigure}
    \caption{Non-adiabatic couplings between the ground and first three excited states of Co(sp)Cl$_2$ calculated with B3LYP and SA-CASSCF. }
    \label{fig:nacv_b3lyp_co}
\end{figure}

\begin{figure}[H]
    \centering
    \begin{subfigure}{0.45\textwidth}
        \centering
        \includegraphics[width=\textwidth]{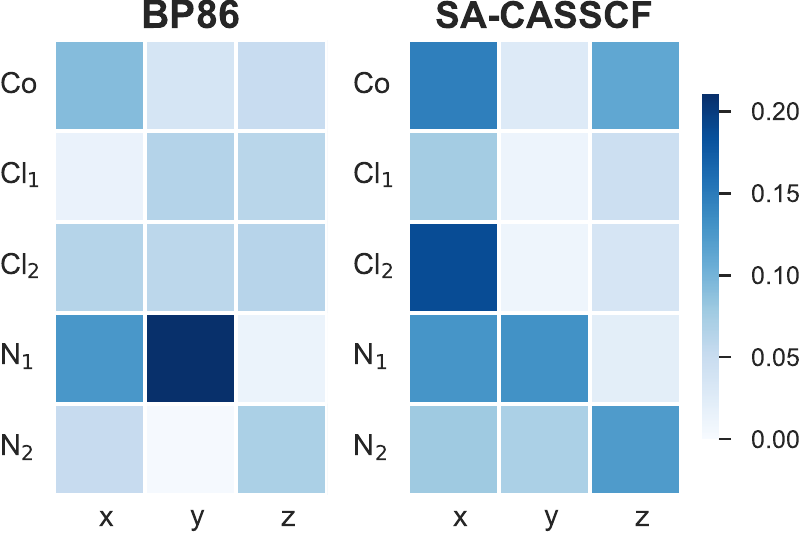}
        \subcaption{NACs between the ground and first excited state}
    \end{subfigure}
    \begin{subfigure}{0.45\textwidth}
        \centering
        \includegraphics[width=\textwidth]{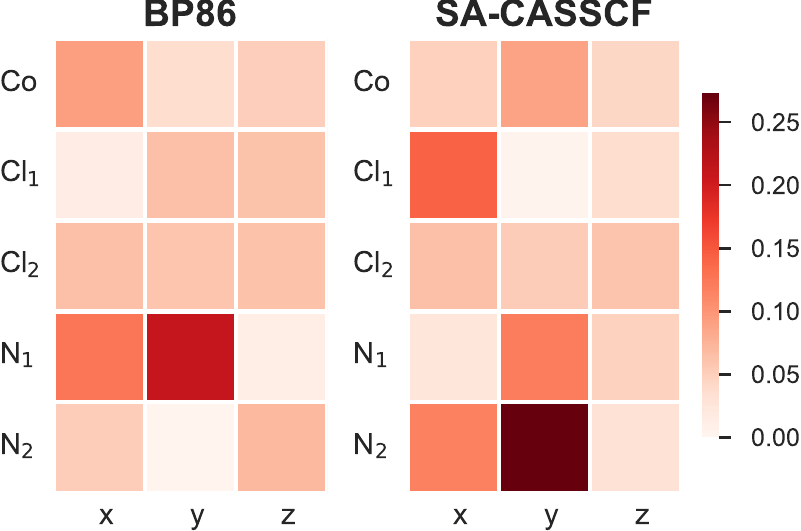}
        \subcaption{NACs between the ground and second excited state}
    \end{subfigure}
        \begin{subfigure}{0.45\textwidth}
        \centering
        \includegraphics[width=\textwidth]{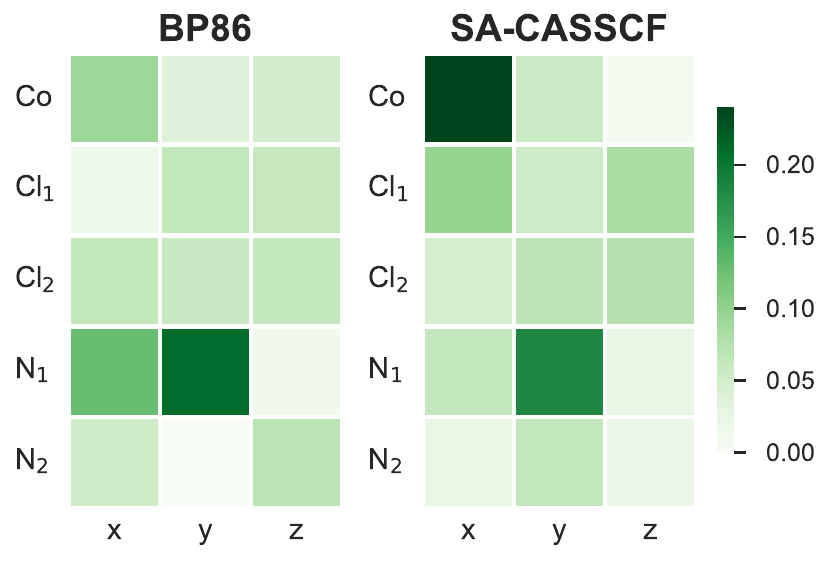}
        \subcaption{NACs between the ground and third excited state}
    \end{subfigure}
    \caption{Non-adiabatic couplings between the ground and first three excited states of Co(sp)Cl$_2$ calculated with BP86 and SA-CASSCF.}
    \label{fig:nacv_bp86_co}
\end{figure}

\begin{figure}[H]
    \centering
    \begin{subfigure}{0.45\textwidth}
        \centering
        \includegraphics[width=\textwidth]{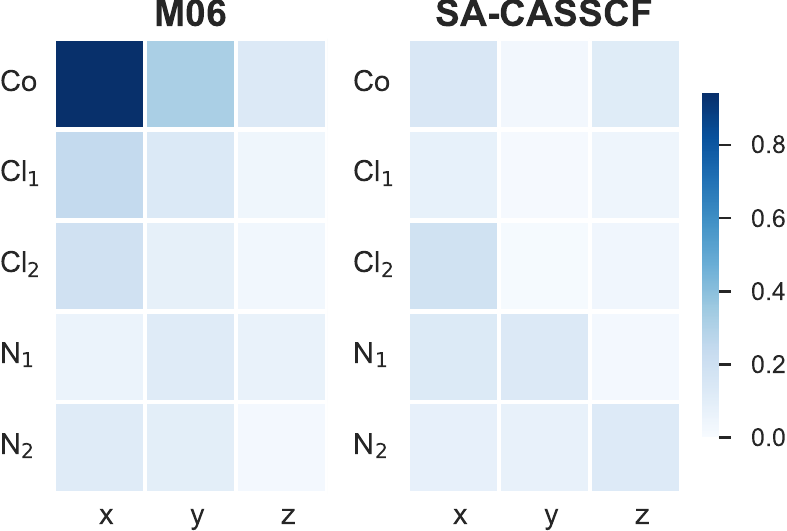}
        \subcaption{NACs between the ground and first excited state}
    \end{subfigure}
    \begin{subfigure}{0.45\textwidth}
        \centering
        \includegraphics[width=\textwidth]{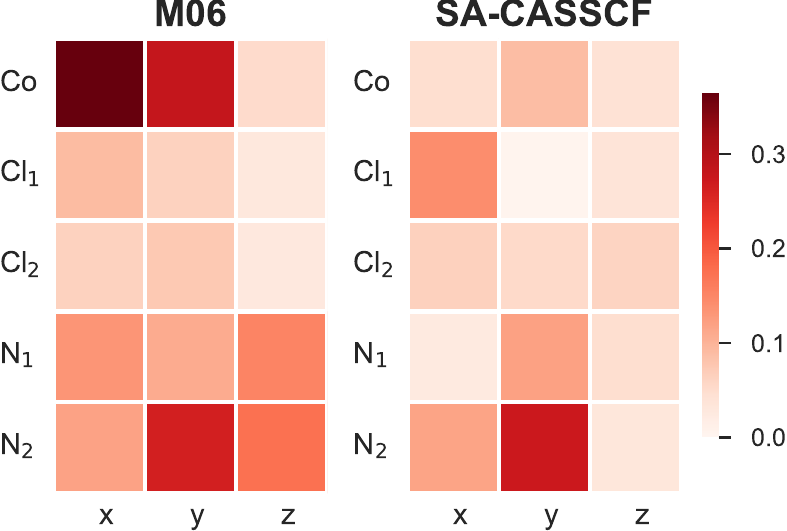}
        \subcaption{NACs between the ground and second excited state}
    \end{subfigure}
        \begin{subfigure}{0.45\textwidth}
        \centering
        \includegraphics[width=\textwidth]{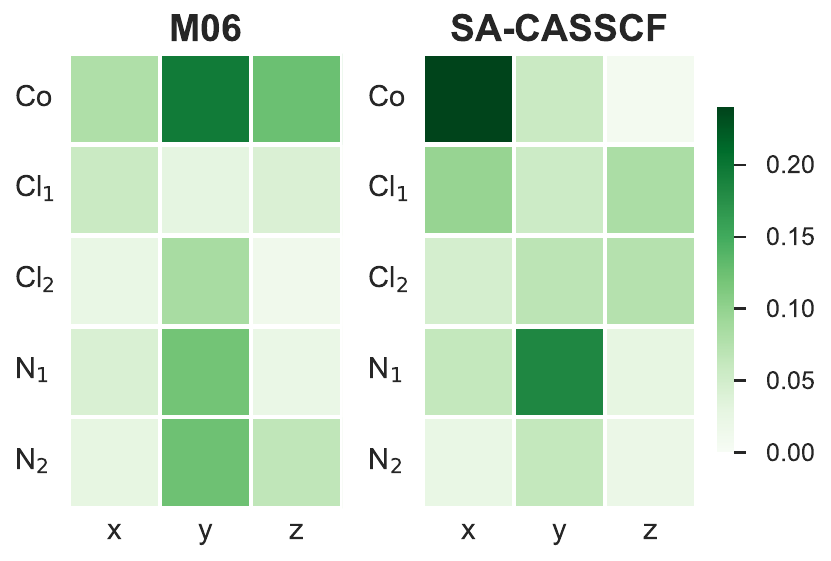}
        \subcaption{NACs between the ground and third excited state}
    \end{subfigure}
    \caption{Non-adiabatic couplings between the ground and first three excited states of Co(sp)Cl$_2$ calculated with M06 and SA-CASSCF.}
    \label{fig:nacv_m06_co}
\end{figure}

\subsection{Ni(sp)Cl$_2$}

\begin{figure}[H]
    \centering
    \begin{subfigure}{0.45\textwidth}
        \centering
        \includegraphics[width=\textwidth]{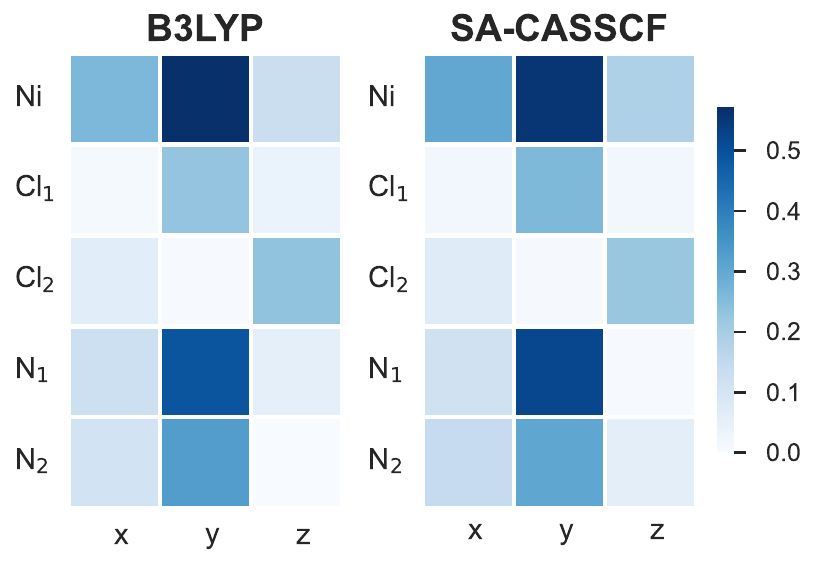}
        \subcaption{NACs between the ground and first excited state}
    \end{subfigure}
    \begin{subfigure}{0.45\textwidth}
        \centering
        \includegraphics[width=\textwidth]{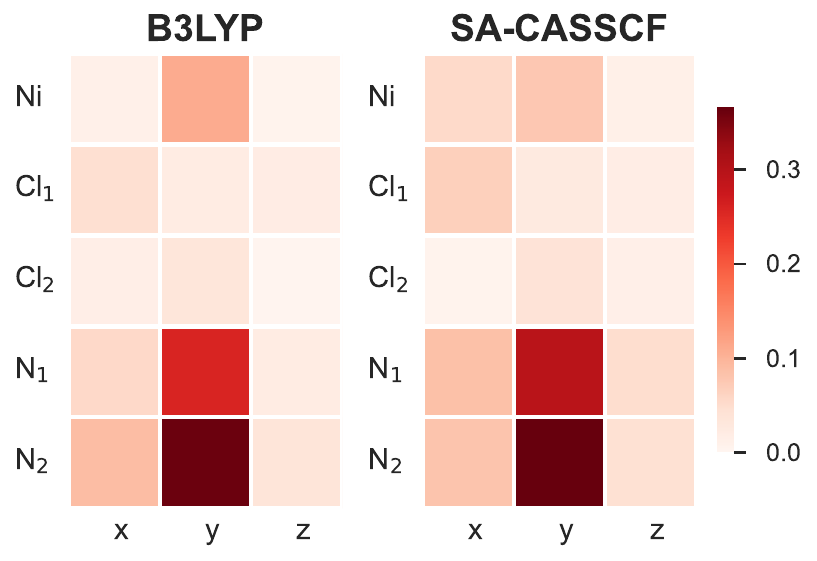}
        \subcaption{NACs between the ground and second excited state}
    \end{subfigure}
        \begin{subfigure}{0.45\textwidth}
        \centering
        \includegraphics[width=\textwidth]{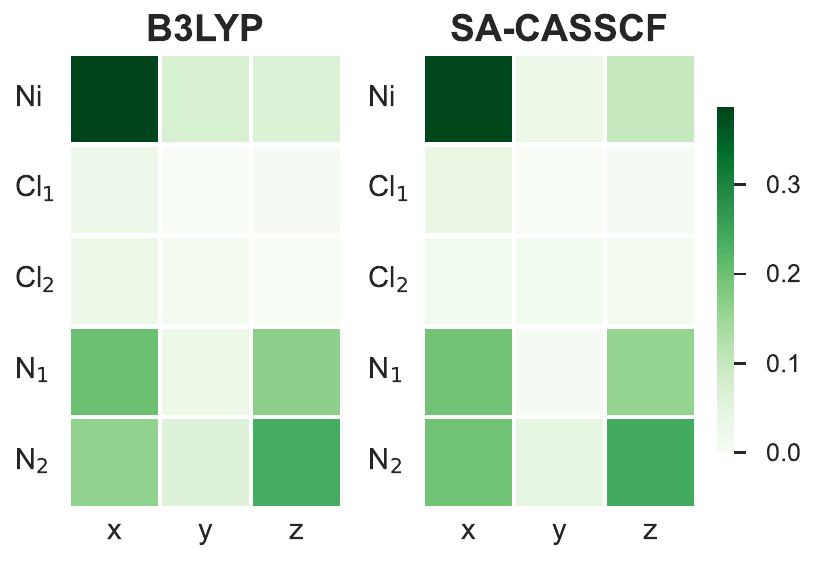}
        \subcaption{NACs between the ground and third excited state}
    \end{subfigure}
    \caption{Non-adiabatic couplings between the ground and first three excited states of Ni(sp)Cl$_2$ calculated with B3LYP and SA-CASSCF.}
    \label{fig:nacv_b3lyp_ni}
\end{figure}

\begin{figure}[H]
    \centering
    \begin{subfigure}{0.45\textwidth}
        \centering
        \includegraphics[width=\textwidth]{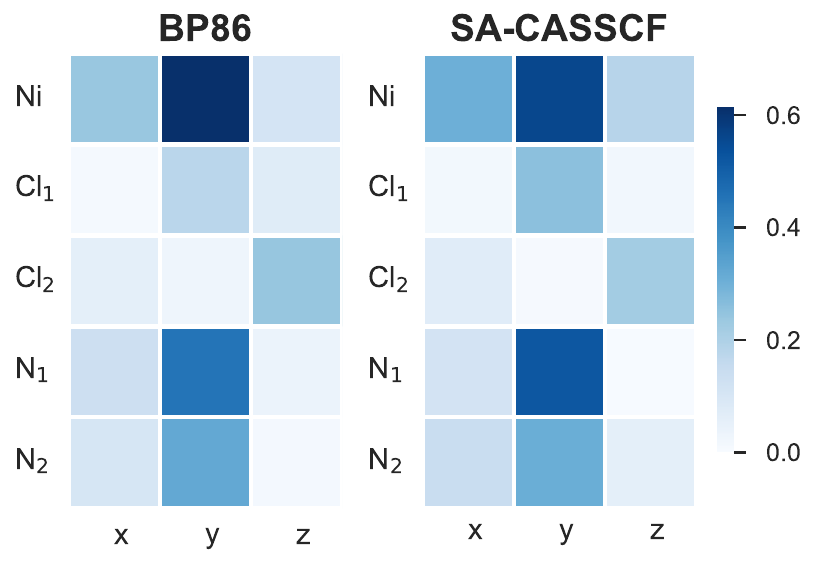}
        \subcaption{NACs between ground and first excited state}
    \end{subfigure}
    \begin{subfigure}{0.45\textwidth}
        \centering
        \includegraphics[width=\textwidth]{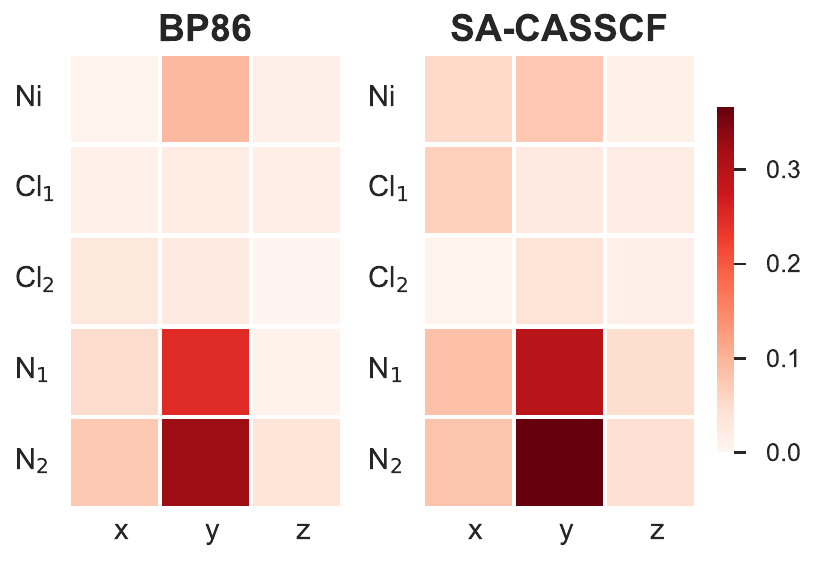}
        \subcaption{NACs between the ground and second excited state}
    \end{subfigure}
        \begin{subfigure}{0.45\textwidth}
        \centering
        \includegraphics[width=\textwidth]{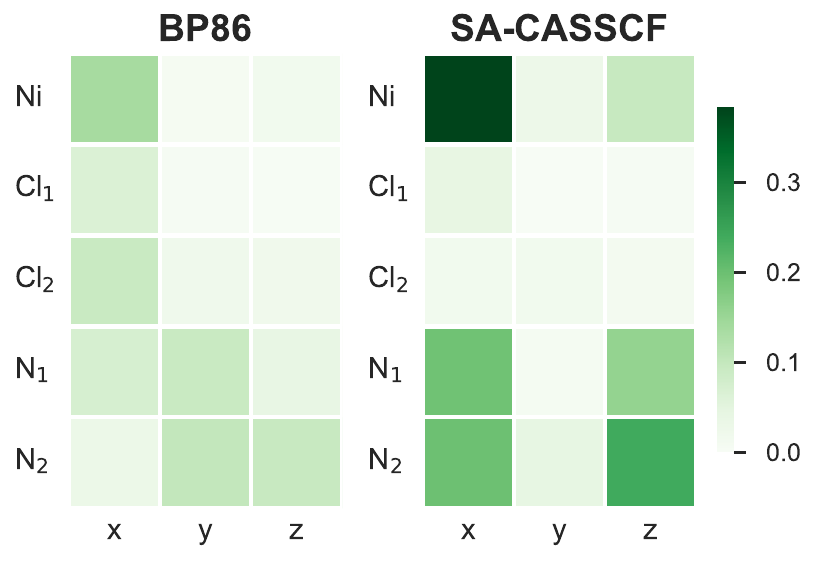}
        \subcaption{NACs between the ground and third excited state}
    \end{subfigure}
    \caption{Non-adiabatic couplings between ground and first three excited states of Ni(sp)Cl$_2$ calculated with BP86 and SA-CASSCF.}
    \label{fig:nacv_bp86_ni}
\end{figure}

\newpage

\subsection{NAC analysis}

REF: Reference value is calculated with SA-CASCCF/ANO-RCC-VTZP (ANO-RCC-MB for C and H atoms)\\ 
MAE: Mean absolute error  \\
MAX: Maximum absolute error  \\
RMSE: Root-mean-square error \\
L: Largest element of the reference NACs for the $n$th excitation \\

\begin{table}[]
 \begin{tabular}{C{0.600cm}|  C{1.800cm}  C{1.800cm} C{1.800cm}| C{1.800cm} C{1.800cm} C{1.800cm} | C{1.800cm}}
    &         & BHandH  &         &         & B3LYP   &         & REF   \\
\#. & MAE     & MAX     & RMSE    & MAE     & MAX     & RMSE    & L     \\ \hline
1   & 0.0075 & 0.127 & 0.0182 & 0.0085 & 0.105 & 0.0185 & 0.187 \\
2   & 0.0035 & 0.047 & 0.0092 & 0.0038 & 0.049 & 0.0089 & 0.272 \\
3   & 0.0095 & 0.151 & 0.0264 & 0.0101 & 0.138 & 0.0265 & 0.239 \\ \hline
    &         & BP86    &         &         & M06     &         & Ref   \\
\#. & MAE     & MAX     & RMSE    & MAE     & MAX     & RMSE    & L     \\
1   & 0.0087 & 0.123 & 0.0198 & 0.0184 & 0.792 & 0.0756 & 0.187 \\
2   & 0.0102 & 0.269 & 0.0309 & 0.0108 & 0.315 & 0.0375 & 0.272 \\
3   & 0.0074 & 0.147 & 0.0189 & 0.0081 & 0.160 & 0.0240 & 0.239
\end{tabular}
\caption{Comparison of non-adiabatic couplings between the ground and first three excited states of Co(sp)Cl$_2$ calculated with TDDFT and SA-CASSCF.}
\end{table}

\begin{table}[]
 \begin{tabular}{C{0.600cm}|  C{1.800cm}  C{1.800cm} C{1.800cm}| C{1.800cm} C{1.800cm} C{1.800cm} | C{1.800cm}}
    &         & BHandH  &         &         & B3LYP   &         & REF   \\
\#. & MAE     & MAX     & RMSE    & MAE     & MAX     & RMSE    & L     \\ \hline
1   & 0.0054 & 0.065 & 0.0126 & 0.0051 & 0.058 & 0.0113 & 0.556 \\
2   & 0.0023 & 0.043 & 0.0060 & 0.0031 & 0.042 & 0.0075 & 0.364 \\
3   & 0.0027 & 0.036 & 0.0071 & 0.0028 & 0.043 & 0.0069 & 0.381 \\ \hline
    &         & BP86    &         &    -     &     -    &     -    & REF   \\
\#. & MAE     & MAX     & RMSE    &     -    &     -    &     -    & L     \\
1   & 0.0073 & 0.081 & 0.0164 &     -    &     -    &     -    & 0.556 \\
2   & 0.0042 & 0.055 & 0.0103 &     -    &     -    &     -    & 0.364 \\
3   & 0.0108 & 0.245 & 0.0347 &     -    &     -    &     -    & 0.381
\end{tabular}
\caption{Comparison of non-adiabatic couplings between the ground and first three excited states of Ni(sp)Cl$_2$ calculated with TDDFT and SA-CASSCF.}
\end{table}

\newpage

\noindent \textbf{Co(sp)Cl$_2$ xyz coordinates optimized BP86-D3BJ-CPCM/def2-TZVP with Orca} \\
\\
46 \\
Coordinates from ORCA-job ./1547094 E -3000.823867165991 \\
  Co          6.99491453281983      2.24406221066966      4.02630250572301 \\
  Cl          6.70330783203854      0.29400169709201      5.08860781728658 \\
  Cl          7.28427211379117      3.93778459369101      5.45458497553286 \\
  N           8.24010279292044      1.89949072769322      2.42310173040006\\
  N           5.43130762330139      2.68527761702432      2.77649028758049\\
  C           9.59073750876111      1.46472620779643      2.88040007767439\\
  C          10.30409641634177      2.56372948955086      3.65580591234117\\
  C          10.41065919317730      3.84083004341960      2.81813409137016\\
  C           9.03099154449875      4.25997785472419      2.30159332955260\\
  C           8.34849295161739      3.11805698078779      1.55347478974527\\
  C           6.98210210742848      3.48611212648167      0.94320256314631\\
  C           6.51398948029109      2.32706054061533      0.06237132877358\\
  C           6.27787853565226      1.13981052424501      1.00045740941212\\
  C           7.60621433458457      0.77124078290542      1.67338627439194\\
  C           5.12817851066320      1.44347467077859      1.97997291889873\\
  C           3.76829162931718      1.50879295036679      1.26503282157835\\
  C           2.62986813668139      1.85651701959809      2.22593310638952\\
  C           2.94856108618489      3.16414956746127      2.95199492540423\\
  C           4.29142131964243      3.04865252030990      3.66996187820806\\
  C           5.86691717652113      3.84279581518216      1.93933659779788\\
  H           9.44743583103578      0.56682090317574      3.49698102891343\\
  H          10.18080855801033      1.17752772470641      1.98812803790420\\
  H          11.30037746469425      2.19754315369848      3.94420014010886\\
  H           9.75064291118618      2.77268922542419      4.58535480861138\\
  H          11.08050473888631      3.65934009510743      1.95986146624239\\
  H          10.85732070580217      4.65258191168023      3.41033504355401\\
  H           9.11642760270600      5.11187566489389      1.60991937149900\\
  H           8.40600987895263      4.58499067443670      3.14768515682741\\
  H           9.00168889180989      2.83552699443736      0.70086745678337\\
  H           7.15207454261539      4.38633958327442      0.33275837935557\\
  H           7.28733659538781      2.07526849667293     -0.67889798541524\\
  H           5.60428149741039      2.59380437461121     -0.49212985383906\\
  H           5.96999705070729      0.25317420411060      0.42507748917769\\
  H           7.47686341804145     -0.05589476407808      2.38481209279967\\
  H           8.31246323224256      0.44213718085385      0.88823432209172\\
  H           5.07901291336170      0.63536735639448      2.72861993441078\\
  H           3.59658505568220      0.53105505815216      0.78973739984077\\
  H           3.79575990840834      2.25479089046548      0.45628943686213\\
  H           2.50740090377449      1.04730244487819      2.96627991495633\\
  H           1.68065091780832      1.93548364129135      1.67568982671273\\
  H           2.17749561757374      3.40401331404449      3.69907312283527\\
  H           2.96707493522768      3.99933958743695      2.23334310146731\\
  H           4.22951563387586      2.25729601944403      4.43379888823564\\
  H           4.56865564251477      3.98016024267931      4.18040951858740\\
  H           5.01382989443260      4.23846473766678      1.36352237746082\\
  H           6.18467183161722      4.63078434414798      2.63658818280902\\

\newpage

\noindent \textbf{Ni(sp)Cl$_2$ xyz coordinates optimized BP86-D3BJ-CPCM/def2-TZVP with Orca} \\
\\
46 \\
Coordinates from ORCA-job ./1575556 E -3126.391533153851\\
  Ni          7.00931485390759      2.28716670001971      4.03639137631787\\
  Cl          6.71655246651106      0.21645203267263      4.87082115106370\\
  Cl          7.19253537463040      4.12869457248019      5.30175142403506\\
  N           8.20857916716811      1.89843429345315      2.42596679904731\\
  N           5.45828402580587      2.67870469047847      2.77204097859377\\
  C           9.54509560849001      1.46294259012908      2.92626177573253\\
  C          10.25471138073396      2.56768248600956      3.69629333596127\\
  C          10.40224969590729      3.82339083892752      2.83336039150064\\
  C           9.04202969822863      4.25647817164119      2.27911947576048\\
  C           8.34359943906593      3.11149220120820      1.55158516729836\\
  C           6.98419970117315      3.48966939682434      0.93344220754402\\
  C           6.51141945097433      2.33581716871398      0.04757110497657\\
  C           6.27963964758585      1.14296553683758      0.98018912043956\\
  C           7.60546429298595      0.77455759434411      1.65166407653581\\
  C           5.14393981813882      1.44249593194074      1.97401802622676\\
  C           3.77588008637925      1.51907618400015      1.27461392402065\\
  C           2.64953881633852      1.84953914219374      2.25461129279115\\
  C           2.97747008738055      3.14617498012383      2.99429135242037\\
  C           4.33241205785945      3.02782123981649      3.68975612628662\\
  C           5.87262545339284      3.84044266484799      1.93379548214214\\
  H           9.37788656973382      0.57608095753627      3.55186961568862\\
  H          10.15122511593420      1.15745464178756      2.05153125794144\\
  H          11.23765716247431      2.19098313631359      4.01530357683689\\
  H           9.68647407253876      2.80473736161278      4.61035836212603\\
  H          11.08771465845713      3.60918334870016      1.99513067402002\\
  H          10.85133555593929      4.64028266476372      3.41657263936425\\
  H           9.15947956623061      5.08249986759869      1.56134470152929\\
  H           8.40890472775527      4.62108054007908      3.10089019879428\\
  H           8.99562088668988      2.80394995875470      0.70700371356922\\
  H           7.15795821894673      4.39248788065211      0.32815735788046\\
  H           7.28072076114341      2.08785057878167     -0.69911846243744\\
  H           5.59900348717027      2.60710842025188     -0.50048883269864\\
  H           5.96670702166156      0.25983257272563      0.40226600137984\\
  H           7.48084985512036     -0.06478866460493      2.34686117741141\\
  H           8.32607399341021      0.47694712664715      0.86725036409570\\
  H           5.10245981050645      0.63314893269764      2.71925574585225\\
  H           3.60186296895349      0.54766201200440      0.78742908764617\\
  H           3.79227705242977      2.27608626865677      0.47595101415557\\
  H           2.53865364591505      1.02913429628648      2.98415403159892\\
  H           1.69265779920500      1.93596278227651      1.71882997682828\\
  H           2.22104246039780      3.37238325280134      3.76037232353575\\
  H           2.98046648910183      3.99332026705759      2.28961714426672\\
  H           4.28744682968110      2.22608387797475      4.44343971831607\\
  H           4.61411941604114      3.95428306930081      4.20494589095729\\
  H           5.00920499329402      4.21891076430716      1.36184521135294\\
  H           6.18384675861024      4.63166266837362      2.62836692129382\\

\newpage

\noindent \textbf{Zn(sp)Cl$_2$ xyz coordinates optimized BP86-D3BJ-CPCM/def2-TZVP with Orca} \\
\\
46\\
Coordinates from ORCA-job ./1547012 E -3397.471564112006\\
  Zn          0.20951328801406     -0.10890216115077      0.97077548287767\\
  Cl         -0.14247522293437     -2.11489760718642      1.95174339162007\\
  Cl          0.47917075142205      1.64711068412979      2.35030112310752\\
  N           1.45479917802831     -0.48083470109146     -0.69437334812239\\
  N          -1.35773968050643      0.33719834879067     -0.36430632301056\\
  C           2.80978122216301     -0.91659427426723     -0.25022670251140\\
  C           3.54217460338996      0.18811111671449      0.49958829145693\\
  C           3.64048775779211      1.45750349869013     -0.35100821770260\\
  C           2.25621465960435      1.87106039784801     -0.85994151280445\\
  C           1.56798775303315      0.72031562894234     -1.58748385304569\\
  C           0.20608657072001      1.08546918047574     -2.20813115370356\\
  C          -0.26699139389147     -0.08864929888836     -3.06568295918427\\
  C          -0.51482623899871     -1.25277822685190     -2.10200392240186\\
  C           0.81087289615103     -1.61739969009043     -1.42117698245451\\
  C          -1.66737419327122     -0.91813884410517     -1.13517343145588\\
  C          -3.02017277400515     -0.86054701116619     -1.86312742502324\\
  C          -4.16464927214199     -0.47500293841755     -0.92396959646040\\
  C          -3.83877149733939      0.84876452021432     -0.23028972270451\\
  C          -2.50667021591834      0.73480564831673      0.50485172721466\\
  C          -0.90954882513805      1.47241303698411     -1.22467117852359\\
  H           2.67267640002061     -1.80531943050993      0.38149805481660\\
  H           3.38368206596333     -1.22045132360512     -1.14709512555511\\
  H           4.54151109560267     -0.17917933173145      0.77552982150241\\
  H           3.00860593887382      0.40861605041102      1.43827077860731\\
  H           4.30307264644813      1.26869021556324     -1.21313427581076\\
  H           4.09234033889193      2.27432752704832      0.23021999109589\\
  H           2.33541687123461      2.71662246952931     -1.55992969768040\\
  H           1.63669684563984      2.20445878562502     -0.01290358390265\\
  H           2.22043728514235      0.41773420144555     -2.43409341576822\\
  H           0.38600209224315      1.97096022311173     -2.83726798517272\\
  H           0.50765278505069     -0.36201272272309     -3.79806205070105\\
  H          -1.17247309276757      0.17231235151488     -3.62972703066922\\
  H          -0.82785828354550     -2.14962407023703     -2.65872761623782\\
  H           0.67421018106860     -2.42917053687522     -0.69350901390853\\
  H           1.51286460982651     -1.96764998902467     -2.20098610294294\\
  H          -1.72662122675468     -1.71107319651494     -0.37062977206633\\
  H          -3.19824997327448     -1.84925437925902     -2.31301208589092\\
  H          -2.97948779484096     -0.13821414398489     -2.69279214879391\\
  H          -4.30253744910006     -1.26196250149924     -0.16243311434213\\
  H          -5.10798674559790     -0.40204847035557     -1.48506636208841\\
  H          -4.61570713606069      1.11709559475840      0.50097955596981\\
  H          -3.80443287939292      1.66378242212526     -0.97110980395163\\
  H          -2.59119727392156     -0.03664189848063      1.28729522465077\\
  H          -2.22417524637777      1.67434361103609      0.99837417991745\\
  H          -1.75524543084545      1.86572748903347     -1.81230710480081\\
  H          -0.58395125579960      2.27184900610761     -0.54387507684456\\

\newpage
\providecommand{\latin}[1]{#1}
\makeatletter
\providecommand{\doi}
  {\begingroup\let\do\@makeother\dospecials
  \catcode`\{=1 \catcode`\}=2 \doi@aux}
\providecommand{\doi@aux}[1]{\endgroup\texttt{#1}}
\makeatother
\providecommand*\mcitethebibliography{\thebibliography}
\csname @ifundefined\endcsname{endmcitethebibliography}  {\let\endmcitethebibliography\endthebibliography}{}